\documentclass[rmp,preprint,aps,nofootinbib.endfloats,11pt]{revtex4}

\def\Tr{\text{Tr}}

\def\fig#1#2{\includegraphics[height=#1]{#2}}
\def\figx#1#2{\includegraphics[width=#1]{#2}}

\newcommand{\s}{\sigma}
\newcommand{\si}{\sigma}
\newcommand{\dg}{^{\dagger}}

\newcommand{\up}{\uparrow}

\newcommand{\la}{\langle}
\newcommand{\ra}{\rangle}

\newcommand{\rarrow}{\rightarrow }

\newlength{\bxwidth}
\newcommand\frm[1]{\epsfig{file=#1,width=\bxwidth}}
\newcommand \renp{\renewcommand{\multirowsetup}{\centering}\multirow}

\newlength{\spear}

\newlength{\fight}
\newcommand{\fg}[3]
{\begin{figure}[tb]
\resizebox{\fight}{!} {\includegraphics{#1}}
\caption{#2}\label{#3}\end{figure}}
\newcommand{\fgb}[3]
{\begin{figure}[b]
\resizebox{\fight}{!}
 {\includegraphics{#1}}
\caption{#2}\label{#3}\end{figure}}
\newcommand{\fgz}[3]
{\begin{figure}[here]
\resizebox{\fight}{!}
{\includegraphics{#1}}
\caption{#2}\label{#3}\end{figure}}
\def\gtappr{{{\lower4pt\hbox{$>$} } \atop \widetilde{ \ \ \ }}}
\def\ltappr{{{\lower4pt\hbox{$<$} } \atop \widetilde{ \ \ \ }}}

\newcommand \narrowboxit[1]{\noindent\marginpar{\mbox{}}
\begin{center}\fbox{\rule{0pt}{\baselineskip}
\parbox{0.5\textwidth}{#1}
}\end{center}\vskip 0.1in}
\newcommand \boxit[1]{\vskip 0.1truein\noindent\marginpar{\mbox{}}
\fbox{\rule{0pt}{\baselineskip}\parbox{0.96\textwidth}{#1}}\vskip 0.1in}
\newlength{\upit}\upit=0.1truein
\newcommand{\raiser}[1]{\raisebox{\upit}[0cm][0cm]{#1}}
\newcommand\frmup[1]{\raiser{\epsfig{file=#1,width=\bxwidth}}}
\newlength{\wx}
\newlength{\wy}
\newlength{\wz}
\usepackage{amsmath}
\usepackage{graphicx}
\usepackage{epsfig}
\usepackage{multirow}
\begin{document}
\newcommand{\vk}{\vec{k}}
\newcommand{\bp}{{\bf p}}
\newcommand{\dw}{\downarrow}
\newcommand{\mat}[1]{\left(\begin{matrix}  #1 \end{matrix} \right)}
\newcommand{\bea}{\begin{eqnarray}}
\newcommand{\eea}{\end{eqnarray}}
\newcommand{\defn}[1]{\textbf{\textit{#1}}}
\bxwidth=1.5 truein
\newcommand{\om}{\omega}
\newcommand{\cg}{{\cal G}}
\newcommand{\dif}[2]{\frac{\delta #1}{\delta #2}}
\newcommand{\ddif}[2]{\frac{\partial #1}{\partial #2}}
\newcommand{\Dif}[2]{\frac{d #1}{d #2}}
\newcommand{\str}{\hbox{Str}}
\newcommand{\Str}{\underline{\hbox{Str}}}
\newcommand{\tr}{{\hbox{Tr}}}
\newcommand{\bk}{{\bf k}}
\newcommand{\bq}{{\bf q}}
\newcommand{\bx}{{\bf x}}
\newcommand{\bR}{{\bf R}}
\newcommand{\bQ}{{\bf Q}}
\newcommand{\vp}{\bf{p}}
\newcommand{\al}{\alpha}
\def\Sfi{{\bf S}_{i}}
\def\Sfj{{\bf S}_{j}}
\def\fig#1#2{\includegraphics[height=#1]{#2}}
\def\figx#1#2{\includegraphics[width=#1]{#2}}
\newlength{\figwidth}
\figwidth=10cm
\newlength{\shift}
\shift=-0.2cm
%\psdraft

\newcommand{\dsp} {\displaystyle}
\setlength{\unitlength}{1mm}
\newcommand{\nmat}[4]{\left[{
\hbox{${
{{\dsp \raisebox{4pt}{$#1$}
 \above1pt \dsp  \raisebox{-4pt}{$#3$} }}}$}
\left| 
{\bf
{{\dsp \raisebox{0pt}{$#2$}
 \above1pt \dsp\raisebox{-0pt}{$#4$} }}}
\right.
} \right]}
%\centerline{\Large\bf 5 }
%\vskip 0.1truein

%%%%%%%%%%%%%%%%%%%%%%%%%%%%%%%%%%%%%%%%%%%%
%% FRONTMATTER
%%%%%%%%%%%%%%%%%%%%%%%%%%%%%%%%%%%%%%%%%%%%

\title{Heavy Fermions: electrons at the edge of magnetism}

\author{P. Coleman{$^1$}}

\affiliation{
$^1$
Center for Materials Theory,
Rutgers University, Piscataway, NJ 08855, U.S.A. }

\vskip 0.1truein
\begin{abstract}
{An introduction to the physics of heavy fermion compounds
is presented, highlighting the conceptual developments and emphasizing 
the mysteries and open questions that persist in 
this active field of research.
\vskip 0.5truein
\noindent {\sl \small
This article is a contribution to volume 1 of the Handbook of Magnetism and
Advanced Magnetic Materials, edited by Helmut Kr\" onmuller and Stuart
Parkin, to be published by John Wiley and Sons, Ltd.
\vskip 1truein
{{\bf  keywords:}  Heavy Fermion, Superconductivity, Local Moments,
Kondo effect, Quantum Criticality. }
}
}
\end{abstract}

%\tableofcontents
\maketitle

\vskip 0.2truein
\tableofcontents

\section{Introduction: ``Asymptotic Freedom''  in  a Cryostat.}\label{}

%% intro3.tex

The term ``heavy fermion '' was  coined by  Steglich, Aarts et al \citep{steglich}
in the late seventies to describe the electronic excitations in a new  class of 
inter-metallic compound 
with an electronic density of states 
as much as 1000 times larger than copper. 
Since the original discovery of heavy fermion behavior in
$CeAl_3$ by  Andres, Graebner and Ott \citep{ott}, 
a diversity of heavy fermion compounds, including
superconductors, antiferromagnets and 
insulators have been discovered. 
In the last ten years, these materials have become the focus of intense 
interest with the discovery
that  inter-metallic 
antiferromagnets can 
be tuned through a quantum phase transition
into a heavy fermion state by pressure, magnetic fields
or chemical doping  \citep{hilbert,hilbert2,mathur}.  The ``quantum critical point'' 
that separates the heavy electron ground state from the antiferromagnet
represents a kind of singularity in the material phase diagram that
profoundly modifies the metallic properties, giving them a 
a pre-disposition towards superconductivity and other
novel states of matter. 

One of the goals of modern condensed matter research is 
to couple magnetic and electronic properties
to develop new classes of material behavior, such as high temperature 
superconductivity or colossal magneto-resistance materials,
spintronics, and the newly discovered multi-ferroic materials. 
Heavy electron materials lie at the very brink of magnetic
instability, in a regime where quantum fluctuations of the magnetic and electronic degrees 
are strongly coupled 
As such, they are an
important test-bed for the development of our understanding about the
interaction between magnetic and electronic quantum fluctuations. 

Heavy fermion materials contain  rare earth or actinide ions
forming  a matrix of localized magnetic moments. 
The active physics of these materials results from the immersion of
these magnetic moments in a quantum sea of mobile conduction electrons.
In most rare
earth metals and insulators, local moments tend to 
order antiferromagnetically, but in heavy electron metals, the 
quantum mechanical jiggling of the local moments induced by 
delocalized electrons is fierce enough to melt the magnetic order.

\fight=0.8\textwidth
%%\fgb{newfigs/heavyfermion.eps}
\fgb{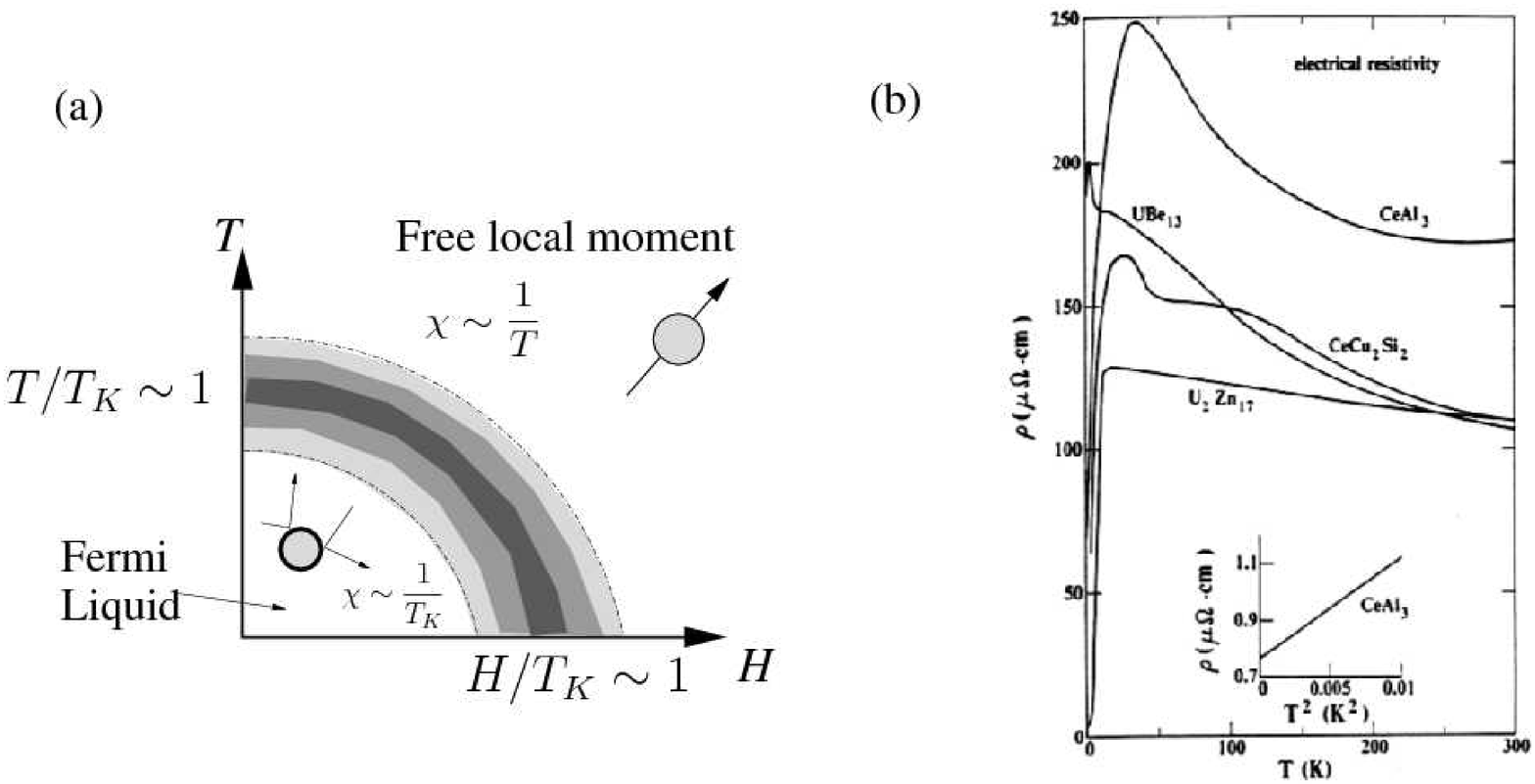}
{
(a) In the Kondo effect, local moments are free at high temperatures
and high fields, but become ``screened'' at temperatures and magnetic
fields that are small compared with the ``Kondo temperature'' $T_{K}$ forming
resonant scattering centers for the electron fluid. The magnetic
susceptibility $\chi $ changes from a Curie law $\chi \sim
\frac{1}{T}$ at high temperature, but saturates at a constant
paramagnetic value $\chi \sim \frac{1}{T_{K}}$ at low temperatures and fields.
(b)The resistivity drops dramatically  at low temperatures in heavy
fermion materials, indicating the development of phase coherence
between the scattering off the lattice of screened magnetic ions.
(After  \citep{rise})}{fig1}

The mechanism by which 
this takes place involves a remarkable piece 
of quantum physics
called the ``Kondo effect'' \citep{kondo,kondo2,jones}. The Kondo
effect describes the process by which a free magnetic ion,
with a Curie magnetic susceptibility at high temperatures,
becomes screened by the spins of the conduction sea, to ultimately
form a spinless scattering center at low temperatures and low magnetic fields.
(Fig. \ref{fig1} a.). In the Kondo effect this screening process is
continuous, and takes place once the magnetic field, or the
temperature drops below a characteristic energy scale called the Kondo
temperature $T_{K}$. Such ``quenched'' magnetic moments act as
strong elastic scattering potentials for electrons, which gives rise
an increase in resistivity produced by isolated magnetic ions.
When the same process takes place inside a heavy electron material, it
leads to a spin quenching at every site in the lattice, but now, the
strong scattering at each site develops coherence, leading to a sudden
drop in the resistivity at low temperatures. (Fig \ref{fig1} (b)).

Heavy electron materials involve the dense lattice analog of the single ion Kondo
effect  and are often called ``Kondo lattice'' compounds  \citep{doniach}. 
In the lattice, the Kondo effect may be alternatively 
visualized as the dissolution of localized, and neutral
magnetic f spins into the quantum 
conduction sea, where they become mobile excitations.
Once mobile, these free spins acquire charge and form  electrons
with a radically enhanced effective mass (Fig. \ref{fig2}). 
The net effect of this process, is an increase in
the volume of the electronic Fermi surface, 
accompanied by a profound transformation
in the electronic masses and interactions. 

\fight=0.8\textwidth
%\fgz{figures/kondolat.eps}
\fgz{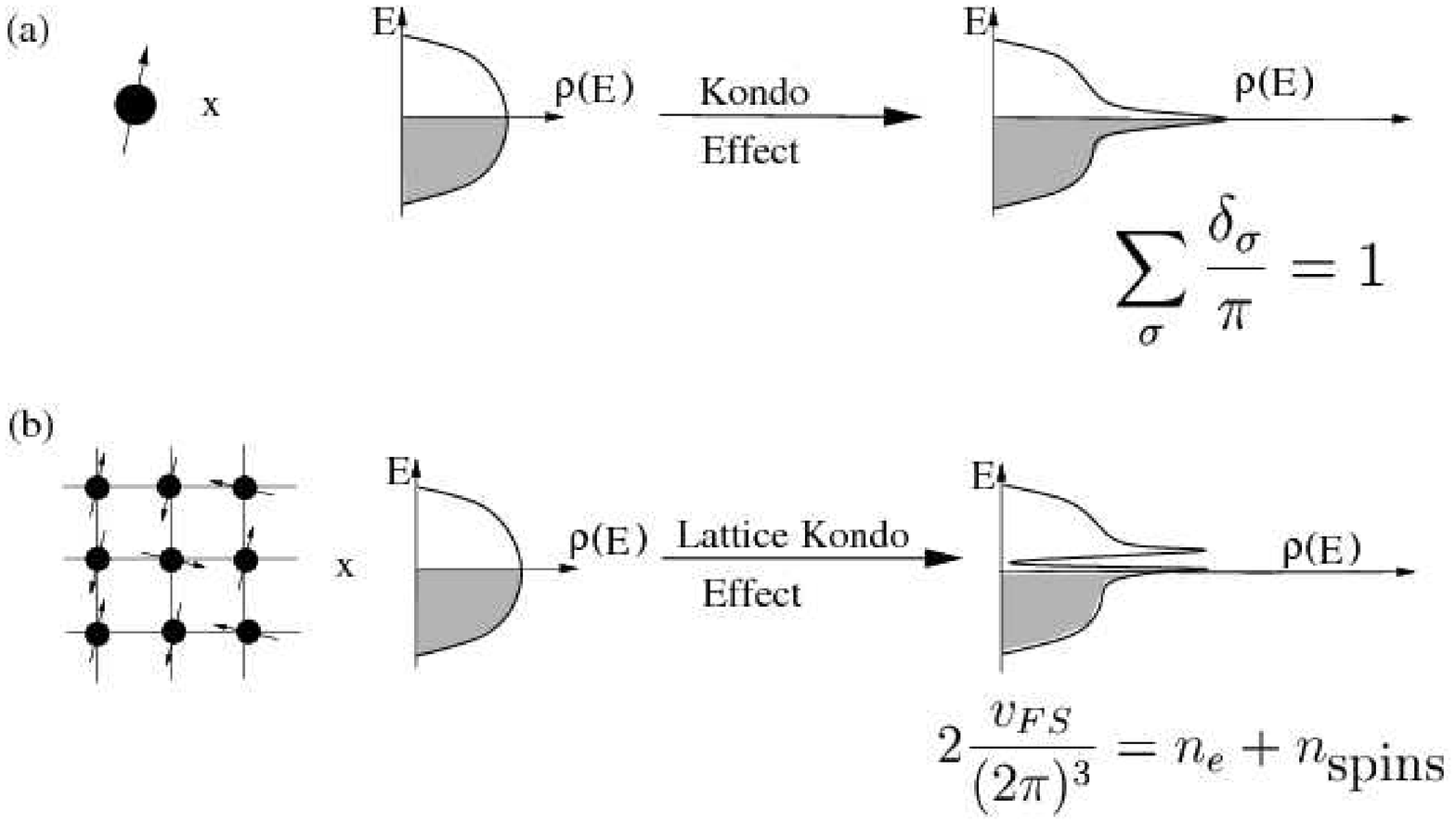}{(a) Single impurity Kondo effect
builds a single fermionic level into the conduction sea, which
gives rise to a resonance in the conduction electron density of states (b)
Lattice Kondo effect builds a fermionic resonance into the conduction
sea in each unit cell. The elastic scattering off this lattice of
resonances
leads to formation of a heavy electron band, of width $T_{K}$. 
}{fig2}

A classic example of such behavior is provided by the inter-metallic
crystal $CeCu_{6}$. Superficially, this material is copper, alloyed 
with 14\% Cerium.  The Cerium $Ce^{3+}$ ions in this
material are $Ce^{3+}$ ions in a  $4f^{1}$ configuration
with a localized magnetic moment with $J=5/2$. 
Yet at low temperatures
they lose their spin, behaving as if they were $Ce^{4+}$ ions with
delocalized f-electrons. 
The heavy electrons that develop in this material are
a  thousand times ``heavier'' than those 
in metallic copper, and move with a group velocity that is slower
than sound.
Unlike copper, 
which has  Fermi temperature of order 10,000K, that of $CeCu_{6}$
is of order $10K$, and above this temperature, 
the heavy electrons disintegrate to reveal the underlying magnetic
moments of the Cerium ions, which manifest themselves as 
a Curie law susceptibility $\chi \sim \frac{1}{T}$.  There are many hundreds
of different variety of heavy electron material, 
many developing new
and exotic phases at low temperatures.  

This chapter is intended as a perspective on the 
the current theoretical and experimental understanding of heavy electron materials.
There are important links between
the material in this chapter, and the proceeding article on the Kondo
effect by  Jones \citep{jones}, the chapter on quantum criticality by
Sachdev  \citep{subirchapter} and the perspective on spin
fluctuation theories of high temperature superconductivity by Norman \citep{normanchapter}.
For completeness, I have included references
to an extensive list of review articles spanning thirty years of
discovery,
% \citep{revhewson,revcox,revhf1,revhf2,revhf3,revhf4,revhf5,revearly1,revearly2,revwilson,revnozieres,revbethekondo,revbethekondo2,revhfsc1,revhfsc2,revfiskaeppli,revki1,revki2,allenrev,revoptical,revqcp1,revqcp2,revvarma2001,revqcp3,revqcp4},
including books on the Kondo effect and heavy fermions \citep{revhewson,revcox}, 
general reviews on heavy fermion physics  \citep{revhf1,revhf2,revhf3,revhf4,revhf5}, early views 
of Kondo and mixed valence physics \citep{revearly1,revearly2},  the solution of the Kondo
impurity model  by renormalization group and the strong coupling
expansion \citep{revwilson,revnozieres}, the
Bethe Ansatz method  \citep{revbethekondo,revbethekondo2},  
heavy fermion superconductivity  \citep{revhfsc1,revhfsc2},  Kondo insulators
 \citep{revfiskaeppli,revki1,revki2},  X-ray spectroscopy
 \citep{allenrev}, optical response in heavy
fermions  \citep{revoptical} and the latest reviews on non-Fermi liquid
behavior and quantum criticality
 \citep{revqcp1,revqcp2,revvarma2001,revqcp3,revqcp4,flouquet-2005}. 
There are inevitable apologies, for this article is
highly selective and  partly
because of lack of lack of space does not cover
dynamical mean field theory approaches to heavy fermion physics \citep{krauth,coxdmft,jarrelldmft,hogandmft}, nor
the extensive literature on the order-parameter
phenomenology of heavy fermion superconductors reviewed in \citep{revhfsc1}.

\subsection{Brief History}\label{}

Heavy electron materials represent a frontier in a journey of
discovery in electronic and magnetic materials that spans more than 70 years.
During this time, the concepts and understanding have undergone
frequent and often dramatic revision. 

\fight= 0.9 \textwidth
%\fg{newfigs/expt2.eps}{(a) Resistance minimum in $Mo_{x}Nb_{1-x}$ after 
\fg{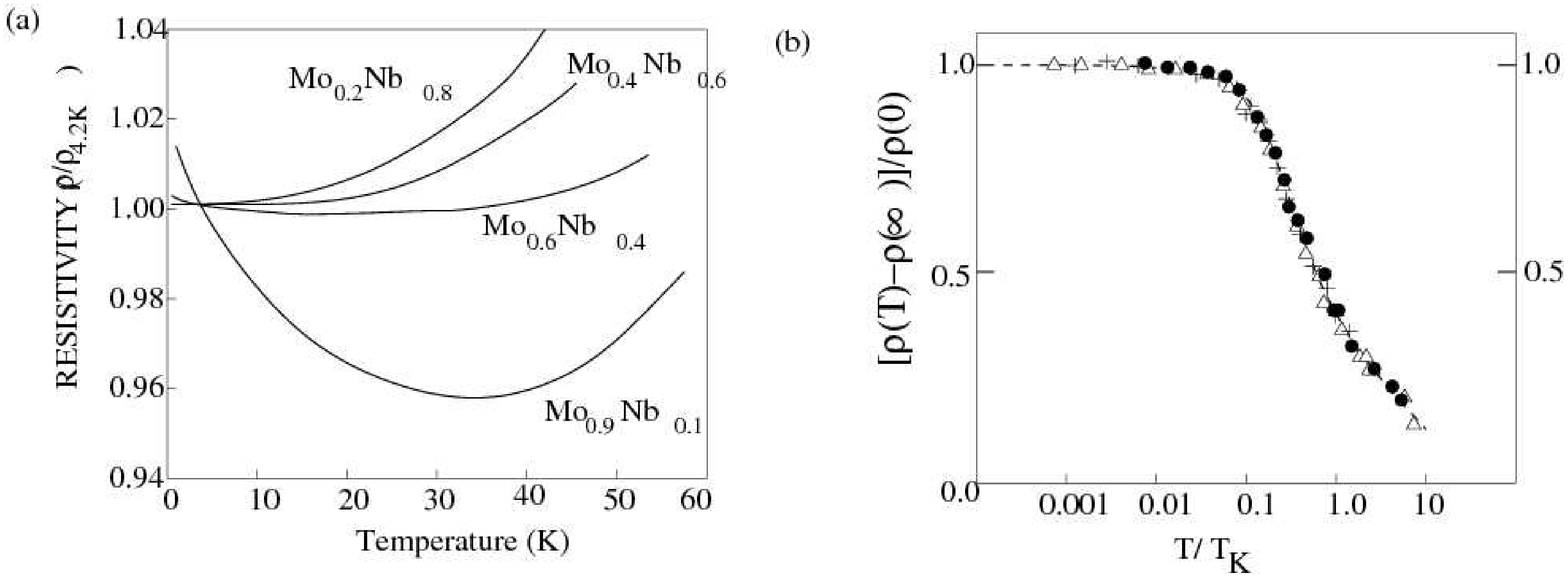}{(a) Resistance minimum in $Mo_{x}Nb_{1-x}$ after 
 \citep{sarachik64}
(b)
Temperature dependence of excess resistivity produced 
by scattering off a magnetic ion, showing universal dependence on
a single scale, the Kondo temperature. Original data from
 \citep{kondo_scaling}}{fig3}

In the early 1930's de Haas {\sl et al.} \citep{resistanceminimum}
%,resistanceminimum2,resistanceminimum3,sarachik64} 
in Leiden, discovered 
a ``resistance minimum'' that develops in the resistivity of copper, gold,
silver and many other metals at low temperatures (Fig. \ref{fig3}).
It took a further 30 years before the purity
of metals and alloys  improved to a point where the resistance
minimum could be linked to the presence of magnetic impurities \citep{clogston62,sarachik64}. 
Clogston, Mathias and collaborators  at Bell Labs  \citep{clogston62}
found they could tune the conditions under which
iron impurities in Niobium were magnetic, by alloying with
Molybdenum. Beyond a certain concentration of Molybdenum, the  iron impurities
become magnetic and a resistance minimum was observed to
develop. 

In  \citeyear{anda}, Anderson formulated
the first microscopic model  for the formation of magnetic moments
in metals. Earlier work by Blandin and Friedel \citep{blandin}  had observed that localized d states
form resonances in the electron sea. Anderson  extended this idea 
and added a new ingredient: the Coulomb interaction between the
d-electrons, which he modeled by 
term 
\begin{equation}
H_{I}= U n_{\uparrow}n_{\downarrow }.
\end{equation}
Anderson showed that  local moments formed once the Coulomb
interaction $U$ became large. 
One of the unexpected consequences of this theory, 
is that local moments develop an antiferromagnetic coupling
with the spin density of the surrounding electron fluid, described by
the interaction \citep{anda,kondo,kondo2,swolf,coqblin} 
\bea
%\label{}
H_I = J \vec \sigma(0)\cdot \vec S
\eea
where $\vec S$ is the spin of the local moment  and  $\vec \sigma(0)$
is the spin density of the electron fluid.
In Japan,  Kondo \citep{kondo}  set out to examine the consequences of
this result. He found that when he calculated the scattering rate $\frac{1}{\tau }
$ of electrons off a magnetic moment to one order higher than Born
approximation, 
\bea
%\label{}
\frac{1}{\tau }
 \propto \left[
J\rho + 2 (J\rho
)^{2 } \ln \frac{D}{T}
\right]^{2},
\eea
where $\rho$ is the
density of state of electrons in the conduction sea and $D$ is the
width of the electron band. As the temperature is lowered, the
logarithmic term grows, and the scattering rate and resistivity
ultimately rises, connecting the resistance minimum with the antiferromagnetic
interaction between spins and their surroundings.

A deeper understanding of the logarithmic term in this scattering rate
required the renormalization group
concept \citep{yuval,yuval2,yuval3,fowler,revwilson,nozieres,revnozieres}. 
The key idea here, is that the physics of a spin inside a metal depends
on the energy scale at which it is probed. 
The ``Kondo'' effect is a manifestation of the phenomenon of ``asymptotic freedom'' that also governs quark physics.
Like the quark, at high energies the local moments inside metals 
are asymptotically free, but
at temperatures and energies below a characteristic scale the Kondo
temperature, 
\begin{equation}
T_{K} \sim D e^{-1/ (2J \rho ) }
\end{equation}
where $\rho $ is the density of electronic states, 
they interact so strongly
with the surrounding electrons 
that they become screened into a singlet state, or
``confined'' at low energies, ultimately forming a Landau Fermi
liquid \citep{nozieres,revnozieres}.

Throughout the 1960s and 1970s, conventional wisdom had it
that magnetism and superconductivity are mutually exclusive. Tiny
concentrations of magnetic  produce a lethal suppression of superconductivity
in conventional metals.   Early work on the interplay of the Kondo
effect and superconductivity by  Maple {\sl et al.}\citep{maplesc}, did suggest
that the Kondo screening suppresses the pair breaking effects of magnetic
moments, but the implication of these results was only slowly digested.
Unfortunately, the belief in the mutual exclusion of local moments and superconductivity
was so deeply ingrained, 
that the first observation of
superconductivity in the ``local moment'' metal $UBe_{13}$
  \citep{bucher} was dismissed by its discoverers as an artifact produced by
stray filaments of  uranium. 
Heavy electron metals were discovered in  \citeyear{ott}
by  \citeauthor*{ott}, 
who observed that the inter-metallic $CeAl_{3}$
forms a metal in which the Pauli susceptibility and linear specific
heat capacity are about 1000 times larger than in conventional metals.
Few believed their speculation that this might be a lattice version of
the Kondo effect, giving rise in the lattice to a narrow band 
of ``heavy'' f-electrons. 
The discovery of superconductivity in $CeCu_{2}Si_{2}$ in a similar
f-electron fluid, a year
later by  Steglich \citep{steglich} , 
was met with widespread disbelief. 
All the measurements of the crystal structure of this material pointed
to the fact that the $Ce$ ions were in a $Ce^{3+}$ or $4f^{1}$ configuration.
Yet this meant one local moment per unit cell - which required an
explanation of how these local moments do not destroy
superconductivity, but rather, are  part of its formation.

Doniach  \citep{doniach},
made the visionary proposal that a heavy electron metal is a dense
Kondo lattice  \citep{rkky2}, in which  every single local moment 
in the lattice undergoes the Kondo effect (Fig. \ref{fig2}). In this
theory, each
spin is magnetically screened by the conduction sea. 
One of the great concerns of the time, raised by  Nozi\` eres \citep{exhaustion},
was whether there could ever be sufficient conduction electrons in a dense
Kondo lattice to screen each local moment.

Theoretical work on this problem was initially stalled, for wont of
any controlled way to compute properties of the Kondo lattice.
In the early 1980's,  Anderson \citep{phillargeN}
proposed a way
out of this log-jam.  
Taking a  cue from the success of the $1/S$ expansion in spin wave
theory, and the  $1/N$ expansion in statistical mechanics and particle
physics, he note that the large magnetic spin degeneracy $N=2j+1$ of
f-moments could could be used to generate an expansion in the small parameter
$1/N$ about the limit where $N\rightarrow \infty
$. Anderson's idea  prompted a renaissance
of theoretical development \citep{ramalargeN,gunnarson,read,nick,me,long,auerbach}, making
it possible to compute the X-ray absorption spectra of these
materials and, for the first time, 
examine how heavy f-bands form within the Kondo lattice.
By the mid eighties, the first de Haas van Alphen experiments  \citep{spring,lonz}
had detected  cyclotron orbits of heavy electrons in
$CeCu_{6}$ and $UPt_{3}$. With
these developments, the heavy fermion concept 
was  cemented. 

On a separate experimental front, in 1983  Ott, Rudiger, Fisk and
Smith \citep{ube13a,ube13b} returned to the material $UBe_{13}$, and by
measuring a large discontinuity in the 
bulk specific heat at the resistive superconducting transition, confirmed it as a bulk
heavy electron superconductor. This 
provided a vital independent  confirmation of Steglich's discovery of heavy
electron superconductivity, assuaging the old doubts 
and igniting a huge new interest in heavy
electron physics.
The number of heavy electron metals
and superconductors 
grew rapidly in the mid 1980s \citep{heavyscrev}.
It became clear  from specific heat, NMR and ultrasound 
experiments on heavy fermion superconductors
that the  gap is 
anisotropic, with lines of nodes
strongly suggesting an electronic, rather than a phonon mechanism of
pairing. These discoveries prompted theorists to return to earlier
spin fluctuation-mediated models of anisotropic pairing. 
In the early summer of 1986, three new theoretical papers were
received 
by Physical Review, 
the first by B\' eal Monod, Bourbonnais and
Emery \Citep{bealmonod} working in Orsay, France, followed closely (six weeks later)
by papers from Scalapino, Loh and Hirsch \citep{scalapino} at UC Santa Barbara, California,
and Varma, Schmitt-Rink and Miyake \citep{miyake} at
Bell Labs, New Jersey. These papers
contrasted heavy electron superconductivity with superfluid $He-3$.
Whereas $He-3$ is dominated by ferromagnetic interactions, which generate
triplet pairing, these works 
showed that in heavy electron systems,
soft antiferromagnetic spin fluctuations resulting from the vicinity
to an antiferromagnetic instability would drive anisotropic d-wave
pairing (Fig. \ref{fig4}). The almost coincident discovery 
of high temperature superconductivity the very same year, 1986, 
meant that these early works on heavy electron superconductivity
were destined to exert huge influence on the evolution of ideas 
about high temperature superconductivity. Both  the RVB
and the spin-fluctuation theory of d-wave pairing in the cuprates are,
in my opinion, close cousins, if not direct descendents 
of these early 1986 papers on heavy electron superconductivity. 
\fight= 0.8 \textwidth
%\fg{newfigs/bealmonodfig.eps}{Figure from  \Citep{bealmonod}, one of three
\fg{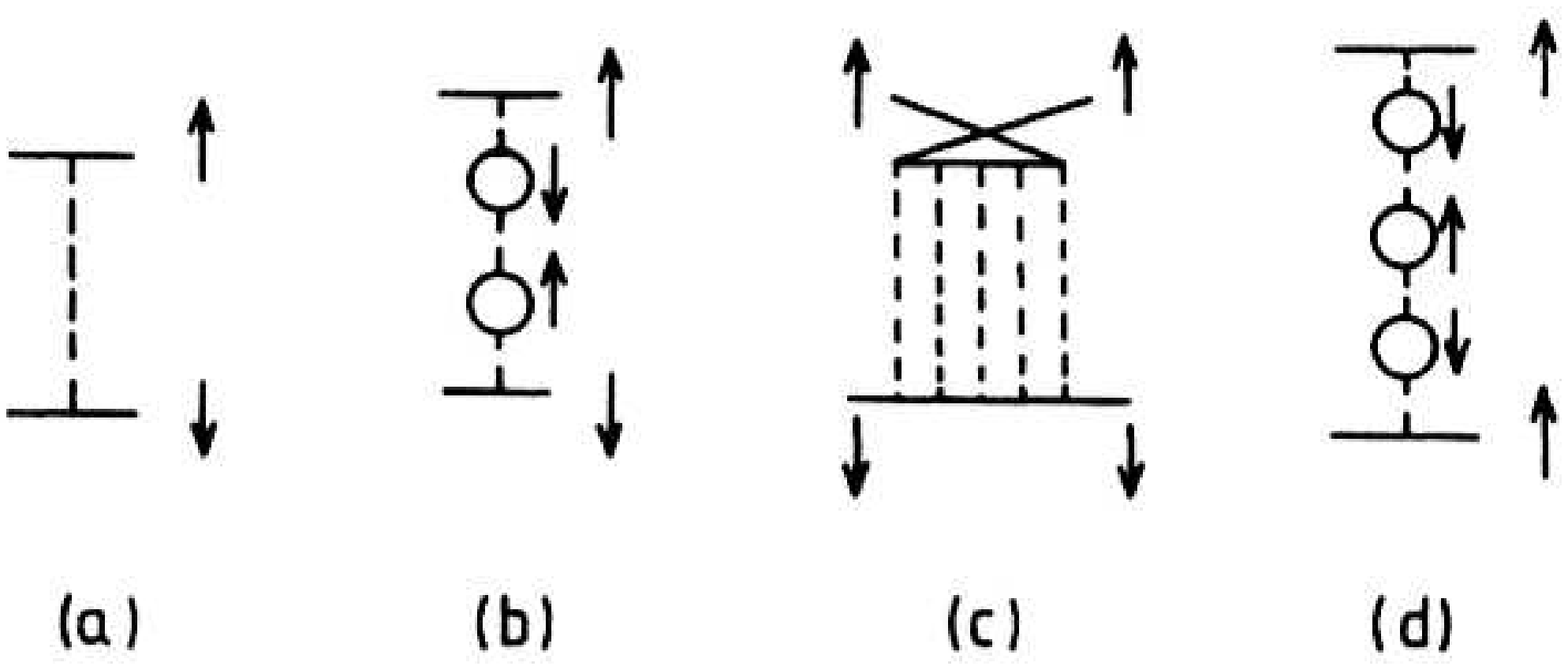}{Figure from  \Citep{bealmonod}, one of three
path-breaking papers in 1986 to link d-wave pairing to
antiferromagnetism. (a) is the bare interaction, (b) and (c)  and (d) the
paramagnon mediated interaction between anti-parallel  or parallel
spins. }{fig4}
\fight= 0.9 \textwidth

After a brief hiatus, interest in heavy electron physics re-ignited
in the mid 1990's with the discovery of quantum critical points in
these materials.  
High temperature superconductivity introduced many  
important new ideas into our conception of electron fluids, including

\begin{itemize}
\item Non Fermi liquid behavior: the emergence of metallic states that
can not be described as fluids of renormalized quasiparticles. 
\item Quantum phase transitions and the
notion that zero temperature quantum critical points 
might profoundly modify finite temperature properties of metal. 
\end{itemize}
Both of these effects are seen in a wide variety of heavy electron
materials, providing an vital alternative venue for research on these
still unsolved aspects of interlinked, magnetic and electronic behavior. 

In 1995,
Hilbert von Lohneyson and collaborators discovered that 
by alloying small amounts of gold into
$CeCu_{6}$ that one can tune $CeCu_{6-x}Au_{x}$ through an
antiferromagnetic quantum critical point, and then reverse the process
by the application of pressure \citep{hilbert2,hilbert}.
These experiments showed that 
a heavy electron metal 
develops ``non-Fermi liquid'' properties at a quantum critical point, 
including a 
linear temperature dependence of the resistivity and  a
logarithmic   dependence of the specific heat coefficient on
temperature. Shortly thereafter,  Mathur {\sl et al. }\citep{mathur}, 
at Cambridge showed that
when pressure is used to drive 
the antiferromagnet  $CeIn_{3}$ through 
through a quantum phase transition, 
heavy electron superconductivity develops
in the vicinity of the quantum phase transition.   
Many new examples of heavy electron system have come to light in the
last few years which follow the same pattern.  In one fascinating development,
 \citep{monthoux}
suggested that if quasi-two dimensional versions
of the existing materials could be developed, then the superconducting
pairing would be less frustrated, leading to a higher transition temperature.
This led experimental groups to explore the effect of introducing layers
into the material $CeIn_{3}$, leading to the discovery of the so called
$1-1-5$ compounds, in which an $XIn_{2}$ layer has been introduced
into the original cubic compound.
 \citep{sarrao,bangetal}. Two notable members of this group are 
$CeCoIn_{5}$
and most recently, $PuCoGa_{5}$ \citep{sarrao2}. The transition temperature rose from
$0.5K$ to $2.5K$ in moving from $CeIn_{3}$ to $CeCoIn_{5}$. Most
remarkably,
the transition temperature rises to 
above 18K in the $PuCoGa_{5}$ material.
This amazing rise in $T_{c}$, and its close connection with quantum
criticality, are very active areas of research, and may hold important
clues  \citep{curro} to the ongoing quest to discover room temperature
superconductivity. 

\subsection{Key elements of Heavy Fermion Metals}\label{}

%%   Phenomenology and the hand-wavy physics in a nutshell 
%%
%%   Local f-physics and the Kondo effect.
%%
%%   Local Physics of f-electrons
%%   4f and 5f wavefunctions
%%   Anderson atomic model of moment formation
%%   Valence fluctuations and the Kondo interaction.
%%   The idea of the Kondo effect.
%%
%%   X-ray inverse X-ray: local physics and Kondo resonance
%%
%%   The Heavy Fermi Liquid 
%%
%%   Key high temperature properties. 
%%
%%   - 1/T susceptibility - Local moment.
%%   - Spin Entropy quenches at LT.
%%
%%   - C=gamma T
%%   - rho = AT^{2}
%%   - chi/gamma ~ 1.
%%   - A/gamma^{2}~1
%%
%%   Spectroscopy.
%%
%%   Heavy quasiparticles. dHvA.
%%   ``Large Fermi surface'' -> Entire zoo: 
%%   Insulators, metals, superconductors.
%%
%%
%%   Subsequent instabilities of the Fermi surface ->
%%   Superconductivity. SDW, DW
%%
%%   Exotic behavior at the very edge of magnetism. 
%%   Quantum Criticality not just density wave formation, but possibly
%%   a deconfinement of the intrinsic spin.
%%
%%
%%   Outline of the article.
%%
%%
%%
%%
%%
%%
%%
%%
%%
Before examining the theory of heavy electron materials, we make a brief
tour of their key properties. Table A. shows a selective list of heavy
fermion compounds

\subsubsection{Spin entropy: a driving force for new 
physics}\label{}

The properties  of heavy fermion compounds derive from
the partially filled f orbitals of rare earth or actinide
ions \citep{revhf1,revhf2,revhf3,revhf4,revhf5}. The large nuclear  charge in these ions 
causes their f orbitals 
to collapse inside the inert gas core of the ion, turning them
into localized magnetic moments. 

\vfill \eject 

\newcommand{\entries}[7]{
 &\renp {2}{27mm}{#1} 
& \renp{2}{20mm}{#2} 
&\renp{2}{18mm}{\small #3} 
&\renp{2}{25mm}{\small #4} 
&\renp{2}{15mm}{#5}
&\renp{2}{25mm}{#6} 
&\renp{2}{15mm}{#7 } 
\\
&&&&&&&\\}
\centerline{\bf Table. A. Selected Heavy Fermion Compounds.}
\vskip 0.1 truein\noindent 
\begin{center}
\begin{tabular}{|l||c|c|l|l|c|c|c|}
\hline
\multirow{2}{20mm}{\bf Type}
 & Material & $T^{*}$ & 
$T_{c},\ x_{c},\ B_{c}$
& Properties & $\rho$ & \renp {2}{20mm}{$\gamma_{n}$\\
 $mJmol^{-1}K^{-2}$
}
& Ref.\\
&&&&&&&\\
\hline
\multirow{2}{20mm}{\bf Metal}
\entries{$CeCu_{6}$
}{$10K$}{-}{\small Simple HF 
Metal}{$T^{2}$}{1600}{[1]}
\hline
\hline
\multirow{6}{20mm}{\bf
Super-\\
conductors}
\entries{$CeCu_{2}Si_{2}$}{$20K$}{ $T_{c}$=0.17K}{First HFSC}{$T^{2}$}{800-1250}{[2]}
\cline{2-8}
\entries{$UBe_{13}$}{2.5K}{ $T_{c}$=0.86K}{\small Incoherent\\
 metal$\rightarrow$HFSC
}{$\rho_{c}\sim 150\mu \Omega \hbox{cm}$}{800}{[3]} 
\cline{2-8}
\entries{$CeCoIn_{5}$}{38K}{ $T_{c}$=2.3K}{\small Quasi 2D HFSC}{$T$}{750}{[4]} 
\hline
\hline
\multirow{4}{20mm}{\bf 
Kondo\\
Insulators}
\entries{$Ce_{3}Pt_{4}Bi_{3}$} {$T_{\chi}\sim 80K$}{-}{\small Fully Gapped KI}{$\sim e^{\Delta /T}$}{-}{[5]}
\cline{2-8}
\entries{$CeNiSn$}{$T_{\chi}\sim 20K$}{-}{\small Nodal KI}{Poor Metal}{-}{[6]}  
\hline
\hline
\multirow{4}{20mm}{\bf Quantum\\
Critical}
\entries{$CeCu_{6-x}Au_{x}$} {$T_{0}\sim 10K$}{$x_{c}=0.1$}{\small Chemically
tuned QCP}{T}{$\sim \frac{1}{T_{0}}\ln \left(\frac{T_{0}}{T} \right)$}{[7]} 
\cline{2-8}
\entries{$YbRh_{2}Si_{2}$}{$T_{0}\sim 24 K$} {$B_{\perp}$=0.06T \\
$B_{\parallel }$=0.66T}{Field-tuned QCP}{T}{$\sim\frac{1}{T_{0}}\ln \left(\frac{T_{0}}{T} \right)$}{[8]} 
\hline
\hline
\multirow{4}{20mm}{\bf 
SC + other\\
Order}\entries{$UPd_{2}Al_{3}$} {110K}{\small $T_{AF}$=14K, \\
$T_{sc}$=2K}{AFM + HFSC}{$T^{2}$}{210}{[9]} 
\cline{2-8}
\entries{$URu_{2}Si_{2}$} {75K}{ $T_{1}$=17.5K, \\
$T_{sc}$=1.3K}{Hidden Order \&  HFSC}{$T^{2}$}{120/65}{[10]} 
\hline
\hline
\end{tabular}
\end{center}

\vskip 0.1in
{\small Unless otherwise stated, $T^{*}$ denotes the
temperature of the maximum in resistivity. $T_{c}$, $x_{c}$ and
$B_{c}$ denote critical temperature, doping and field. $\rho $ denotes
the temperature temperature dependence in the normal
state.$\gamma_{n}=C_{V}/T$ is the specific heat coefficient in the
normal state. 
[1]  \citep{cecu6a,cecu6}, 
[2]  \citep{steglich,steglich2,steglich3},
[3]  \citep{ube13a,ube13b}, 
[4]  \citep{sarrao,bangetal}, 
[5]  \citep{ce3bi4pt3,ce3bi4pt3optical}, 
[6]  \citep{cenisn,cenisn-semimetal,CeNiSn-thermal1}, 
[7] \citep{hilbert,hilbert2}, 
[8]  \citep{trovarelli,silke,custers,custers2}, 
[9] \citep{steglich3,upd2al3pairing,upd2al3nqr}, 
[10]  \citep{uru2si2,nexus}.
}

%\vfill \eject

\noindent Moreover, the large
spin-orbit coupling in f-orbitals combines the spin and angular momentum
of the f-states into a state of definite $J$ and 
it is these large quantum spin degrees of freedom 
that lie at the heart of heavy fermion physics.

Heavy fermion materials display properties which change qualitatively,
depending on the temperature, so much so, that 
the room temperature  and low temperature behavior almost resemble 
two different materials. 
At room temperature, high magnetic fields and high 
frequencies, they behave as local moment
systems, with a Curie law susceptibility
\begin{equation}\label{chi}
\chi ={  M^2\over 3T } \qquad \qquad M^2 = (g_J\mu_B)^2J(J+1) 
\end{equation}
where $M$ is the magnetic moment of an f state with total
angular momentum $J$ and 
the gyromagnetic ratio $g_J$. 
However, at temperatures  beneath a characteristic scale we
call $T^{*}$ (to distinguish it from the single-ion Kondo
temperature $T_{K}$), 
the localized spin degrees of freedom
melt  into the conduction sea, releasing their spins as 
as mobile, conducting f-electrons.  
\fight=0.7\textwidth
%\fg{newfigs/ube13fig.eps}
\fg{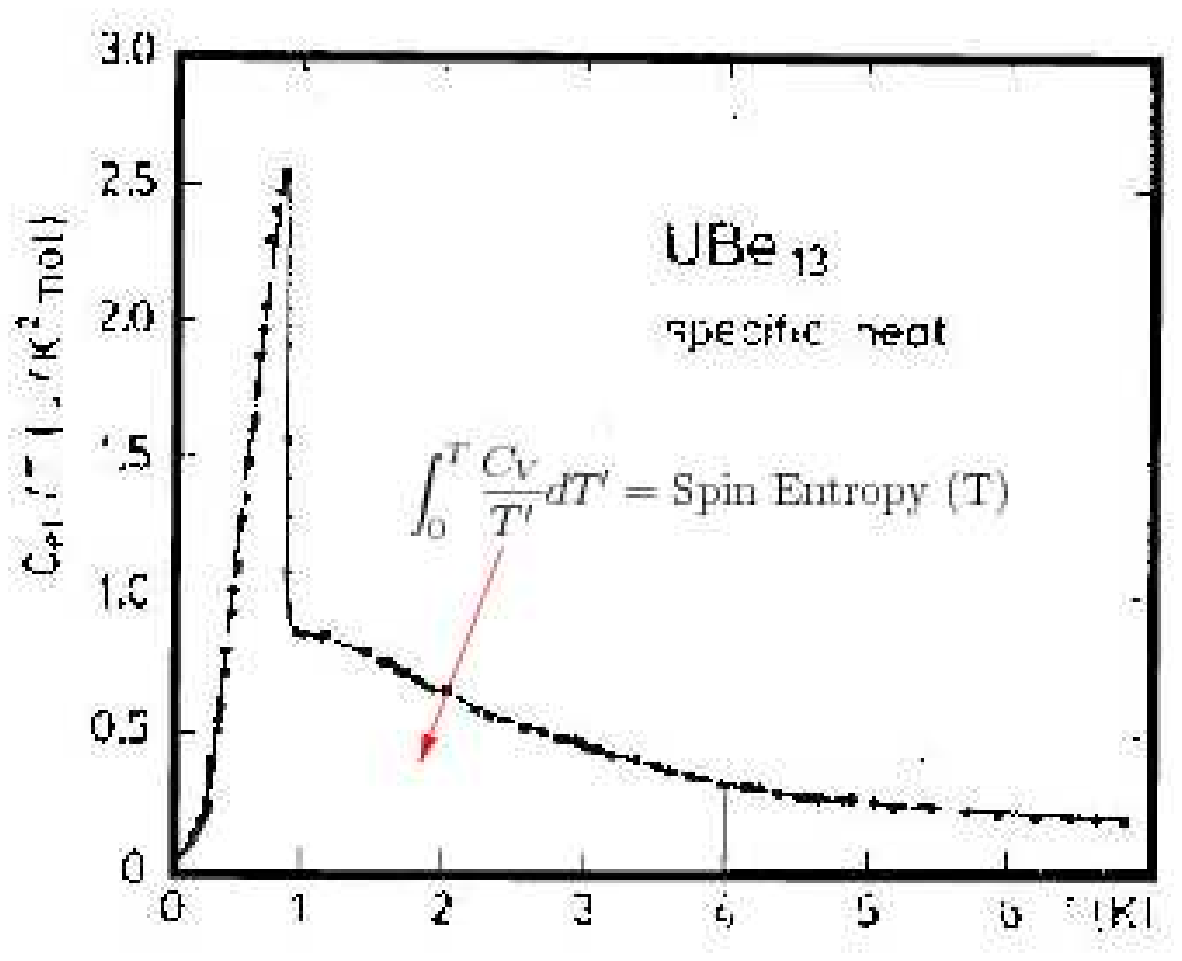}
{Showing the specific heat coefficient of $UBe_{13}$ after
\citep{ube13c}. The area under the $C_{V}/T$ curve up to a
temperature $T $ provides a measure of 
the amount of unquenched spin entropy at that temperature.  The
condensation entropy of heavy fermion superconductors is
derived from the spin-rotational degrees of freedom of the local
moments, and the large scale of the condensation entropy 
indicates that spins partake in the formation of the order
parameter.
}{fig5}

A Curie susceptibility 
is the hallmark of the 
decoupled, rotational
dynamics of the f-moments, associated with an unquenched entropy of
${\cal S}=k_{B}\ln N$ per spin, 
where $N=2J+1$ is the spin-degeneracy of an isolated magnetic moment of angular momentum $J$. For
example, in a Cerium heavy electron material, the $4f^{1}$ ($L=3$)
configuration of the $Ce^{3+}$ ion is spin-orbit coupled into a state
of definite $J=L-S = 5/2$ with $N=6$.
Inside the crystal, the full rotational symmetry of each magnetic f-ion 
is often reduced by crystal fields to a quartet ($N=4$) or a Kramer's
doublet $N=2$.
At the characteristic
temperature $T^{*}$, as the Kondo effect develops, 
the spin entropy is rapidly lost from the material, and large quantities
of heat are lost from the material. 
Since the area under the specific heat curve determines the entropy,
\begin{equation}
S (T) = \int_{0}^{T}\frac{C_{V}}{T'}dT',
\end{equation}
a rapid loss of spin entropy at low temperatures
forces a sudden rise in the specific
heat capacity. 
Fig. \ref{fig5} illustrates this phenomenon with
the specific heat capacity of $UBe_{13}$. Notice how the specific heat
coefficient $C_{V}/T$ rises to a value of order $1J/mol/K^{2}$, and
starts to saturate at about $1K$, indicating the formation of a Fermi
liquid with a linear specific heat coefficient. Remarkably, just as
the linear specific heat starts to develop, $UBe_{13}$
becomes superconducting, as indicated by the large
specific heat anomaly.

\subsubsection{``Local'' Fermi liquids with a single scale }\label{localfl}

The standard theoretical framework for describing metals 
is Landau Fermi liquid theory \citep{landaufl}, 
according to which, the excitation spectrum  of a metal
can be adiabatically connected to  those of a non-interacting 
electron fluid. 
Heavy Fermion metals are extreme examples of Landau Fermi liquids which
push the  idea of adiabaticity into an regime
where the bare electron interactions, on the scale of electron volts, 
are hundreds, even thousands of times larger than the millivolt 
Fermi energy scale of the heavy electron quasiparticles.
The Landau Fermi liquid that develops in these materials 
shares much in common with the Fermi liquid that develops around 
an isolated magnetic impurity \citep{nozieres,revnozieres}, once it is quenched by the conduction
sea as part of the Kondo effect. There are three key features  of this
Fermi liquid:
\begin{itemize}

\item  {\bf Single scale $T^{*}$} The 
quasiparticle density of states $\rho^{*}\sim 1/T^{*}$ 
and scattering amplitudes 
$A_{\bk \sigma  ,\bk '\sigma '}\sim T^{*}$ scale approximately 
with a single scale $T^{*}$. 

\item  {\bf Almost incompressible}. 
Heavy electron fluids  are ``almost incompressible'', in the sense 
that the  
charge susceptibility $\chi_{c}= dN_{e}/d\mu<< \rho^{*}$ is 
unrenormalized and typically more than an order
magnitude smaller than the quasiparticle  density of states
$\rho^{*}$.
This is because the lattice of
spins severely modifies  the quasiparticle density of states,
but leaves the charge density of the fluid $n_{e} (\mu)$, 
and its dependence on the chemical potential $\mu$ unchanged. 

\item {\bf Local.} Quasiparticles scatter when in the vicinity of a
local moment,  giving rise to a small momentum dependence to the
Landau scattering amplitudes. \citep{yamadalocal,yamada2,bedell} .

\end{itemize}

Landau Fermi liquid theory relates the properties of a Fermi liquid to
the density of states of the quasiparticles and 
a small number of interaction parameters \citep{pethickbaym} 
If  $E_{\bk \sigma }$ is the
energy of an isolated quasiparticle, then the
quasiparticle density of states  $\rho^{*} = \sum_{\bk\sigma}
\delta (E_{\bk\sigma}- \mu)$
determines the linear specific heat
coefficient 
\begin{equation}
\gamma =\hbox{Lim}_{T\rightarrow 0}
\left(\frac{C_{V}}{T}  \right)
=\frac{\pi^{2}k_{B}^{2}}{3} \rho^{*}.
\end{equation}
In conventional metals the linear specific heat coefficient is
of order $1-10\ mJ \ mol^{-1}K^{-2}
$. 
In a system with quadratic dispersion, $E_{\bk}=
\frac{\hbar^{2}k^{2}}{2 m^{*}}$, the quasiparticle density of states
and effective mass $m^{*}$ are directly proportional
\begin{equation}
\rho^{*} = \left(\frac{k_{F}}{\pi^{2}\hbar^{2}} \right)m^{*},
\end{equation}
where $k_{F}$ is the Fermi momentum. In heavy fermion compounds, the scale of $\rho ^{*}$ varies widely, 
and specific heat coefficients in the range $100- 1600 \ mJ \ mol^{-1}K^{-2}
$ have been observed.
From this simplified perspective,
the quasiparticle effective masses in heavy electron materials are two
or three orders of magnitude ``heavier'' than in conventional metals.

In Landau Fermi liquid theory, a change $\delta n_{\bk' \sigma' }$ in
the quasiparticle occupancies causes a shift in the quasiparticle energies
given by
\begin{equation}
\delta E_{\bk \sigma } = \sum_{\bk'\sigma '} 
f_{\bk \sigma ,\bk\sigma'}
\delta n_{\bk ' \sigma '}
\end{equation}
In a simplified model with a spherical Fermi 
surface, the Landau interaction parameters only depend on the relative
angle $\theta_{\bk ,\bk '} $
between the quasiparticle momenta,
and are expanded in terms of Legendre Polynomials as 
\begin{equation}
f_{\bk \sigma ,\bk\sigma'}
= 
\frac{1}{\rho^{*}}\sum_l (2l+1)P_{l} ( \theta_{\bk ,\bk '})
[F_{l}^{s}+\sigma \sigma ' F_{l}^{a}].
\end{equation}
The dimensionless ``Landau parameters'' $F_{l }^{s,a}$  parameterize
the detailed quasiparticle interactions. The s-wave ($l=0$) 
Landau parameters
determine the 
magnetic and charge susceptibility of a Landau Fermi liquid
are given by \citep{landaufl,pethickbaym}
\begin{eqnarray}\label{renorm}
\chi_{s} &=& 
\mu_{B}^{2}
\frac{\rho^{*}}{1 + F_{0}^{a}}= \mu_{B}^{2}\rho^{*}\left[
1- A_{0}^{a} \right]
\cr
\chi_{c} &=& e^{2}\frac{\rho^{*}}{1 + F_{0}^{s}}= e^{2}\rho^{*}\left[1- A_{0}^{s} \right]
\end{eqnarray}
where, the quantities 
\begin{equation}
A_{0}^{s,a} = \frac{F_{0}^{{s,a}}}{1+F_{0}^{{s,a}}}
\end{equation}
are the s-wave Landau scattering amplitudes in the charge (s) and spin (a)
channels, respectively \citep{pethickbaym}.

The assumption of local scattering and incompressibility in heavy
electron fluids simplifies the situation, for in this case
only the $l=0$ components of the
interaction remain and the quasiparticle scattering amplitudes become
\begin{equation}
A_{\bk \sigma, \bk '\sigma '} = \frac{1}{\rho^{*}}
\left(A^{0}_{s} + \sigma  \sigma ' 
A^{0}_{a} \right).
\end{equation}
Moreover, in local scattering
the Pauli principle dictates that quasiparticles scattering at the
same point can only scatter when in 
in opposite spin states, so
that 
\begin{equation}
A^{(0)}_{\uparrow\uparrow }= A^{0}_{s} + A^{0}_{{a}} = 0.
\end{equation}
and hence $A^{0}_{s }= - A^{0}_{a}$.  The additional
the assumption of incompressibility forces $\chi_{c}/
(e^{2}\rho^{*})<<1$, so that now $A_{0}^{s}= - A_{0}^{a}\approx 1$ and
all that remains is a single parameter $\rho^{*}$.

This line of reasoning, first developed for the single impurity
Kondo model by Nozi\` eres \citep{revnozieres,nozieres}, later extended to a bulk Fermi liquid
by Bedell and Engelbrecht \citep{bedell}, enables us to understand two important
scaling trends amongst heavy electron systems.
The first consequence, deduced from 
(\ref{renorm}), 
is that the dimensionless Sommerfeld ratio, or ``Wilson ratio''
$W =\left(\frac{\pi^{2}k_B^2}{\mu_{B}^{2}}
\right)\frac{{\chi}_{s}}{{\gamma}} \approx 2$. Wilson \citep{revwilson}, found that this ratio
is almost exactly equal to two in the numerical renormalization 
group treatment of the  impurity Kondo model.  
The connection between this
ratio and the local Fermi liquid theory was first identified by
Nozi\` eres,  \citep{nozieres,revnozieres}. In real heavy electron systems, the effect of
spin orbit coupling slightly modifies the precise numerical form for
this ratio, 
nevertheless, the observation that 
 $W\sim 1$ over a wide range of materials in which the density of
 states vary by more than a factor of 100, is an indication of the
 incompressible and local character of heavy Fermi liquids (Fig. \ref{fig6}).

\fight=0.5 \textwidth
%\fg{newfigs/wilson.eps}
\fg{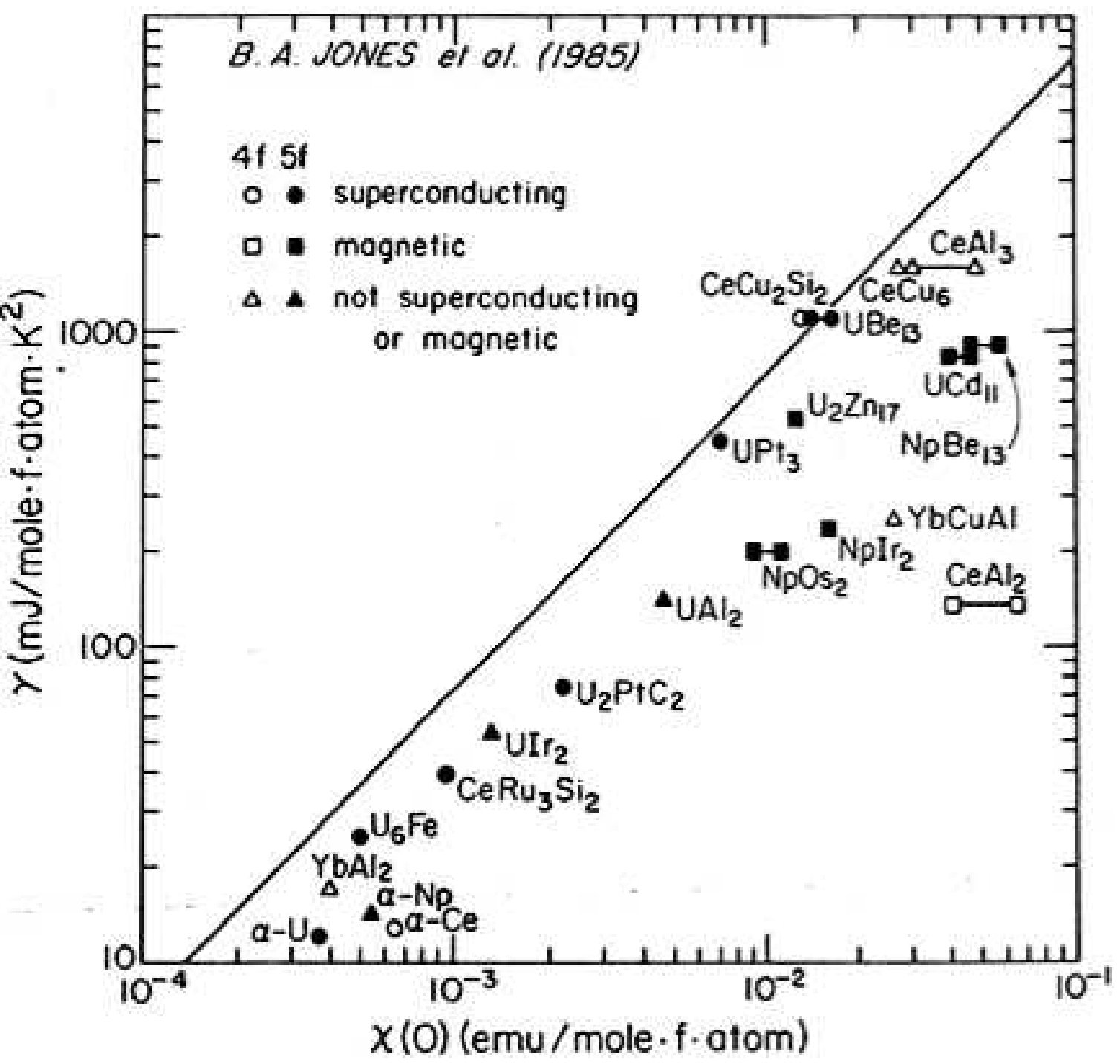}
{Plot of linear specific heat coefficient vs 
Pauli susceptibility to show approximate constancy of Wilson ratio.
(After B. Jones \citep{revhf2}).}{fig6}

A second consequence of locality appears
in the transport properties. In a Landau Fermi liquid, inelastic 
electron-electron
scattering produces  a quadratic temperature dependence 
in the resistivity
\begin{equation}
\rho (T)= \rho_{0}+ AT^{2}.
\end{equation}
In conventional metals, resistivity is dominated by electron-phonon
scattering, and the ``$A$'' coefficient  is generally too
small for the electron-electron contribution to the resistivity to be observed.
In strongly interacting metals, the $A$ coefficient becomes large, and
in a beautiful piece of phenomenology, 
Kadowaki and Woods \citep{kadowaki}, 
observed that the ratio of $A$ to the square of the specific heat
coefficient $\gamma^{2}$
\begin{equation}
\alpha_{KW}= \frac{A}{\gamma^{2}}\approx (1 \times 10^{-5})\mu
\Omega\hbox{cm}[\hbox{mol K}^{2}/\hbox{mJ}]
\end{equation}
is approximately constant,  over a range of $A$ spanning four 
orders of magnitude.  This too, can be simply understood from local
Fermi liquid theory, 
where the local scattering amplitudes 
give rise to 
an electron mean-free path given by 
\begin{equation}
\frac{1}{k_{F}l^{*}} \sim \hbox{constant} + \frac{T^{2}}{( T^{*})^{2}}.
\end{equation}
The ``$A$'' coefficient in the electron resistivity 
that results from the second-term satisfies
$A\propto \frac{1}{(T^{*})^{2}}\propto \tilde{\gamma}^{2}$. A
more detailed calculation is able to account for the magnitude of the
Kadowaki Woods constant, and its weak residual dependence on the spin
degeneracy $N=2J+1$ of the magnetic ions (see Fig. \ref{fig7}.).

The approximate validity of the scaling relations 
\begin{equation}
\frac{\chi}{\gamma}\approx \hbox{cons}, \qquad \frac{A}{\gamma^{2}}\approx \hbox{cons}
\end{equation}
for a wide range of heavy electron compounds, constitutes
excellent support for the  Fermi liquid picture of heavy electrons.
\fight=0.5 \textwidth
%\fg{newfigs/kadowaki.eps}
\fg{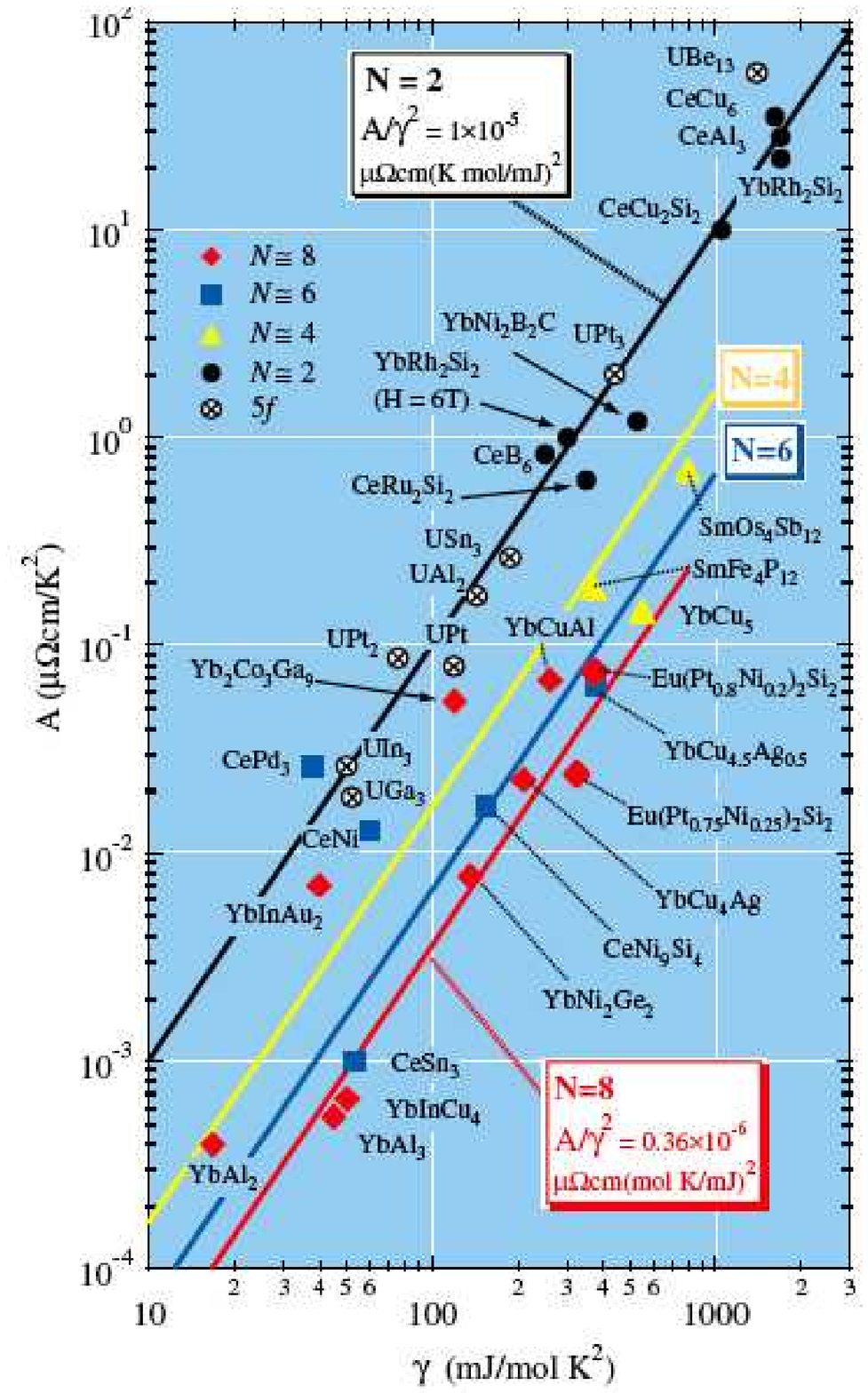}
{Approximate constancy of the Kadowaki Woods ratio, 
for a wide range of heavy electrons, after  \citep{kontani}.
When spin-orbit effects are taken into account, 
the Kadowaki Wood ratio 
depends on the effective degeneracy $N=2J+1$
of the magnetic ion, which when taken into account leads to a far more
precise collapse of the data onto a single curve. 
}{fig7}

A classic signature
of heavy fermion behavior is
the dramatic change in transport properties that accompanies the development
of a  coherent heavy fermion band structure(Fig. [6]). 
At high temperatures heavy fermion compounds  exhibit 
a 
large  saturated resistivity, induced 
by incoherent spin-flip  scattering of the conduction electrons
off the  local f-moments. This scattering 
{\sl grows} as the temperature is lowered, but at the same time, it
becomes increasingly elastic at low temperatures.
This leads to the
development of phase coherence.
%\fg{newfigs/coherence_onuki.eps}{ Development of coherence in
\fg{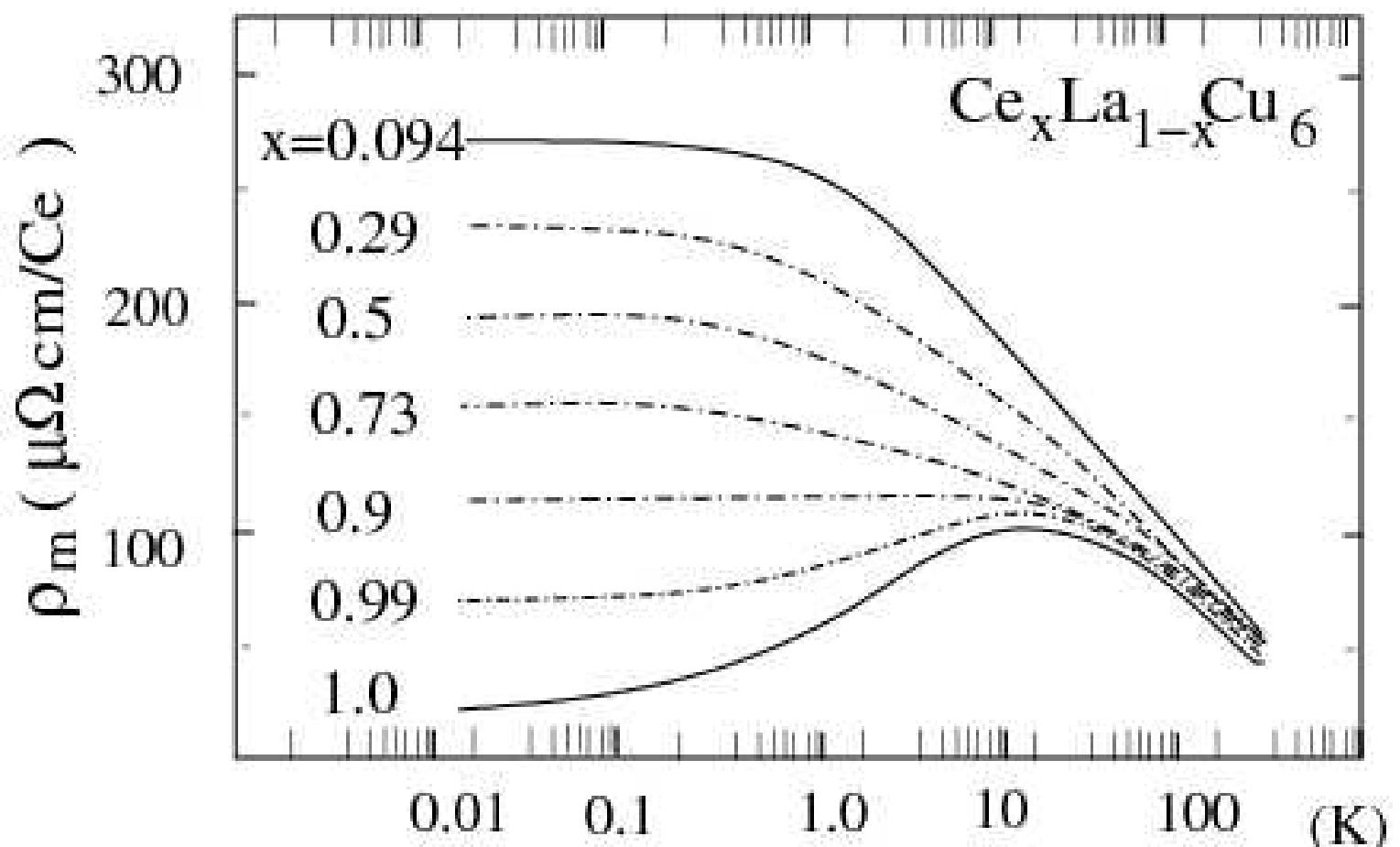}{ Development of coherence in
in  $Ce_{1-x}La_xCu_6$ after Onuki and Komatsubara \citep{cecu6}.
}{fig8}
\noindent the f-electron spins.
In the case of heavy fermion metals,
the development of coherence is marked 
by a rapid reduction in the resistivity, but 
in a remarkable class of 
heavy fermion or ``Kondo insulators'',
the development of coherence
leads to a filled band with a tiny insulating gap of order
$T_K$. 
In this case coherence is marked by a sudden
exponential rise in the resistivity and Hall constant.

The classic  example of coherence 
is provided by metallic $CeCu_6$, which develops ``coherence''
and a maximum in its resistivity around $T=10\  K$. 
Coherent heavy electron propagation is readily destroyed by
substitutional impurities. 
In $CeCu_6$, $Ce^{3+}$  ions can be continuously 
substituted with non magnetic $La^{3+}$ ions, producing a continuous
cross-over from coherent Kondo lattice to 
single impurity behavior (Fig. \ref{fig8} ).

One of the important principles of the Landau Fermi liquid 
is the Fermi surface counting rule, or Luttinger's
theorem\citep{luttinger}.
In non interacting electron band theory, the volume of the Fermi surface
counts the number of conduction electrons. For interacting
systems this rule survives \citep{martin82,oshikawa}, 
with the unexpected corollary that 
the spin states of the screened local moments are also included in the sum
\begin{eqnarray}\label{luttinger}
{2 V_{\rm FS}\over (2 \pi)^3} = [n_e+n_{spins}]
\end{eqnarray}
Remarkably, even though f-electrons are localized as magnetic moments
at high temperatures, in the heavy Fermi liquid, they contribute to
the Fermi surface volume.  

The most most direct evidence for the large heavy f- Fermi surfaces
derives from de Haas van Alphen and
Shubnikov de Haas experiments that measure the oscillatory
diamagnetism or and resistivity produced by coherent
quasiparticle orbits (Fig. \ref{fig9}). 
These experiments provide a direct measure of the heavy electron mass,
the Fermi surface geometry and volume. Since
the pioneering measurements on $CeCu_{6}$ and $UPt_{3}$ by Reinders
and Springford, Taillefer and Lonzarich in the
mid-eighties \citep{spring,lonz,lonz_early}, an
extensive number of such measurements have been
carried out \citep{cecu6,julianetal,onuki3,mccollam}. 
Two key features are observed:
\begin{itemize}
\item A Fermi surface volume which counts the f-electrons 
as itinerant quasiparticles.
\item Effective masses often in excess of one hundred free electron
masses. Higher mass quasiparticle orbits, 
though inferred from thermodynamics, can not be observed with current
measurement techniques. 

\item Often, but not always, the Fermi surface geometry is in accord with
band-theory, despite the huge renormalizations of the electron mass.
\end{itemize}
\fight=\textwidth
%\fg{newfigs/orbits.eps}{(a) Fermi surface of $UPt_{3}$ calculated from
\fg{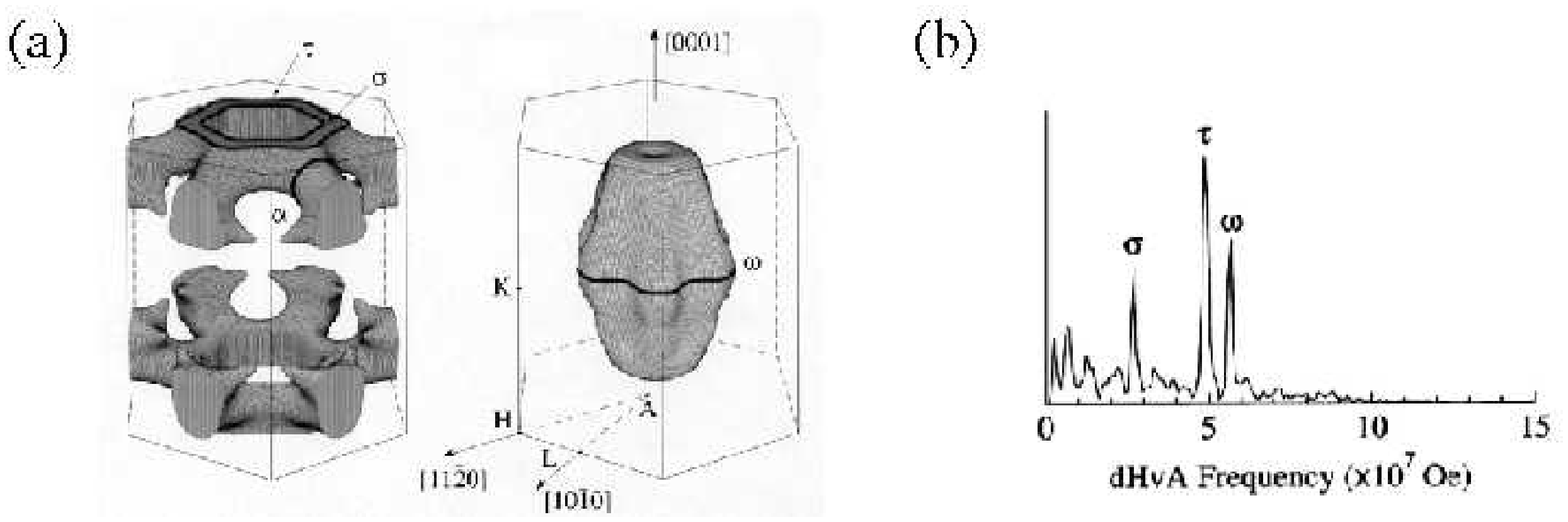}{(a) Fermi surface of $UPt_{3}$ calculated from
band-theory assuming itinerant $5f$ electrons  \citep{oguchi,norman1,freeman}, showing
three orbits ($\sigma $, $\omega$ and $\tau $) that are identified by
dHvA measurements, after  \citep{onuki3}. (b) Fourier transform of dHvA oscillations
identifying $\sigma $, $\omega$ and $\tau $ orbits shown in (a).
}{fig9}

Additional confirmation of the itinerant nature of the f-quasiparticles
comes from the observation of a Drude peak in the optical conductivity.
At low temperatures, in the coherent regime, an extremely narrow
Drude peak  can be observed in the optical conductivity of
heavy fermion metals. 
The weight under the Drude peak is a measure of the plasma frequency:
the diamagnetic response of the heavy fermion metal. This is found
to be extremely small, depressed by the large mass enhancement of
the quasiparticles \citep{millislee,revoptical}.  
$$
\int_{\vert \omega \vert \ltappr T_K} {d\om \over \pi}\si _{qp} ( \omega) = {ne^2 \over m^*}
\eqno(6)$$
Both the optical and dHvA experiments indicate that the presence of
f-spins depresses both the spin and diamagnetic response of the electron
gas down to low temperatures.  

\section{Local moments and the Kondo lattice}\label{}

\subsection{Local Moment Formation }\label{}

\subsubsection{The Anderson Model}\label{}
We begin with 
a discussion of  how magnetic moments form at high temperatures, and
how they are screened again at low temperatures to form a Fermi liquid.  The basic model for local moment
formation is the Anderson model \citep{anda}
\begin{eqnarray}\label{l}
H= \overbrace {\sum_{k,\sigma}\epsilon_k n_{k\sigma } + 
\sum_{k,\sigma}
V (k)\left[c\dg_{k\sigma }f_{\sigma }+ f\dg_{\sigma }c_{k\sigma } \right]
}^{{H_{resonance}}}
+\underbrace {E_{f}n_{f}+ U n_{f\uparrow
}n_{f\downarrow}}_{H_{atomic}}
\end{eqnarray}
where $H_{atomic}$ describes the atomic limit of an isolated magnetic
ion and $H_{resonance}$ describes the hybridization of the localized f-electrons
in the ion with the Bloch waves of the conduction sea.  
For pedagogical reasons, our discussion will initially focus on the case 
where the f-state is a Kramer's
doublet. %%The
%%Anderson model has been extensively studied over the decades, and a
%%variety of complimentary techniques are available for its
%%detailed study, including
%%\begin{itemize}
%%
%%\item [-]self-consistent diagrammatic expansions in 
%%$U$ \citep{yosida,yamadalocal,horvatic} and $1/N =1/ (2j+1)$ \citep{gunnarson,read,nick,me,long}
%%
%%\item [-]canonical transformations \citep{swolf,coqblin} and perturbative renormalization group \citep{haldane}

%%\item [-]Monte Carlo methods \citep{fye,jarell}.

%%\item [-]numerical renormalization group \citep{krish,numericalrg,jones}.

%%\item [-]Bethe Ansatz solution of its thermodynamics and many-body
%%spectrum \citep{wiegman,okiji,okiji2,schlottmann},  
%%\end{itemize} 
%%
%Strictly, for 
%f-systems, the spin quantum number $\sigma $ runs over $2j+1$ fold 
%values,  $\sigma\in [-j, \dots , j]$ 
% and the local
%interaction is then $U\sum_{\sigma>\sigma '}n_{\sigma }n_{\sigma '}$.
%However, in many cases, 
%where the crystal field splitting is larger than
%the Kondo temperature, the f-state is indeed reduced to a Kramer's doublet
%described by the above model (see Fig. \ref{fig7} )
There are two key elements to the Anderson model:
\begin{itemize}
\item {\bf Atomic limit.}
The atomic physics of an isolated ion with a single $f$ state,
described by the model 
\begin{equation}\label{}
H_{atomic} = E_{f}n_{f}+ U n_{f\uparrow }n_{f\downarrow}.
\end{equation}
Here $E_{f}$ is the  energy of the $f$ state
and $U$
is the  Coulomb energy associated with two electrons in
the same orbital. The atomic physics contains the basic mechanism for 
local moment formation, valid for f-electrons, but also seen in a
variety of other contexts, such as transition metal atoms and 
quantum dots.  

The four quantum states of the atomic model are
\begin{eqnarray}\label{}
\begin{array}{rll}
\begin{array}{c}
\vert f^2 \ra \cr
\vert f^0 \ra 
\end{array}
&
\left. 
\begin{array}{l}
\qquad E (f^{2}) = 2E_f+U\cr
\qquad E (f^{0}) =0\cr
\end{array}
\right\}&\hbox{non-magnetic}\cr\cr
\vert f^1 \up\ra, \quad \vert f^1\downarrow\ra &\ \qquad  E (f^{1}) =E_f.&\hbox{magnetic}.
\end{array}
\end{eqnarray}
In a magnetic ground-state, the 
cost of inducing a ``valence fluctuation'' by 
removing or adding an electron to the $f^{1}$ state is 
positive, i.e. 
\begin{eqnarray}\label{}
{\hbox{removing:}}\qquad E (f^{0})-E (f^{1})=-E_{f}>0&\Rightarrow& U/2 > E_{f}+U/2,
\cr
\hbox{adding:}\qquad E (f^{2})-E (f^{1})=E_{d}+U>0&\Rightarrow& E_{d}+U/2 > - U/2,
\end{eqnarray}
or  (Fig. \ref{fig10}).
\begin{equation}
U/2 >  E_d+U/2 > -U/2.
\end{equation}
Under these conditions, a local moment is  well-defined
provided the temperature is smaller than the valence fluctuation scale
$T_{VF}=\hbox{max} (E_{f}+U, -E_{f})$. 
At lower temperatures, the atom behaves exclusively as a
quantum top.

\fight = 0.5 \textwidth
%\fg{newfigs/nandmod.eps}{Phase diagram for Anderson Impurity Model in the
\fg{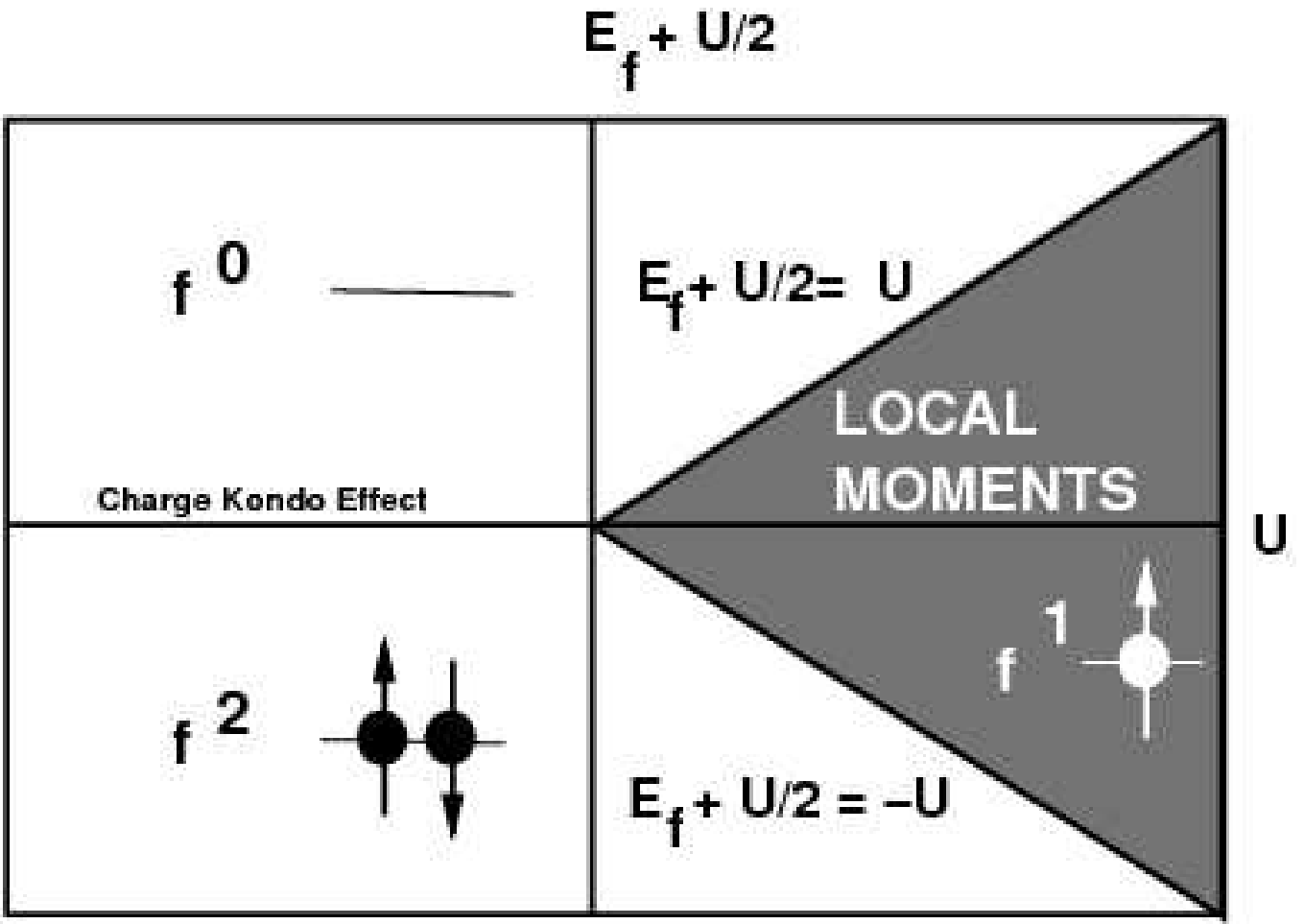}{Phase diagram for Anderson Impurity Model in the
Atomic Limit.
}{fig10}
\fight = \textwidth

\item {\bf Virtual bound-state formation.}  When the magnetic ion is immersed in a
sea of electrons, 
the f-electrons within the core of the atom hybridize
with the Bloch states of surrounding electron sea \citep{blandin} to
form a resonance described by 
\begin{equation}
H_{resonance}= 
\sum_{k,\sigma}\epsilon_k n_{k\sigma } + 
\sum_{k,\sigma}
\left[V (\bk)c\dg_{k\sigma }f_{\sigma }+ V (\bk)^{*}f\dg _{\sigma }c_{k\sigma } \right]
\end{equation}
where the hybridization matrix element $V (\bk)= \langle \bk\vert V_{atomic}\vert f\rangle $ 
is the overlap of the atomic
potential between a localized f-state and a  Bloch wave.
In the absence of any interactions, the hybridization broadens the
localized f-state, producing a resonance of width 
\begin{equation}
\Delta = \pi \sum_{\bk} \vert V (k)\vert^{2}\delta
(\epsilon_{\bk}-\mu)= \pi V^{2}\rho 
\end{equation}
where $V^{2}$  is the average of the hybridization around the Fermi surface.

\end{itemize}

There are two complimentary ways to approach the physics of the Anderson model 
\begin{enumerate}
\item [-]the 
``atomic picture'',  which 
starts with 
the interacting, but isolated atom ($V (k)=0$), 
and considers  the effect of immersing it in an electron sea by slowly
dialing up the hybridization.

\item [-]the ``adiabatic picture'' which starts 
with the non-interacting resonant ground-state ($U=0$), and 
then considers the effect of dialing up the 
interaction term $U$.   

\end{enumerate}
These approaches paint a contrasting and, at first sight,
contradictory picture of a local moment in a Fermi sea. 
From the adiabatic perspective, the
ground-state is always a  Fermi liquid (see \ref{localfl}),
but from atomic perspective, provided the hybridization is smaller
than  $U$ one expects a local magnetic moment,
whose low lying degrees of freedom are purely rotational.
How do we resolve this paradox? 

Anderson's original work provided a mean-field treatment of the
interaction.  He
found  that at interactions larger than
$U_{c}\sim \pi\Delta $ 
local moments develop with  a finite magnetization $M=\langle n_{\uparrow}\rangle -\langle n_{\downarrow }\rangle$.
The mean field theory provides an approximate guide to the conditions 
required for moment formation, but it doesn't account
for the restoration of the 
singlet symmetry of the ground-state at low temperatures. 
The resolution of the adiabatic and the atomic picture 
derives from 
quantum spin fluctuations which cause the
local moment to ``tunnel'' on a slow time-scale $\tau_{sf}$
between the two degenerate ``up'' and
``down'' configurations. 
\begin{equation}
e^{-}_{\downarrow }+f^{1}_{\uparrow}
\rightleftharpoons
e^{-}_{\uparrow }+f^{1}_{\downarrow}
\end{equation}
These fluctuations are the origin of the Kondo effect.
From the energy uncertainty principle, below a temperature $T_{K}$ 
at which the thermal excitation energy $k_{B}T$
is of order the characteristic tunneling rate 
$\frac{\hbar }{\tau_{sf}}$, a paramagnetic state with a Fermi liquid
resonance will form. 
The characteristic width  of the 
resonance is then determined by the Kondo energy 
$k_{B}T_{K}\sim \frac{\hbar }{\tau_{sf}}$.
The existence of this resonance was
first deduced by Abrikosov and Suhl \citep{abrikosov,suhl}, but it is more 
frequently called the ``Kondo resonance''. 
From perturbative renormalization group reasoning
 \citep{haldane}and  the
Bethe ansatz solution of the Anderson model \citep{wiegman,okiji,okiji2}
we know that for large 
$U>>\Delta$, 
the Kondo scale depends 
exponentially on $U$. In the symmetric Anderson model,
where $E_{f}=-U/2 $, 
\begin{equation}\label{kondotemp}
T_{K}= \sqrt{\frac{ 2 U\Delta }{\pi^{2}}} \exp \left(
{- \frac{\pi U}{8\Delta }} \right).
\end{equation}
The temperature $T_{K}$ marks 
the crossover from a 
a high temperature Curie law $\chi \sim \frac{1}{T}$ susceptibility to 
 a low temperature paramagnetic susceptibility $\chi \sim 1/T_{K}$. 

\subsubsection{Adiabaticity  and the Kondo resonance}

A central quantity in the physics of f-electron systems
is the f-spectral function, 
\begin{equation}
A_{f} (\omega)= \frac{1}{\pi}{\rm Im}G_{f} (\omega -i\delta )
\end{equation}
where $G_{f} (\omega)=
-i\int_{-\infty}^{\infty}dt
\langle  {\rm T}f_{\sigma }(t) f_{\sigma }\dg (0)\rangle e^{i\omega t}
$ is the Fourier transform of the time-ordered f-Green's function.
When an f-electron is added, or removed from the f-state, the
final state  has a distribution of
energies described by the f-spectral function.  
From a spectral decomposition of the f-Green's
function, the 
positive energy part of the f-spectral function
determines the energy distribution for electron addition, 
while the negative energy part measures the energy
distribution of for electron removal:
\begin{eqnarray}\label{l}
A_{f} (\omega)= \left\{
\begin{array}{lr}
\overbrace {\sum_{\lambda}
\left\vert\langle\lambda\vert
f\dg_{\sigma }\vert \phi_0\rangle \right\vert^{2} \delta (\omega - [E_{\lambda}-E_{0}]),}^{\hbox{\tiny Energy distribution of state
formed by adding one f-electron.   }}& (\omega>0)\cr
\underbrace {\sum_{\lambda}
\left\vert\langle\lambda\vert
f_{\sigma }\vert \phi_0\rangle \right\vert^{2} \delta (\omega -
[E_{0}-E_{\lambda}]),}_{\hbox{\tiny Energy distribution of state
formed by removing an f-electron}} & (\omega<0)\cr
\end{array}
 \right.
\end{eqnarray}
where $E_{0}$ is the energy of the ground-state, and $E_{\lambda}$ is
the energy of an excited state $\lambda$, formed by adding or
removing an f-electron. 
For negative energies, this spectrum 
can be measured by measuring the energy distribution of
photo-electrons produced by X-ray photo-emission, while for positive energies,
the spectral function can be measured from inverse X-ray
photo-emission \citep{allenrev,allen83}. 
The weight
beneath the Fermi energy
peak, determines the f-charge of the ion
\begin{equation}
\langle n_{f}\rangle = 2 \int_{{-\infty}}^{0}d\omega A_{f} (\omega)
\end{equation}
In a magnetic ion, such as a Cerium atom in a $4f^{1}$ state, this
quantity is just a little below unity.

\fight = 0.7\textwidth
%\fg{newfigs/adiabatic_gray.eps}{Schematic illustrating the evaluation of the f-spectral function $A_{f}
\fg{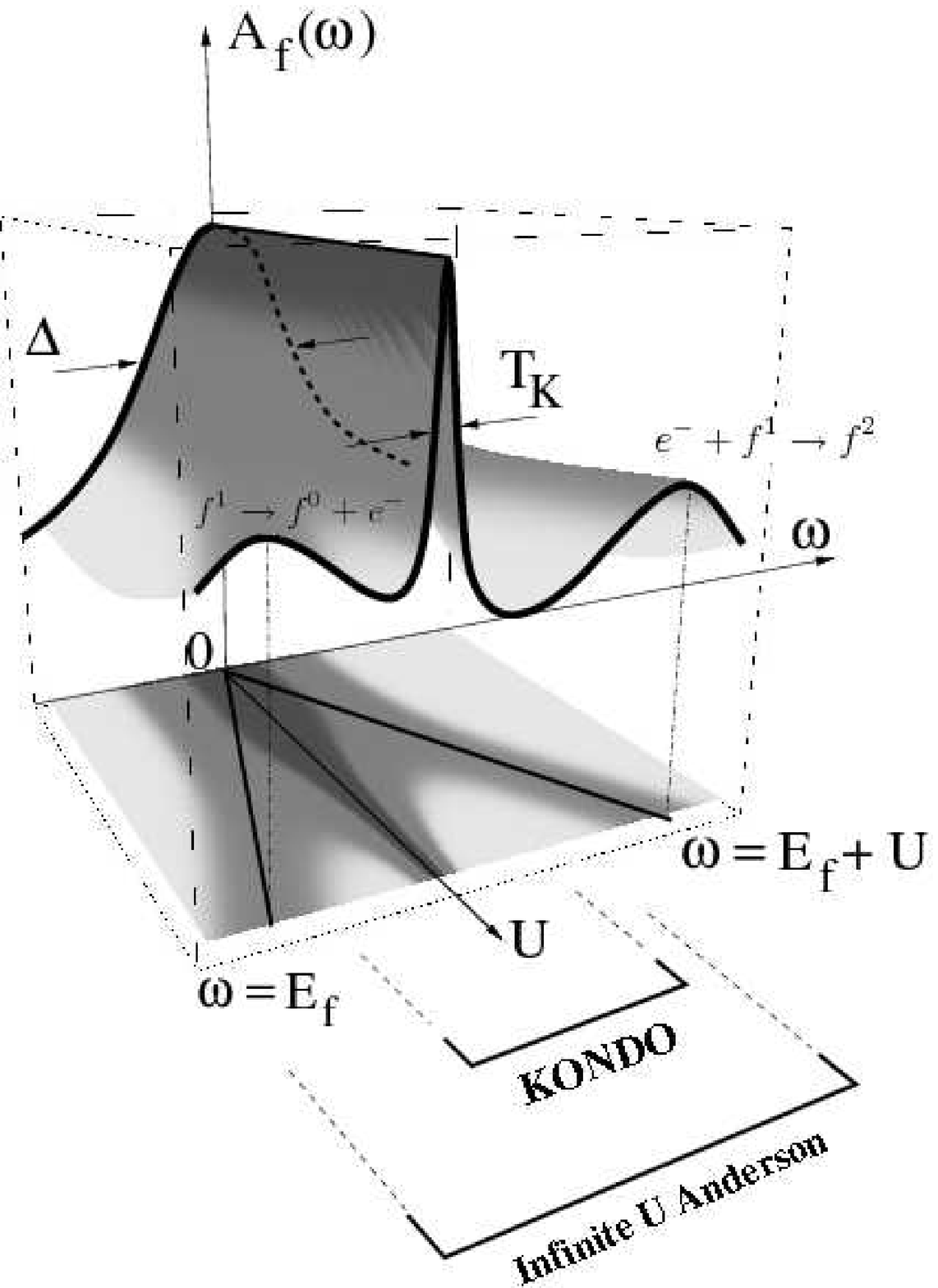}{Schematic illustrating the evaluation of the f-spectral function $A_{f}
(\omega)$ as interaction strength $U$ is turned on continuously,
maintaining a constant f-occupancy by shifting the bare f-level
position
beneath the Fermi energy. 
The lower
part of diagram is the density plot of f-spectral function, showing
how the non-interacting resonance at $U=0$ splits into 
an upper and lower atomic peak at $\omega=E_{f}$ and 
$\omega=E_{f}+U$.
}{fig11}
\fight = \textwidth

Fig. (\ref{fig11}.)  illustrates the effect of the interaction on the
f-spectral function. 
In the non-interacting limit ($U=0$), the f-spectral function is a Lorentzian
of width $\Delta $.  If we 
turn on the interaction $U$, being careful 
to shifting  the f-level position beneath the Fermi
energy
to maintain a constant occupancy, the resonance splits into
three peaks, two 
at energies $\omega=E_{f}$
and $\omega=E_{f}+U$ corresponding to the energies for a valence
fluctuation,  plus an additional central ``Kondo resonance''
associated with the spin-fluctuations of the local
moment. 

At first sight, once the interaction is much larger than the
hybridization width $\Delta $, one might expect there to be no spectral
weight left at low energies. But this violates the idea of
adiabaticity.
In fact, there are always certain adiabatic invariants that do not change,
despite the interaction.  One such quantity is the
phase shift $\delta_{f}$
associated with the scattering of conduction electrons
off the ion; another is the height of the f-spectral function at
zero energy, and it turns out that these two quantities are related.
A rigorous result due to
Langreth \citep{langreth}, tells
us that the spectral function at $\omega=0$ is diretly determined by
the f-phase shift, so that its
non-interacting value
\begin{equation}
A_{f} (\omega =0)= \frac{\sin^2\delta_{f}}{\pi \Delta },
\end{equation}
is preserved by adiabaticity.  Langreth's result can be heuristically
derived by noting that $\delta_{f}$ is the phase of the f-Green's
function  at the Fermi energy, so that $G_{f} (0-i\delta )^{-1}= \Vert
G_{f}^{-1} (0)\vert e^{-i\delta_{f}}$. Now in a Fermi liquid, the
scattering at the Fermi energy is purely elastic, and this implies
that $Im G_{f}^{-1} (0-i\delta )= \Delta $, the bare hybridization
width. From this it follows that $Im G_{f}^{-1} (0)=\vert G_{f}^{-1} (0)\vert \sin
\delta_{f}= \Delta $, so that $G_{f} (0)= e^{i\delta_{f}}/(\Delta \sin
\delta)$, and the above result follows. 

The phase shift $\delta_{f}
$ is set via the Friedel sum rule, according to which the sum of the 
up and down scattering phase shifts, 
gives the total number of f-bound-electrons, or
\begin{equation}\label{friedel}
\sum_{\sigma }\frac{\delta_{f\sigma}}{\pi} = 2 \frac{\delta_{f}}{\pi} = n_{f}.
\end{equation}
for a two-fold degenerate $f-$ state.
At large distances, the wavefunction of scattered electrons 
$\psi_{f} (r)\sim{\sin (k_{F} r
+ \delta_{f})}{/r}$ is ``shifted inwards'' by 
by a distance $\delta_{l}/k_{F}= (\lambda_{F}/2)\times
(\delta_{l}/\pi)$. 
This sum rule is sometimes called a ``node counting'' rule, because
if
you think about a large sphere enclosing the impurity, 
then each 
time the phase shift passes through $\pi$,
a node
crosses the spherical boundary and one more electron per channel is
bound beneath the Fermi sea. 
Friedel's sum rule 
holds for interacting electrons, providing
the ground-state is adiabatically accessible from the non-interacting
system \citep{langer,langreth}. Since $n_{f}=1$ in an $f^{1}$ state,
the Friedel sum rule tells us that the phase shift is  $\pi/2$ for a
two-fold degenerate $f-$ state. 
In other words, adiabaticity tell us that the electron is {\sl
resonantly scattered} by the quenched local moment. 

Photo-emission studies do reveal 
the three-peaked structure characteristic of the Anderson model 
in many 
$Ce$ systems, such as $CeIr_{2}$ and $CeRu_{2}$ \citep{allen83} (see
Fig. \ref{fig12}). Materials in which 
the Kondo resonance is wide enough to be resolved 
are more 
``mixed valent''
materials in which 
the f- valence departs significantly from unity.
Three peaked structures have also been observed
in 
certain $U$ 5f materials such as $UPt_{3}$ and
$UAl_{2}$ \citep{allen85}materials, but it has not yet been resolved in $UBe_{13}$.
A three peaked structure has recently been observed 
in 4f $Yb$  materials, such as $YbPd_{3}$, where the $4f^{13}$
configuration contains a single $f$ hole, so that the positions of the
three peaks are reversed relative to Ce \citep{allen92}.

\fight = 0.4\textwidth
%\fg{newfigs/three_peak.eps}{Showing spectral functions for three
\fg{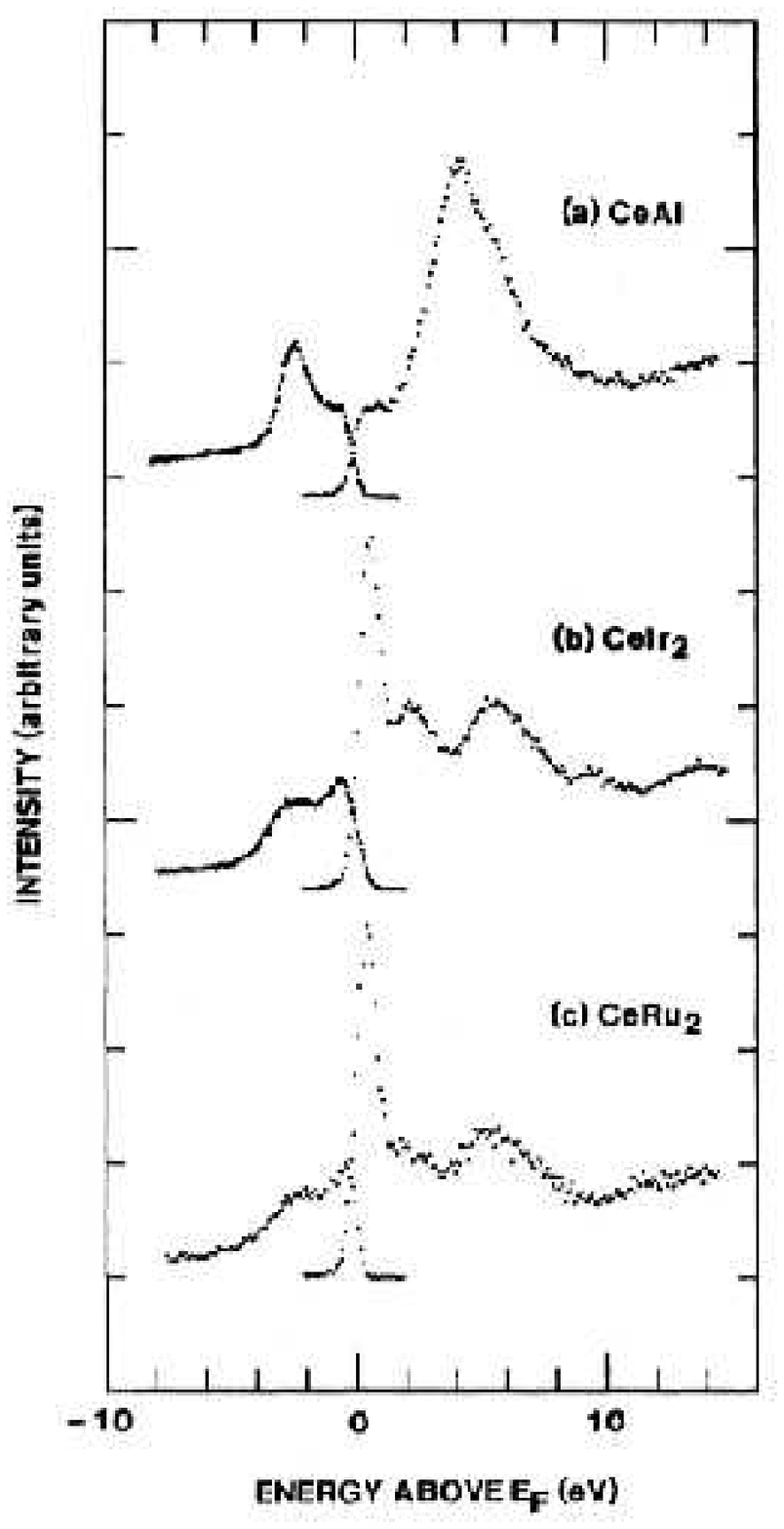}{Showing spectral functions for three
different Cerium f-electron materials, measured using
X-ray photoemission (below the Fermi energy ) and inverse X-ray
photoemission
(above the Fermi energy) after  \citep{allen83}. $CeAl$ is
an antiferromagnet and does not display a Kondo resonance.
}{fig12} 
\fight = 0.4\textwidth

\subsection{Hierachies of energy scales}\label{localmoments}
\subsubsection{Renormalization Concept}\label{}
\fight=0.9\textwidth
%\fg{newfigs/heirachyx.eps}{(a) Cross-over energy scales for the Anderson
\fg{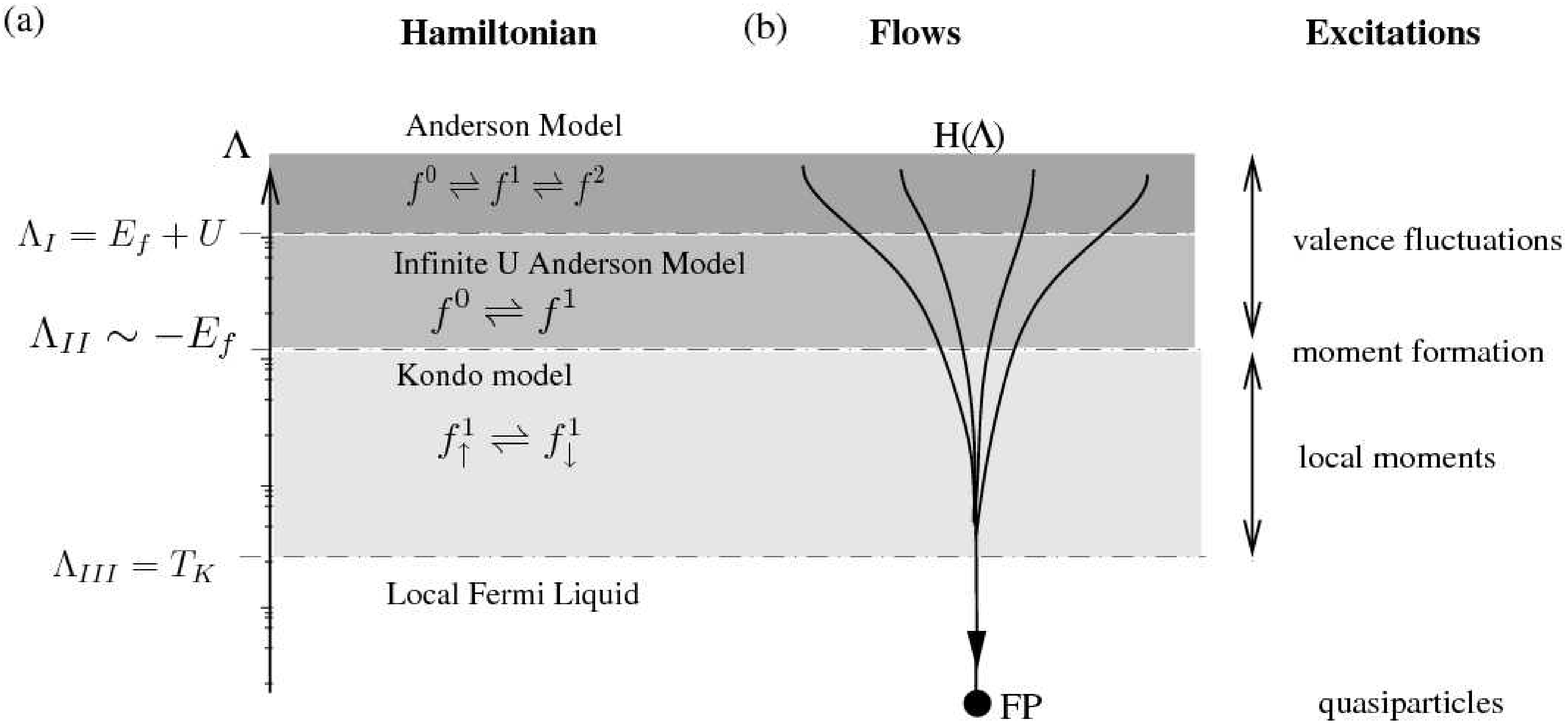}{(a) Cross-over energy scales for the Anderson
model.  
At scales below $\Lambda_{I}$, valence fluctuations into the doubly
occupied state are suppressed. All lower energy physics 
is described by the infinite $U$ Anderson model. 
Below 
$\Lambda_{II}$, all valence fluctuations are suppressed, and the
physics involves purely the 
spin degrees of freedom of the ion, coupled to the conduction sea via
the Kondo interaction. The Kondo scale renormalizes to strong coupling 
below $\Lambda_{III}$ , 
and the local moment becomes screened to form a local Fermi liquid. 
(b) Illustrating the idea of
renormalization group flows towards a Fermi liquid fixed point. 
}{fig13}

To understand how a Fermi
liquid emerges when a local moment is immersed in a quantum
sea of electrons,  theorists had to 
connect physics on on several widely spaced energy scales. 
Photoemission shows that the characteristic energy to produce
a valence fluctuation is of the order of volts, or tens of thousands
of Kelvin, yet the the characteristic physics we are interested in
occurs at scales hundreds, or thousands of times smaller. 
How can we distill the essential effects of the atomic
physics at electron volt scales on the low energy physics at millivolt
scales? 

The essential tool for this task is the ``renormalization group''
 \citep{yuval,yuval2,yuval3,poor,poor2,revwilson,revnozieres,nozieres}, 
based on the idea that the physics 
at low energy scales only depends on a small subset of ``relevant'' variables from
the original microscopic Hamiltonian. The extraction of these relevant
variables is accomplished by ``renormalizing'' the Hamiltonian
by systematically eliminating the high
energy virtual excitations and adjusting the low energy Hamiltonian to take care
of the interactions that these virtual excitations induce in the low
energy Hilbert space. This leads to a 
 family of Hamiltonian's $H
(\Lambda)$, 
each with a different 
high-energy cut-off $\Lambda$,  which share the same 
low energy physics. 

The systematic 
passage from a Hamiltonian $H (\Lambda)$  to a renormalized Hamiltonian $H
(\Lambda')$ with a smaller cutoff $\Lambda' = \Lambda/b$ is
accomplished by dividing the the eigenstates of  $H$ into a 
a  low energy subspace $\{L \}$ and  a high energy subspace $\{H \}$, with energies $\vert \epsilon\vert  <
\Lambda '=\Lambda/b$ and a 
$\vert \epsilon\vert \in [\Lambda
',\Lambda]$ respectively. 
The Hamiltonian is then broken up into terms that are
block-diagonal in these subspaces, 
\begin{equation}\label{bdiag}
H=\nmat{H_{L}}{V\dg }{V}{H_{H}},
\end{equation}
where  $V$ and
$V\dg $ provide the matrix elements between $\{L \}$ and $\{H \}$. 
The effects of the
$V$ are then taken into account by 
carrying  out a  unitary (canonical) transformation  
that block-diagonalizes the Hamiltonian, 
\begin{equation}\label{cantrans}
H (\Lambda)\rightarrow  U H (\Lambda) U \dg 
= \nmat{\tilde{H}_{L}} {0}{0}{\tilde{H}_{H}}
\end{equation}
The renormalized Hamiltonian is then given by  
$H (\Lambda')=\tilde{H}_{L}= H_{L}+ {\delta  H}$.  
The flow of key parameters in the Hamiltonian resulting from
this process is called a renormalization group flow.

At certain important cross-over energy scales, large tracts of the Hilbert space
associated with the Hamiltonian are projected out by the
renormalization process, and the character of the Hamiltonian changes
qualitatively. 
In the Anderson model, there are three such important energy scales,(\ref{fig13})
\begin{itemize}

\item $\Lambda_{I} = E_{f}+U$ where valence fluctuations
$e^{-}+f^{1}\rightleftharpoons f^{2}$ into the doubly occupied $f^{2}$
state are eliminated. For $\Lambda<< \Lambda_{I}$, 
the physics is described by the  infinite $U$ Anderson model
\begin{equation}\label{}
H= {\sum_{k,\sigma}\epsilon_k n_{k\sigma } + 
\sum_{k,\sigma}
V (k)\left[c\dg_{k\sigma }X_{0\sigma }+ X_{\sigma 0}c_{k\sigma } \right]
}
+E_{f}\sum_{\sigma} X_{\sigma \sigma },
\end{equation}
where $X_{\sigma \sigma }=\vert f^{1}:\sigma \rangle \langle
f^{1}:\sigma \vert $,
$X_{0\sigma }= \vert f^{0}\rangle \langle f^{1}{\sigma }\vert $
and 
$X_{\sigma 0}= \vert f^{1}:\sigma \rangle \langle f^{0}\vert$
are ``Hubbard operators'' that connect the states 
in the projected Hilbert space with no double occupancy.

\item $\Lambda_{II} \sim\vert E_{f}\vert= -E_{f}$, where valence fluctuations into the
empty state $f^{1}\rightleftharpoons f^{0}+e^{-}$ are eliminated to
form a local moment. Physics below this scale is described by the
Kondo model 

\item $\Lambda=T_{K}$, the Kondo temperature below which the local
moment is screened to form a resonantly scattering local Fermi liquid.
\end{itemize}
In the symmetric Anderson model, $\Lambda_{I}=\Lambda_{II}$, and the
transition to local moment behavior occurs in a one-step crossover process.

\subsubsection{Schrieffer Wolff transformation}\label{}

The unitary, or canonical transformation that eliminates the charge
fluctuations at scales $\Lambda_{I}$ and $\Lambda_{II}$
was first
carried out by Schrieffer and Wolff \citep{swolf,coqblin}, who showed how this 
model gives rise to a residual antiferromagnetic interaction between
the local moment and conduction electrons. 
The emergence of this antiferromagnetic interaction 
is associated with  a process called ``superexchange'': the virtual process 
in which an electron or hole briefly migrates off the ion, to be
immediately replaced by another with a different spin.
When these processes are removed by the canonical
transformation, they induce an antiferromagnetic interaction between the
local moment and the conduction electrons. This can be seen by considering  the 
two possible spin exchange processes 
\begin{eqnarray}\label{}
e^{-}_{\uparrow}+ f^{1}_{{\downarrow }}&\leftrightarrow& f^{2}
\leftrightarrow e^{-}_{\downarrow }+f^{1}_{{\uparrow}}
\qquad \Delta E_{I} \sim
U + E_{f}\cr
h^{+}_{\uparrow}+ f^{1}_{{\downarrow }}&\leftrightarrow& 
f^{0}\leftrightarrow h^{+}_{\downarrow }+f^{1}_{{\uparrow}}
\qquad \Delta E_{II} \sim
-E_{f}
\end{eqnarray}
Both process requires that the f-electron and  
incoming particle are in a
spin-singlet. From 
second order perturbation theory, 
the energy of the singlet is  lowered by 
an amount $-2J$ where
\begin{eqnarray}
J =  V^{2}\left[\frac{1}{\Delta E_{1}}+ \frac{1}{\Delta E_{2}}
\right] ,
\end{eqnarray}
and the  factor of two derives from the two ways a singlet 
can emit an electron or hole into the continuum
\footnote{
To calculate the matrix elements associated with valence fluctuations,
take 
\[
\vert f^{1}c^{1}\rangle = \frac{1}{\sqrt{2}}
(
f\dg_{\uparrow}c\dg_{\downarrow }-
c\dg_{\uparrow}f\dg_{\downarrow })\vert 0\rangle 
,\qquad 
\vert f^{2}\rangle = f\dg_{\uparrow}f\dg_{\downarrow }\vert
0\rangle 
\qquad 
\hbox{and} \qquad 
\vert c^{2}\rangle = c\dg_{\uparrow}c\dg_{\downarrow }\vert
0\rangle 
\]
 then 
$\langle c^{2}\vert \sum_{\sigma}Vc\dg_{\sigma}f_{\sigma}\vert
 f^{1}c^{1}\rangle = \sqrt{2}V
$ 
and 
$\langle f^{2}\vert \sum_{\sigma}Vf\dg_{\sigma}c_{\sigma}\vert
 f^{1}c^{1}\rangle = \sqrt{2}V
$
}
and $V\sim V (k_{F})$ is the hybridization matrix element near
the Fermi surface.  
For the symmetric Anderson model, where $\Delta
E_{1}=\Delta E_{II}= U/2$, $J = 4V^{2}/U $. 

 If we introduce the electron spin density operator
$\vec{\sigma } (0)= \frac{1}{{\cal N}}\sum_{k,k'}c\dg _{k\alpha }\vec{\sigma
}_{\alpha \beta }c _{k'\beta }
$, where ${\cal N}$
is the number of sites in the lattice, then  the effective interaction
will have the form
\begin{equation}\label{}
H_{K} = - 2J P_{S=0}
\end{equation}
where $P_{S=0}= \left[\frac{1}{4}- \frac{1}{2}\vec{\sigma } (0)\cdot
\vec{S}_{f}\right]$ is the singlet projection operator.
If we drop the constant term, then the 
effective interaction induced by the virtual charge fluctuations must 
have the form
\begin{equation}
H_{K}= J \vec{\sigma} (0)\cdot \vec{S}_{f}
\end{equation}
where $\vec{S}_{f}$ is the spin of the localized moment.   The
complete ``Kondo Model'', $H = H_{c}+H_{K} $
describing the conduction electrons and their
interaction with the local moment is 
\begin{equation}\label{}
H = 
\sum_{\bk \si}\epsilon_{\bk}
c\dg _{\vk  \sigma  }c _{\vk  \sigma  }
+ J \vec{\sigma} (0)\cdot \vec{S}_{f}.
\end{equation}

\subsubsection{The Kondo Effect}\label{}

The  anti-ferromagnetic sign of the super-exchange interaction $J$  in the Kondo
Hamiltonian is the origin of the spin screening physics of the Kondo
effect.  The bare interaction
is weak, but  the spin fluctuations it induces have the effect of
{\sl antiscreening } the 
interaction at low energies, renormalizing it to larger and larger values.
To see this, we follow a Anderson's ``Poor Man's'' scaling
procedure \citep{poor,poor2}, which takes advantage of the observation  that
at small $J$
the renormalization  in the 
Hamiltonian associated with 
the block-diagonalization process 
$\delta H = \tilde{H}_{L}- H_{L}$ is given by second-order
perturbation theory:
\begin{equation}
\delta H_{ab}= \langle a\vert \delta H\vert b\rangle = 
\frac{1}{2}\left[T_{ab} (E_{a})+T_{ab} (E_{b})
\right]
\end{equation}
where  
\begin{equation}
T_{ab} (\omega) = \sum_{\vert \Lambda\rangle  \in \{ H\}}
\left[\frac{V\dg _{a\Lambda
}V_{\Lambda b}}{\omega-E_{\Lambda }}\right] 
\end{equation}
is the many body ``t-matrix'' associated with virtual transitions into
the high-energy subspace $\{ H\}$. 
For the Kondo model, 
\begin{equation}
V = {\cal  P}_{H} J\vec{ S} (0)\cdot \vec{S}_{d}{\cal P}_{L} 
\end{equation}
where ${\cal P}_{H}$ projects the intermediate state into the high energy
subspace while ${\cal P}_{L}$ projects the initial state into the
low energy subspace.  There are two virtual scattering
processes that contribute the the antiscreening effect, 
involving a high energy electron (I) or a high energy
hole (II).

\underline{Process I} is denoted by the diagram
\vskip 0.1in
\bxwidth = 3in
%\centerline {\frm{figures/scat1.eps}}
\centerline {\frm{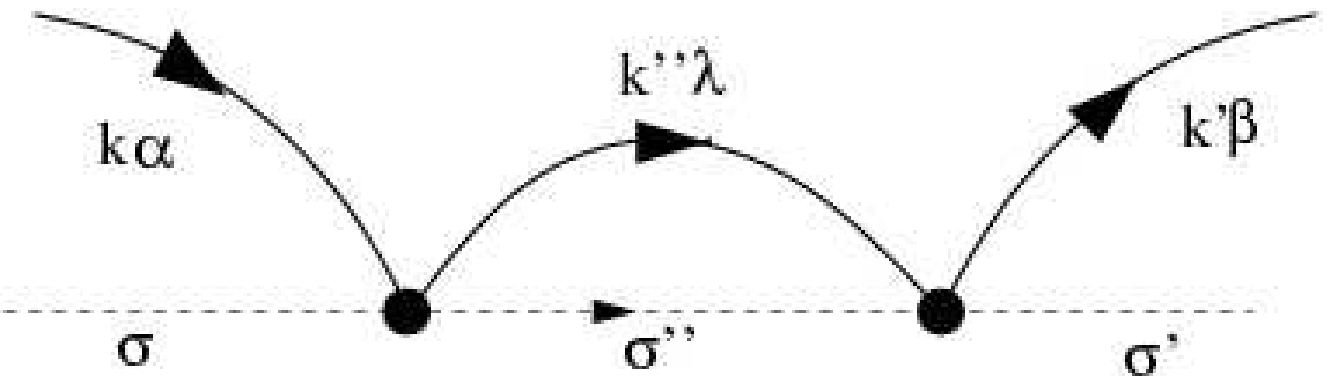}}

\noindent 
and starts in state $\vert
b\rangle = \vert  k \alpha , \sigma \rangle$ , passes through a
virtual state
$\vert \Lambda\rangle = \vert  c\dg_{k''}\alpha \sigma ''\rangle $
where $\epsilon_{k''}$ lies at high energies in the range
$\epsilon_{k''}\in [\Lambda/b,\Lambda]$ and ends in state
$\vert a\rangle = \vert  k' \beta , \sigma'\rangle $. 
The resulting renormalization
\begin{eqnarray}\label{proca}
\langle k' \beta , \sigma ' \vert T^{I} (E)\vert  k\alpha , \sigma \rangle = 
&=& 
\sum_{\epsilon _{k''} \in [\Lambda-\delta \Lambda,\Lambda]}
\left[\frac{1}{E-\epsilon _{k''}}
\right]
J^{2 }\times 
(\sigma ^{a}_{\beta \lambda}\sigma ^{b}_{\lambda \alpha })
( S^{a}_{\sigma ' \sigma ''}S^{b}_{\sigma'' \sigma })\cr
&\approx &J^{2}\rho  \delta \Lambda
\left[\frac{1}{E-\Lambda}
\right]
(\sigma ^{a}\sigma ^{b})_{\beta \alpha }
(S^{a}S^{b})_{\sigma' \sigma }
\end{eqnarray}

In \underline{Process II}, denoted by \\
%\centerline{\frm{newfigs/scat2.eps}}
\centerline{\frm{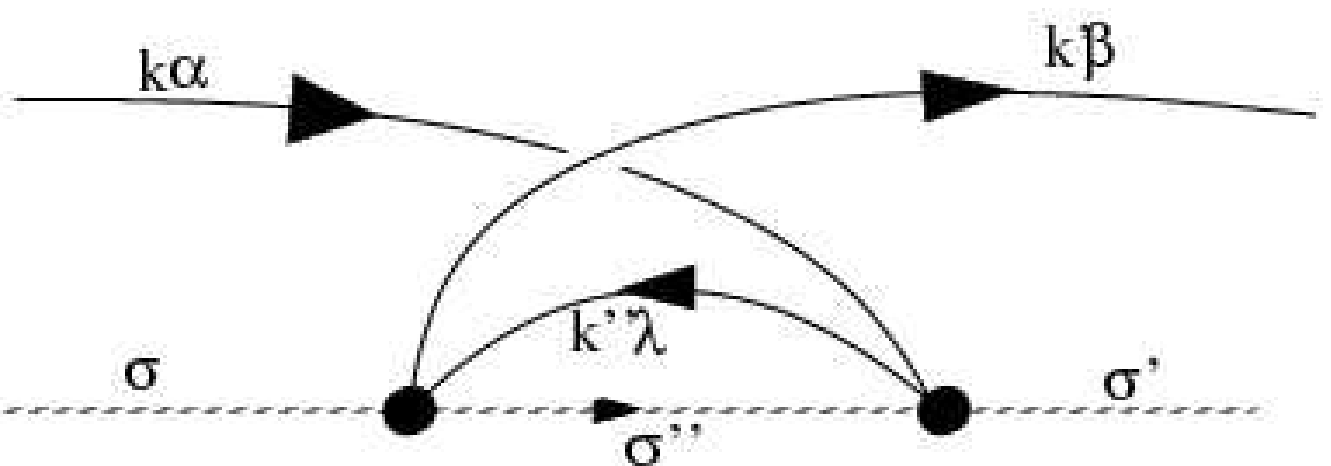}}

\noindent 
the formation of a virtual hole excitation $\vert \Lambda \rangle=
c_{k''\lambda}\vert \sigma ''\rangle  $ introduces an electron line that crosses itself, introducing a
negative sign into the scattering amplitude.
The spin operators of the conduction sea and antiferromagnet
reverse their relative order in process II, which introduces a
relative minus sign into the T-matrix
for scattering into a high-energy hole-state, 
\begin{eqnarray}\label{procb}
\langle k'\beta \sigma' \vert T^{(II)} (E)
\vert 
k\alpha \sigma \rangle 
&=& 
-\sum_{\epsilon _{k''} \in [-\Lambda,-\Lambda+\delta \Lambda]}
\left[\frac{1}{E- (\epsilon _{k}+\epsilon_{k'}-\epsilon _{k''})}
\right]
J^{2 }(\sigma ^{b}\sigma ^{a})_{\beta \alpha }
(S^{a}S^{b})_{\sigma' \sigma }\cr
&=&-J^{2}\rho  \delta \Lambda
\left[\frac{1}{E-\Lambda}
\right]
(\sigma ^{a}\sigma ^{b})_{\beta \alpha }
(S^{a}S^{b})_{\sigma' \sigma }
\end{eqnarray}
where we have assumed that 
the energies $\epsilon _{k}$ and $\epsilon _{k'}$  are negligible
compared with $\Lambda$. 

Adding (Eq. \ref{proca}) and
(Eq. \ref{procb}) gives 
\begin{eqnarray}\label{}
\delta H^{int}_{k'\beta \sigma';k\alpha \sigma }&=& \hat T^{I} + T^{II}= 
-\frac{J^{2}\rho \delta \Lambda}{\Lambda}
 [\sigma ^{a}, \sigma ^{b}]_{\beta \alpha } S^{a}S^{b}\cr
&=&2\frac{J^{2}\rho \delta \Lambda}{\Lambda}
\vec{ \sigma }_{\beta \alpha }\vec{ S}_{\sigma
'\sigma }.
\end{eqnarray}
so the high energy virtual spin fluctuations
enhance or ``anti-screen'' 
the Kondo coupling constant
\begin{equation}
J (\Lambda')= J (\Lambda) + 2 J^{2}\rho \frac{\delta \Lambda}{\Lambda}
\end{equation}
If we introduce the coupling constant $g = \rho J$, recognizing that 
$d \ln  \Lambda= - \frac{\delta \Lambda}{\Lambda}$we see that it
satisfies
\narrowboxit{
\begin{equation}
\frac{\partial g}{\partial \ln  \Lambda} =\beta (g)= - 2 g^{2} + O (g^{3}).
\end{equation}}
This is an example of a \underline{negative} $\beta $ function:  
a signature of an interaction which grows with the renormalization
process. At high energies, the
weakly coupled local moment
is said to be \underline{asymptotically
free}.
The solution to the scaling equation is
\begin{equation}\label{}
g (\Lambda')= \frac{g_{o}}{1 - 2 g_{o} \ln (\Lambda/\Lambda')}
\end{equation}
and if we introduce the ``Kondo temperature'' 
\begin{equation}\label{}
T_{K} = D\exp\left[-\frac{1}{2 g_{o} } \right]
\end{equation}we see that this can be written
\begin{equation}
2g (\Lambda')= \frac{1}{\ln  (\Lambda/T_{K})}
\end{equation}
so that once $\Lambda'\sim T_{K}$, the coupling constant becomes of order
one - at lower energies, one reaches ``strong coupling'' where the
Kondo coupling can no longer be treated as a weak perturbation.
One of
the fascinating things about this flow to strong coupling, is that
in the limit $T_{K}<<D$, 
all explicit dependence on the bandwidth $D$ disappears and the Kondo
temperature $T_{K}$ is the only intrinsic energy scale in the physics. Any physical
quantity must depend in a universal way on ratios of energy to
$T_{K}$, thus 
the universal part of the 
Free energy must have the form
\begin{equation}
F (T) = T_{K } \Phi  (T/T_{K}), 
\end{equation}
where $\Phi (x)$ is universal. We can also understand the
resistance created by spin-flip scattering off a magnetic impurity in
the same way. The resistivity is given by $\rho_{i} = \frac{ne^{2}}{m}\tau
(T,H)$ where the scattering rate must also have a scaling form
\begin{equation}
\tau (T,H) = \frac{n_{i}}{\rho } \Phi_{2} (T/T_{K}, H/T_{K})
\end{equation}
where $\rho $ is the density of states (per spin) of electrons and 
$n_{i}$ is the
concentration of magnetic impurities and 
the function $\Phi_{2}(t,h)$ is universal.
To leading order in the Born approximation,
the scattering rate is given by $\tau = 2 \pi \rho  J^{2} S (S+1). =
\frac{2 \pi S (S+1)}{\rho } 
(g_0)^{2}
$ where $g_{0}=g (\Lambda_{0})$ is the bare coupling at the energy
scale that moments form.
We can obtain the behavior at a finite temperature by replacing
$g_{0}\rightarrow g (\Lambda=2 \pi T)$, where upon
\begin{equation}
\tau (T) =  \frac{2 \pi S (S+1)}{\rho } \frac{1}{4\ln^{2} (2 \pi T/T_{K})}
\end{equation}
gives the leading high temperature growth of the resistance associated
with the Kondo effect. 

The kind of perturbative analysis we have gone through here takes us
down to the Kondo temperature. The physics at lower energies
requires corresponds to the strong coupling limit 
of the Kondo model. 
Qualitatively, once $J\rho >>1$, the local moment is 
bound into a spin-singlet with a conduction electron. 
The number of bound-electrons is $n_{f}=1$, so that by the Friedel sum
rule (eq. \ref{friedel}) in a paramagnet the phase shift 
$\delta_{\uparrow}=\delta_{\downarrow }=\pi/ 2$,
the unitary limit of scattering.   
For
more details about the local Fermi liquid that forms, we refer the reader to the accompanying chapter on the
Kondo effect by Barbara Jones \citep{jones}.

\subsubsection{Doniach's Kondo Lattice Concept}\label{}

The discovery of heavy electron metals
prompted Doniach \citep{doniach} to make the radical proposal that 
heavy electron materials derive from a dense lattice version of the Kondo
effect, described by 
the {\bf Kondo Lattice model} \citep{rkky2} 
{\begin{equation}\label{}
H=\sum_{\bk\sigma }\epsilon_{\bk}c\dg _{\bk\si
}c_{\bk\si}
+ J\sum_{j} \vec{S}_{j}\cdot c\dg _{\bk\alpha
}{\vec{\sigma }} _{\alpha \beta }c_{\bk'\beta
}e^{i (\bk'- \bk)\cdot {\bR }_{j}}
\end{equation}
In effect, Doniach was implicitly proposing that the key
physics of heavy electron materials 
resides in the interaction of neutral local moments
with a charged conduction electron sea. 

Most local moment systems develop  antiferromagnetic order at
low temperatures. 
A  magnetic moment 
at location ${\bx}_{0}$
induces a wave of ``Friedel'' oscillations in the 
electron spin density (Fig. \ref{fig16})
\begin{equation}
\langle \vec{\sigma } (\bx)\rangle = -J\chi ({\bx}-{\bx}_{0})
\langle \vec{S} ({\bx}_{0})\rangle 
\end{equation}
where 
\begin{eqnarray}\label{}
\chi (x)=2 \sum_{\bk, \vec{ k}'}
\left(
\frac{f (\epsilon _{\bk})-f (\epsilon
_{\bk'})
}{\epsilon _{\bk'}-
\epsilon _{\bk}} \right)e^{i
(\bk-\bk')\cdot {\bx}}
\end{eqnarray}
is the non-local susceptibility of the metal. 
\fight=0.8\textwidth
%\fg{newfigs/friedel.eps}{Spin polarization around 
\fg{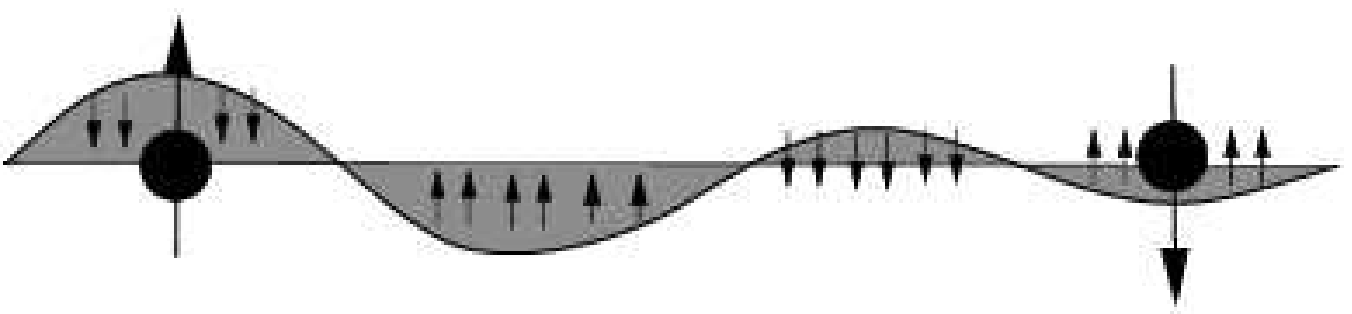}{Spin polarization around 
magnetic impurity contains Friedel oscillations and induces an RKKY
interaction
between the spins
}{fig16}
The sharp discontinuity in the occupancies $f (\epsilon_{k})$
at the Fermi surface 
is responsible for Friedel oscillations in induced spin density 
that decay with a power-law
\begin{equation}\label{}
\langle \vec\sigma (r)\rangle 
 \sim -J\rho  \frac{\cos 2 k_{F}r  }{\vert
k_{F}r\vert ^{3}
}
\end{equation} 
where $\rho $ is the conduction electron density of states and $r$ is
the distance from the impurity. 
If a second local moment is introduced at location ${\bx}$, 
it couples to this Friedel oscillation with energy $J\langle 
\vec{S} (\bx)\cdot \vec{\sigma } (x)\rangle $ giving rise to 
the ``RKKY''  \citep{rkky,rkky2,rkky3} magnetic  interaction,
\begin{equation}\label{}
H_{RKKY} = \overbrace {- J^{2 }\chi ({\bx}- {\bx}')}^{{J_{RKKY}
({\bx}-{\bx}')}} \vec{S} (\bx)\cdot
 \vec{S} (\bx').
\end{equation}
where 
\begin{equation}
J_{RKKY} (r)\sim -J^{2}\rho  \frac{\cos 2 k_{F}r  }{k_{F}r}.
\end{equation}
In alloys containing a dilute concentration of
magnetic transition metal ions, the oscillatory RKKY interaction gives rise 
to a frustrated, glassy magnetic state known as a ``spin glass''. 
In dense systems, the RKKY interaction typically 
gives rise to an ordered antiferromagnetic state with a N{\'e}el temperature
$T_{N}$ of order $J^{2}\rho $. Heavy electron metals narrowly escape
this fate.

%\fight=0.6\textwidth \fg{newfigs/doniach.eps}{Doniach diagram, illustrating the antiferromagnetic
\fight=0.6\textwidth \fg{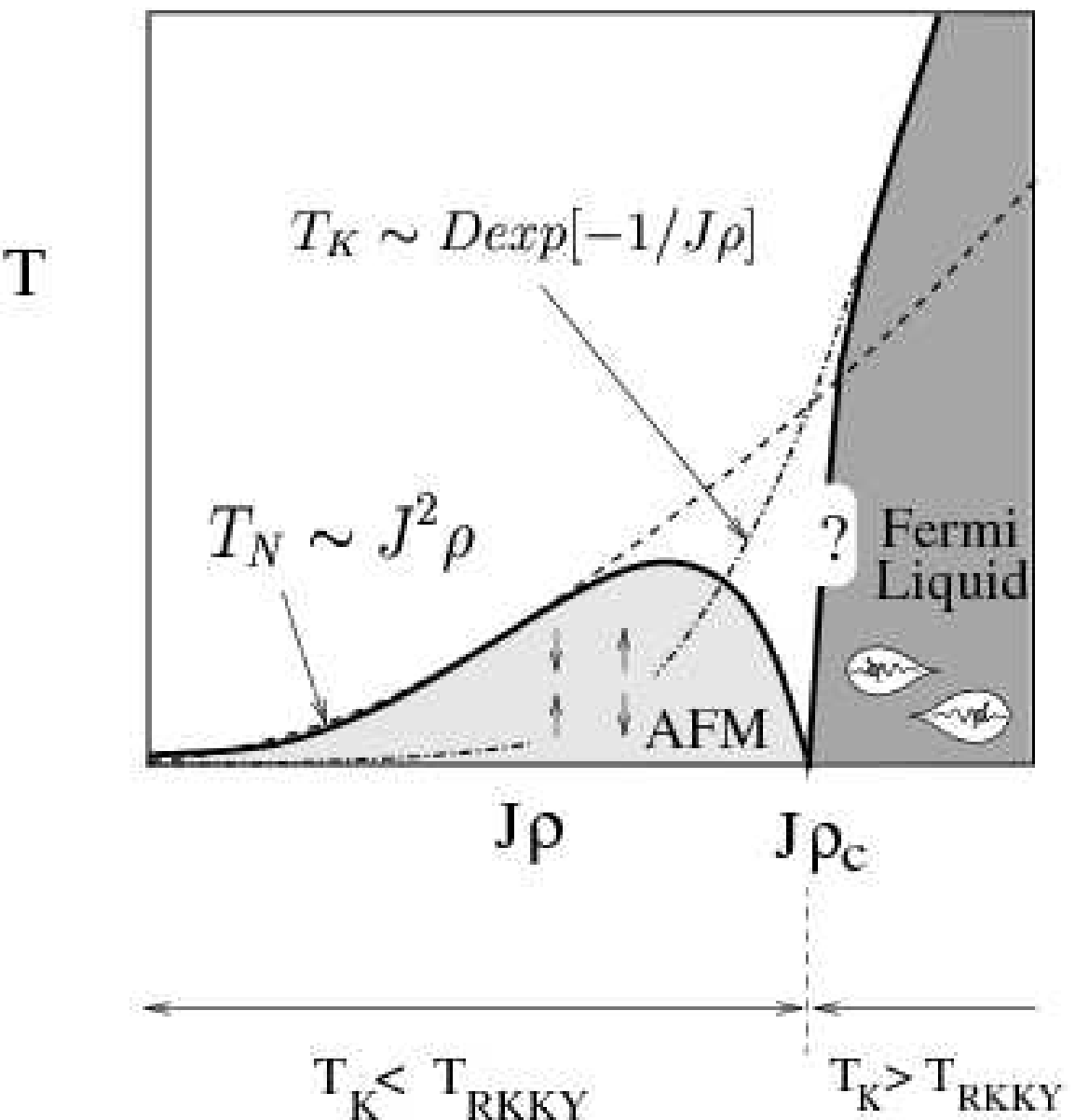}{Doniach diagram, illustrating the antiferromagnetic
regime,
where $T_{K}<T_{RKKY}$ and the heavy fermion regime, where $T_{K}>
T_{RKKY}$. Experiment has told us in recent times that the transition
between these two regimes is a quantum critical point.  The effective 
Fermi temperature of the heavy Fermi liquid is indicated as a solid
line. Circumstantial experimental evidence suggests that this scale
drops to zero at the antiferromagnetic quantum critical point, but 
this is still a matter of controversy. 
}{fig17}

Doniach argued that there are two scales in the Kondo lattice, the
single ion Kondo
temperature $T_{K} $ and $T_{RKKY}$, given by
\begin{eqnarray}\label{}
T_{K}&=& D e^{-1 / ( 2 J \rho) }\cr
T_{RKKY}&=& J^{2}\rho 
\end{eqnarray}
When $J\rho  $ is small, then $T_{RKKY}$ is the largest scale and an antiferromagnetic
state is formed, but when 
the $J\rho$ is large, the 
Kondo temperature is the largest scale so 
a dense Kondo lattice ground-state becomes stable. 
In this paramagnetic state, 
each site resonantly scatters electrons with a phase  shift $\sim \pi/ 2$.
Bloch's theorem then insures
that the resonant elastic scattering at each site will act 
coherently, forming 
a renormalized band of width $\sim T_{K}$ (Fig. \ref{fig17}). 

As in the impurity model,  one can identify
the Kondo lattice ground-state with the large $U$ limit of the
Anderson lattice model. By appealing to adiabaticity, one can then link the
excitations to the small $U$ Anderson lattice model. According to
this line of argument, the quasiparticle Fermi surface volume must
count the number of conduction and f-electrons \citep{martin82} even 
in the large $U$ limit, where it corresponds to 
the number of electrons {\sl plus} the number of spins
\begin{equation}
2 \frac{{\cal V}_{FS}}{(2 \pi )^{3}}= n_{e }+n_{\hbox{spins}}.
\end{equation}
Using topology, and certain basic
assumptions about the response of a Fermi liquid to a flux, 
Oshikawa \citep{oshikawa}  has been able to short-circuit this tortuous
path of reasoning, proving that the Luttinger relationship 
holds for the Kondo lattice model without reference to
its finite $U$ origins.

\fight=0.7\textwidth
%\fg{newfigs/contra.eps}{Contrasting (a) the ``screening cloud'' picture of the
\fg{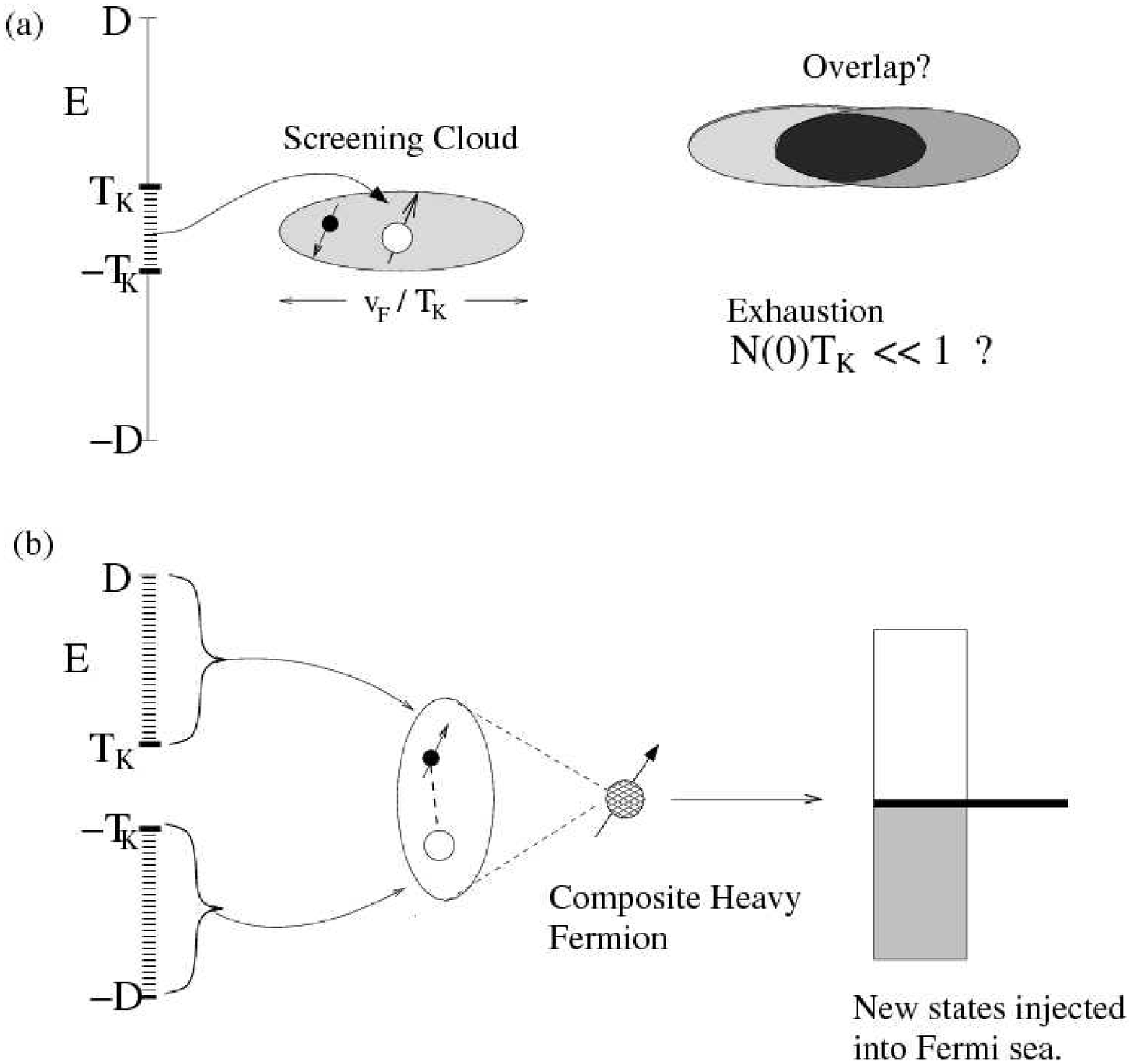}{Contrasting (a) the ``screening cloud'' picture of the
Kondo effect with (b) the composite fermion picture. In (a), 
low energy electrons form the Kondo singlet, leading to
the exhaustion problem. 
In (b) the composite heavy electron is a highly localized
bound-state between local moments and high energy electrons which
injects new electronic states into the conduction sea at the chemical
potential.  Hybridization of these states with conduction electrons 
produces a singlet ground-state, forming 
a Kondo resonance in the single impurity, and a coherent heavy
electron band in the Kondo lattice.
}{fig18}
\fight=3 truein

There are however, aspects to the Doniach argument that leave
cause for concern:
\begin{itemize}

\item it  is purely a comparison
of energy scales and does not provide a detailed mechanism 
connecting the heavy fermion phase to the 
local moment antiferromagnet. 

\item simple estimates of the value
of $J\rho $ required for heavy electron behavior give an artificially
large  value of the coupling constant $J\rho
\sim 1$.  
This issue was later resolved 
by the observation that large spin degeneracy $2j+1$
of the 
spin-orbit coupled moments, which can be as large as $N=8$ in $Yb$
materials,  enhances the rate of scaling to strong coupling,
leading to a Kondo temperature \citep{me}
\begin{equation}
T_{K } = D (NJ \rho )^{\frac{1}{N}}\exp \left[ - \frac{1}{N J \rho } \right]
\end{equation}
Since the scaling enhancement effect stretches out across decades of
energy, it is largely robust against crystal fields  \citep{sato}. 

\item Nozi\` eres' exhaustion paradox \citep{exhaustion}. If one considers
each local moment to be magnetically screened by a cloud of low energy
electrons within an energy  $T_{K}$ of the Fermi energy one arrives at
an ``exhaustion paraodox''.
In this interpretation, the number of electrons 
available to screen each local moment is of order 
$T_{K}/D<<1$
per unit cell. Once the concentration of magnetic impurities
exceeds  $\frac{T_{K}}{D}\sim 0.1 \%$ for ($T_{K}=10K$,
$D=10^{4}K$), the
supply of screening electrons 
would be exhausted, logically excluding any sort of dense Kondo effect.
Experimentally, features of single ion  Kondo behavior persist to much higher
densities.  The resolution to the exhaustion paradox lies in the more modern
perception that spin-screening  of local moments 
extends {\sl up } in energy , from the 
Kondo scale $T_{K}$ out to  the bandwidth. In this respect,
Kondo screening is reminiscent of Cooper pair formation, which 
involves electron states that extend upwards from the gap energy to
the Debye cut-off.  From  this perspective, the Kondo length scale
$\xi \sim v_{F}/T_{K}$
is analogous to the coherence length of a superconductor \citep{burdin}, defining 
the length scale over which the conduction spin and local moment
magnetization are coherent without setting any limit on the degree to
which the correlation clouds can overlap ( Fig. \ref{fig18}).

\end{itemize}

%%

%%
%%      Kondo lattice
%%      Doniach argument
%% 	Coherence
%\item Large N idea
%\item  Large N Kondo lattice
%%      
%%     

\subsection{The Large $N $ Kondo Lattice}\label{}

\subsubsection{Gauge theories, Large N and strong correlation.}\label{}

The ``standard model'' for metals 
is built upon the expansion to high orders in the strength of the
interaction. This approach, pioneered by Landau, and later 
formulated in the language of finite temperature perturbation theory
by Pitaevksii, Luttinger, Ward,  Nozi\` eres and others \citep{landaufl,pitaevskii,luttingerward,nozieresfl},  provides the foundation
for our understanding of metallic behavior in most conventional
metals.

The development of a parallel formalism and approach for
strongly correlated electron systems is still in its infancy, and
there is no universally accepted approach. At the heart of the problem
are the large interactions which effectively  remove large
tracts of Hilbert space and impose strong constraints on the low-energy
electronic dynamics. 
One way to describe these highly constrained Hilbert spaces, is
through the use of gauge theories. 
When written as a field theory, local constraints manifest themselves 
as locally conserved quantities.  General principles link
these conserved quantities  with a set of gauge symmetries.
For example, in the Kondo lattice, if a spin $S=1/2$ operator
is represented by fermions,
\begin{equation}
\vec{ S}_{j} = f\dg_{j\alpha } \left(\frac{\vec\sigma }{2} \right)_{\alpha \beta }f_{j\beta },
\end{equation}
then the representation must be supplemented 
by the constraint
$n_{f} (j) =1$ on the conserved f-number at each site. 
This constraint 
means one can change the phase of each
f-fermion at each site arbitrarily
\begin{equation}
f_{j}\rightarrow e^{i\phi_j}f_{j},
\end{equation}
without changing the spin operator 
$\vec{S}_{j}$ or the Hamiltonian.
This is the local gauge symmetry.

Similar issues also arise in  the infinite $U$ Anderson or Hubbard
models where the ``no
double occupancy'' constraint can be established by using a slave boson 
representation \citep{slaveb,barnes} of Hubbard operators:
\begin{equation}
X_{\sigma 0} (j) = f\dg_{j\sigma}b_{j}, \qquad X_{0\sigma } (j) = b\dg_{j} f_{j\sigma }
\end{equation}
where $f\dg_{j\sigma }$ creates a singly occupied f-state, $f\dg_{j\sigma }\vert
0\rangle \equiv \vert f^{1},j\rangle $ while $b\dg $ creates an empty $f^{0}$
state, 
$b\dg_{j}\vert  0\rangle =
\vert f^{0},j\rangle $.  In the slave boson, the gauge charges
\begin{equation}
Q_{j}=
\sum_{\sigma }f\dg_{j\sigma }f_{j\sigma }+ b\dg_{j}
b_{j}
\end{equation}
are conserved and the physical Hilbert space corresponds
to $Q_{j}=1$ at each site.  The gauge symmetry is now
$f_{j\sigma}\rightarrow e^{i\theta_{j}}f_{j\sigma}$, 
$b_{j}\rightarrow e^{i\theta_{j}}b_{j}$. 
These two examples illustrate the link between strong correlation
and gauge theories.
\begin{equation}
\hbox{\bf 
strong correlation}\leftrightarrow \hbox{\bf constrained Hilbert Space}
\leftrightarrow \hbox{\bf gauge theories}
\end{equation}
A key feature of these gauge theories, is the appearance
of ``fractionalized fields''   which carry either spin or charge, but not
both. How then, can  a 
Landau Fermi liquid emerge within a Gauge theory with fractional
excitations ?

Some have suggested that  Fermi liquids can not reconstitute
themselves in such strongly constrained gauge theories.  
Others have advocated against gauge theories, arguing that
the only reliable way forward is to return to  ``real world'' models
with a full fermionic
Hilbert space and a finite interaction strength.  
A third possibility is that the gauge theory approach is valid, but 
that heavy quasiparticles emerge as bound-states of gauge
particles. 
Quite independently of one's position on the importance of gauge
theory approaches, the Kondo lattice poses a severe computational
challenge, in no small part, because of
the absence of any small parameter for resumed perturbation theory.
Perturbation theory in the Kondo coupling constant $J$ always fails below
the Kondo temperature. How then, 
can one develop a controlled computational tool to
explore the transition from local moment magnetism to the heavy fermi liquid?

One route forward is to seek a family
of models that interpolates between the models of physical interest,
and a limit where the physics can be solved exactly. 
One approach, as we shall discuss later, is to consider Kondo
lattices in variable dimensions $d$, and expand in powers of $1/d$
about the limit of infinite dimensionality \citep{krauth,jarrelldmft}. 
In this limit, electron self-energies become momentum independent, the
basis of the dynamical mean-field theory.
Another approach, with the advantage that it can be married with gauge theory, is the use of 
large $N$ expansions.  The idea here
is to generalize the problem
to a family of models in which the f-spin degeneracy $N=2j+1$
is artificially driven to infinity. In this extreme
limit, the key physics is captured as 
a mean-field theory, and 
 finite $N$
properties
are obtained through an expansion in the small parameter $1/N$. 
Such large $N$ expansions have played an important role 
in the context of the spherical
model of statistical mechanics \citep{berlin} and  in field theory
 \citep{witten}. 
The next section discusses
how the gauge theory of the Kondo lattice model can be treated in a large $N$ expansion.

\subsubsection{Mean field theory of the Kondo lattice}\label{}

Quantum large $N$ expansions are a kind of
semi-classical limit where  $1/N\sim \hbar $ plays the role of a synthetic
Planck's constant.
In a Feynman path integral 
\begin{equation}\label{}
\langle x_{f} (t) \vert  x_{i},0\rangle  = 
\int {\cal D}[ x]
\exp \left[ \frac{i}{\hbar}S[x,\dot x]\right]
\end{equation}
where $S$ is the classical action and 
the quantum action $A=\frac{1}{\hbar }S$ is 
``extensive'' in the variable $\frac{1}{\hbar }$. When 
$\frac{1}{\hbar }\rightarrow \infty $, fluctuations around the 
classical trajectory vanish and the transition amplitude is entirely
determined by the classical action to go from $i$ to $f$.
A large $N$ expansion for the partition function $Z$ of a  quantum system
involves a path integral in imaginary time over the fields $\phi $
\begin{equation}\label{}
Z= \int {\cal D}[ \phi ] e^{- NS[\phi ,\dot \phi ]}
\end{equation}
where $NS$ is the action (or free energy) associated with the field
configuration in space and time.
By comparison, we see 
that the large $N$ limit of quantum systems
corresponds to an alternative  classical mechanics where 
$1/N\sim \hbar  $ emulates Planck's constant 
and new types of collective behavior 
not pertinent to strongly interacting
electron systems, start to appear. 

Our model for a Kondo lattice of spins
localized at sites $j$ is
\begin{equation}\label{}
H=\sum_{\bk\sigma }\epsilon_{\bk}c\dg _{\bk\si
}c_{\bk\si}
+ \sum_{j} H_{I} (j)
\end{equation}
where 
\begin{equation}\label{coqblins}
H_{I} (j)= \frac{J}{N} S_{\alpha \beta } (j)
c \dg _{j\beta } c _{j\alpha } 
\end{equation}
is the Coqblin Schrieffer form of the Kondo interaction Hamiltonian \citep{coqblin}
between an f-spin with $N=2j+1$ spin components
and the conduction sea. 
The spin of the local moment at site  $j$ is 
represented as a bilinear of Abrikosov pseudo-fermions
\begin{equation}\label{sunspin}
S _{\alpha \beta } (j)=f\dg _{j\alpha } f_{j\beta } - \frac{n_{f}}{N}\delta_{\alpha \beta }
\end{equation}
and
\begin{equation}\label{}
c \dg _{j\alpha } = \frac{1}{\sqrt{{\cal N}_{s}}}\sum_{\bk}c\dg _{\bk\alpha
}e^{-i\bk\cdot \vec{R}_{j}}
\end{equation}
creates an electron localized at site $j$, where ${\cal N}_{s}$ is the
number of sites.

Although this is a theorists' idealization - a ``spherical cow
approximation'',  it nevertheless captures key aspects of the physics.
This model ascribes a spin degeneracy of $N=2j+1$
to the both the f-electrons {\sl and} 
the conduction electrons. While this 
is justified for a single impurity, 
a more realistic
lattice model requires the introduction of Clebsch Gordon coefficients
to link the spin-$1/2$  conduction electrons with the spin-$j$
conduction electrons.  

To obtain a mean-field theory, each term in the Hamiltonian must
scale as $N$. 
Since the interaction contains two sums over the spin variables, 
this criterion is met by rescaling 
the coupling constant 
replacing $J\rightarrow \frac{\tilde{J}}{N}$. 
Another important aspect to this model, is the constraint on charge
fluctuations, 
which in the Kondo limit imposes the constraint 
$n_{f}=1$. 
Such a constraint can be imposed in a path integral with a 
Langrange multiplier term $\lambda (n_{f}-1)$.  However, with $n_{f}=1$,
this is not extensive in $N$, and can not be treated using a
mean-field value for $\lambda$. 
The resolution is to 
generalize the constraint to
$n_{f}=Q$, where $Q$ is an integer chosen
so that as $N$ grows, $q=Q/N$ remains fixed.   Thus, for instance, if we are interested in $N=2$, this
corresponds to $q= n_{f}/N = \frac{1}{2}$.  
In the large $N$
limit, it is then sufficient to apply the constraint on the average
$\langle n_{f}\rangle = Q$, through a static Lagrange
multiplier coupled to the difference $( n_{f}-Q)$. 

The next step 
is to carry out a ``Hubbard Stratonovich''
transformation on the interaction
\begin{equation}\label{}
H_{I} (j) = -\frac{J}{N}\left(c \dg _{j\beta }f_{j\beta } \right)
\left(f\dg _{j\alpha } c_{j\alpha } \right).
\end{equation}
Here we have absorbed the term
$-\frac{J}{N}n_{f}c\dg_{j\alpha }c_{j\alpha}$
derived from the
spin-diagonal part of (\ref{sunspin}) 
by a shift
$\mu\rightarrow \mu-\frac{Jn_{f}}{N^{2}}$ in the chemical potential.
This interaction has the form $-gA\dg  A$, with $g=\frac{J}{N}$ and
$A = f\dg _{j\alpha } c_{j\alpha }$, which we  factorize 
using a Hubbard
Stratonovich transformation,
\begin{equation}
- g A\dg A \rightarrow A\dg V + \bar V A + \frac{\bar V V}{g}
\end{equation}
so that  \citep{lacroix,read}
\begin{equation}\label{interaction}
H_{I} (j)\rightarrow H_{I}[V,j]=
\bar V_{j}\left(c \dg _{j\alpha  }f_{j\alpha  } \right)
+
\left( f\dg _{j\alpha } c_{j\alpha } \right)V_{j}
 +N\frac{\bar V_{j}V_{j}}{J}.
\end{equation}
This is an exact transformation, provided the 
 $V_{j} (\tau )$ are treated as fluctuating
variables inside a path integral. The $V_{j}$ can be regarded as a spinless
exchange boson for the Kondo effect. In the parallel treatment of
the infinite Anderson model \cite[]{long}, $V_{j}= V b_{j}$ is the ``slave boson'' field
associated with  valence fluctuations. 

In diagrams:
\vskip 0.2cm
\begin{equation}\label{}
\epsfig {file=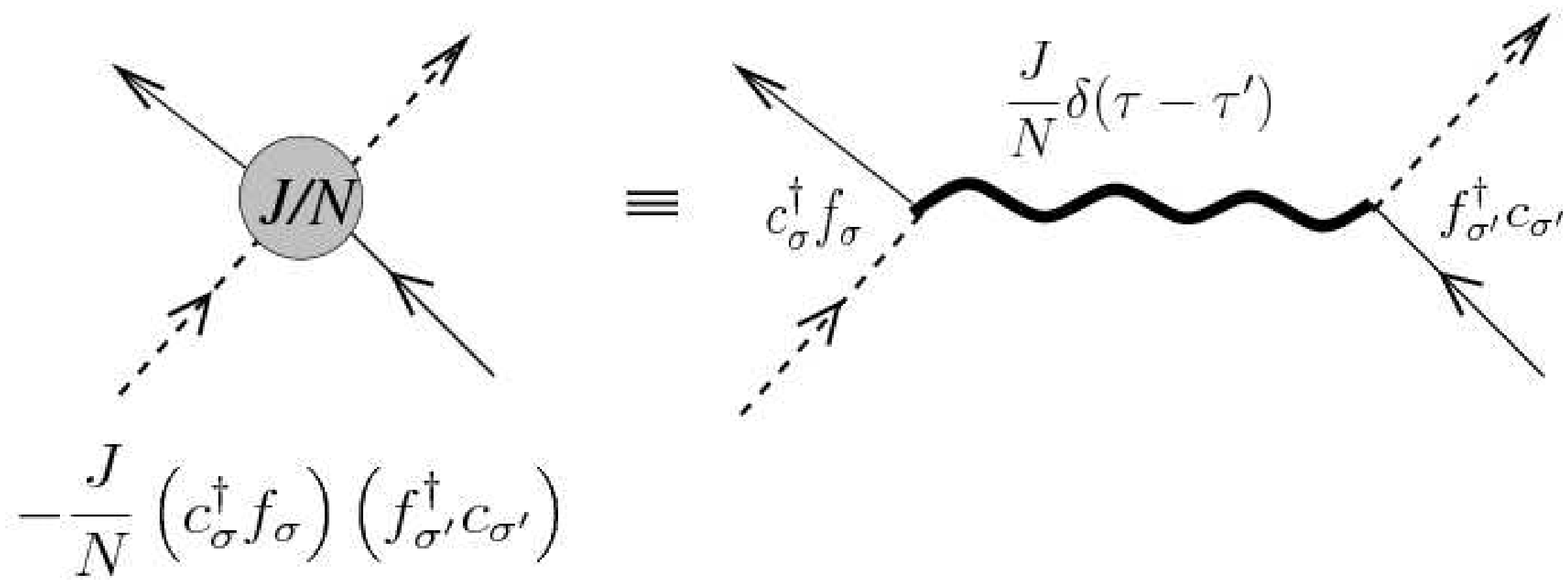,width=10cm}
\end{equation}

The path integral for the Kondo lattice is then
\begin{eqnarray}\label{}
Z= \int {\mathcal{D}}[V,\lambda ] 
\overbrace {
\int \mathcal{D}[c,f]
\exp \left[-\int_{0}^{\beta }\left(\sum_{k \sigma } c\dg_{\bk\sigma
}\partial_{\tau }c_{\bk\sigma }+ \sum_{j\sigma}f\dg_{j\sigma}\partial_{\tau}f_{j\sigma}
+
 H[V,\lambda ]\right)
 \right]}^{
= {\rm Tr}\bigl[T exp \left(
-\int_{0}^{\beta }H[V,\lambda]d\tau  \right)\bigr]}
\end{eqnarray}
 where
\begin{eqnarray}\label{pathintegral}
H[V,\lambda ]=\sum_{\bk\sigma }\epsilon_{\bk}c\dg _{\bk\si
}c_{\bk\si}
+ \sum_{j} \left(H_{I}[V_{j},j]+\lambda _{j} [n_{f} (j)-Q] \right),
\end{eqnarray}
 This is the ``Read Newns'' path integral formulation \citep{read,auerbach} of 
the Kondo lattice model. 
The path integral contains an outer integral 
$\int {\mathcal{D}}[V,\lambda] $ over the gauge fields
$V_{j}$ and $\lambda _{j} (\tau )$,  and an inner integral
$\int \mathcal{D}[c,f]$ over the fermion fields 
moving in the environment of the gauge fields. 
The inner path integral is equal to
a trace over the time-ordered exponential of $H[V,\lambda]$.

Since the action in this path integral
grows extensively with $N$, 
the large $N$ limit
is saturated by the saddle point configurations of $V$ and
$\lambda$, eliminating the
the outer integral in 
(\ref{pathintegral}). We seek a  translationally invariant, static,
saddle point where $\lambda_{j} (\tau ) =\lambda$ and
$V_{j} (\tau ) =V$.  
Since the Hamiltonian is static, the 
interior path integral can be written as the trace over the
Hamiltonian evaluated at the saddle point, 
\begin{equation}\label{}
Z= {\rm  Tr}e^{-\beta H_{MFT}}, \qquad \qquad \qquad (N\rightarrow \infty )
\end{equation}
where 
 \boxit{
\begin{equation}\label{mfham}
H_{MFT}=H[V,\lambda]=\sum_{\bk\sigma }\epsilon_{\bk}c\dg _{\bk\si
}c_{\bk\si}
+ \sum_{j,\alpha }
\left(
\bar Vc \dg _{j\beta }f_{j\beta }+
V
f\dg _{j\alpha } c_{j\alpha }
 + \lambda f\dg _{j\alpha
}f_{j\alpha }\right)
 +N n\left(\frac{\bar VV}{J}- \lambda _{o}q
 \right).
\end{equation}}
\noindent 

The saddle point is determined by the condition  that the Free
energy
$F= - T \ln Z$ is stationary with respect to variations in $V$ and
$\lambda$. To impose this condition we need to diagonalize $H_{MFT}$
and compute the Free energy.
First we rewrite the  mean field Hamiltonian in momentum space, 
\begin{equation}\label{}
H_{MFT}= \sum_{\bk\sigma }\left(c\dg _{\bk\sigma },f\dg
_{\bk\sigma } \right)
\left[
\begin{array}{cc}\epsilon _{\bk}& \bar  V\\
V&\lambda \end{array} \right]\left(
\begin{array}{c}
c_{\bk\sigma }\\
f_{\bk\sigma }
\end{array} \right)
+Nn\left(\frac{\bar VV}{J}- \lambda q\right),\end{equation}
where
\begin{equation}
f\dg _{\vec{ k}\sigma }= \frac{1}{\sqrt{\cal  N}}\sum_{j}f\dg _{j\sigma }
e^{i \vec{ k}\cdot \vec{ R}_{j}}
\end{equation}
is the Fourier transform of the $f-$electron field. 
This Hamiltonian can then be diagonalized in the form
\begin{equation}
H_{MFT}= \sum_{\bk\sigma }\left(a\dg _{\bk\sigma },b\dg
_{\bk\sigma } \right)\begin{bmatrix}
E_{\bk+}&0\cr
0&E_{\bk-}\end{bmatrix}
\begin{pmatrix}a_{\bk\sigma }\cr b_{\bk\sigma }
\end{pmatrix}
+N{\cal N}_s\left(\frac{\vert V\vert^{2}}{J}- \lambda q\right),\end{equation}
where
$a\dg _{\bk\sigma }
$ and $b\dg _{\bk\sigma }$
are linear combinations of $c\dg _{\bk\sigma }$ and $f\dg _{\vec{
k}\sigma} $ which describe the 
quasiparticles of the theory. The 
momentum state eigenvalues $E=E_{\vec{k\pm}}$ are the roots of the equation
\begin{equation}\label{roots}
{\rm  Det}
\left[E
\underline{1}-\begin{pmatrix}\epsilon _{\bk}& \bar V\cr
V&\lambda \end{pmatrix}
 \right]= (E-\epsilon_{\bk}) (E-\lambda)-\vert V\vert ^{2}= 0,
\end{equation}
so 
\begin{equation}\label{}
E_{\bk\pm } = \frac{\epsilon _{\bk}+\lambda}{2}\pm
\left[\left(\frac{\epsilon _{\bk}-\lambda }{2}
\right)^{2}+ \vert V\vert ^{2} \right]^{\frac{1}{2}}
\end{equation}
are the energies of the upper and lower bands. The dispersion
described by these energies is shown in Fig. \ref{fig20} .
\fight=4 truein
%\fg{newfigs/dispersion2.eps}{(a) Dispersion produced by the injection of a
\fg{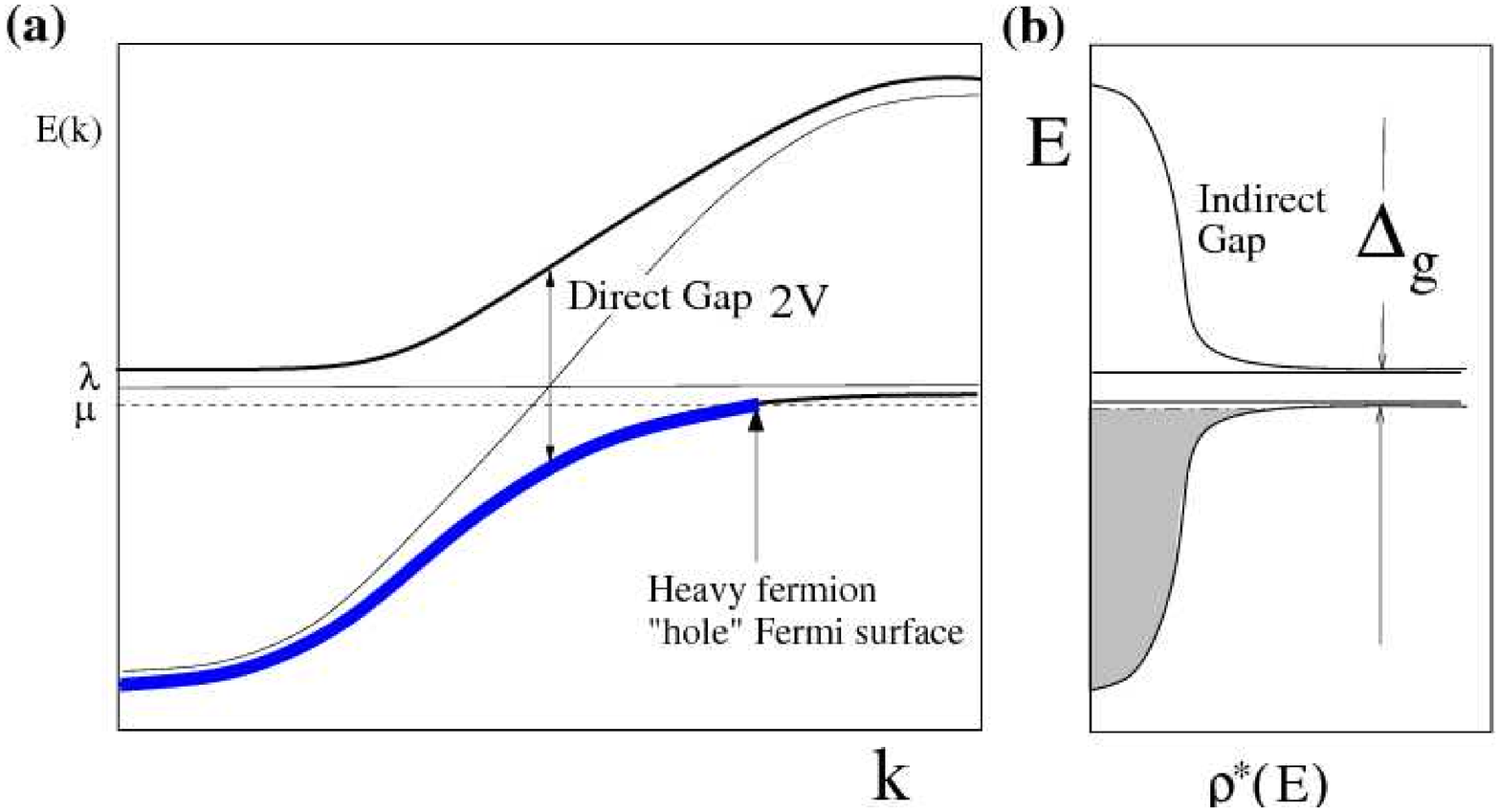}{(a) Dispersion produced by the injection of a
composite fermion into the conduction sea. 
(b) Renormalized density of states, showing ``hybridization gap''
($\Delta _{g}$).
}{fig20}
Notice that: 
\begin{itemize}
\item 
hybridization between the f-electron states and
the conduction electrons builds an upper and lower Fermi band, 
separated by an indirect ``hybridization gap'' of width 
$\Delta _{g}=E_{g} (+)-E_{g} (-)\sim T_{K}$, where
\begin{equation}\label{}
E_{g} (\pm)=\lambda \pm\frac{V^{2}}{D_{\mp}}.
\end{equation}
and $\pm D_{\pm}$ are the top and bottom of the
conduction band. The ``direct'' gap between the upper and lower bands
is $2|V|$.

\item 
From 
(\ref{roots}),
the relationship between the energy of the heavy electrons ($E$) and the
energy of the conduction electrons ($\epsilon $) is given by
$\epsilon= E- {\vert V\vert^{2}}/{( E-\lambda) }$,
so that the  density of heavy electron states
$\rho^{*} (E)=\sum_{\bk,\pm}\delta
(E-E^{(\pm)}_{\bk})$ 
 is related to the conduction
electron density of states $\rho (\epsilon) $ by
\begin{equation}\label{dens2}
\rho^{*} (E)= \rho \frac{d\epsilon }{dE} = 
\rho (\epsilon) \left(1 +
\frac{\vert V\vert ^{2}}{(E-\lambda )^{2}} \right)\sim
\left\{ 
\begin{array}{lr}
\rho \left(1 +
\frac{\vert V\vert ^{2}}{(E-\lambda )^{2}} \right)&\hbox{outside\
hybridization gap,}\cr
0&\hbox{inside
hybridization gap,}\
\end{array}
\right.
\end{equation}
so the ``hybridization gap'' is flanked by two sharp peaks
of approximate width  $T_{K}$.

%\item The effective mass of the Fermi surface has the
%\underline{opposite sign} to the original conduction sea from which it
%is built, so naively, the Hall constant would change sign as coherence
%develops. 

\item The Fermi surface volume 
\underline{expands} in response to the injection 
of heavy electrons into the conduction sea,
\begin{equation}\label{fsvolume}
N a^{D}\frac{V_{FS}}
{(2\pi)^{3}}
=
\langle \frac{1}{{\cal N}_{s}}
\sum_{\bk\sigma }n_{\bk\sigma }
\rangle = Q + n_{c}
\end{equation}
where $a^{D}$ is the unit cell volume, $n_{\bk\sigma }=a\dg _{\bk \sigma}a_{\bk \sigma }+
b\dg _{\bk \sigma}b_{\bk \sigma }
$ is the quasiparticle number operator and 
$n_{c}$ is the number of conduction electrons per unit
cell. More instructively, if $n_{e}=n_{c}/a^{D}$ is the electron  density,
\begin{equation}\label{}
\overbrace{\phantom{\frac{V_{FS}}{(2 )^{3}}}
\hskip -0.3truein
n_{e}}^{\hbox{$e^{-}$ density}}= \overbrace {N
\frac{V_{FS}}{(2\pi )^{3}}
}^{\hbox{q.particle density}}-\underbrace
{\phantom{\frac{V_{FS}}{(2\pi )^{3}}}\hskip -0.4 truein \frac{Q}{a^{D}}
}_{\hskip -0.5truein\hbox{+ve background}},
\end{equation}
so the  electron  density
$n_{c}$ 
divides into a contribution carried
by the enlarged Fermi sea, whose enlargement is 
compensated by the development of a positively charged background.
Loosely speaking, each neutral
spin in the Kondo lattice has ``ionized'' to produce
$Q$ negatively charged heavy fermions, leaving behind a
Kondo singlet of charge $+Qe$ (Fig. \ref{fig21}.).
\end{itemize}

\fight=7.2 truein
%\fg{newfigs/charge.eps}{Schematic diagram from  \citep{indranil05}.(a) High temperature state: small Fermi surface
\fg{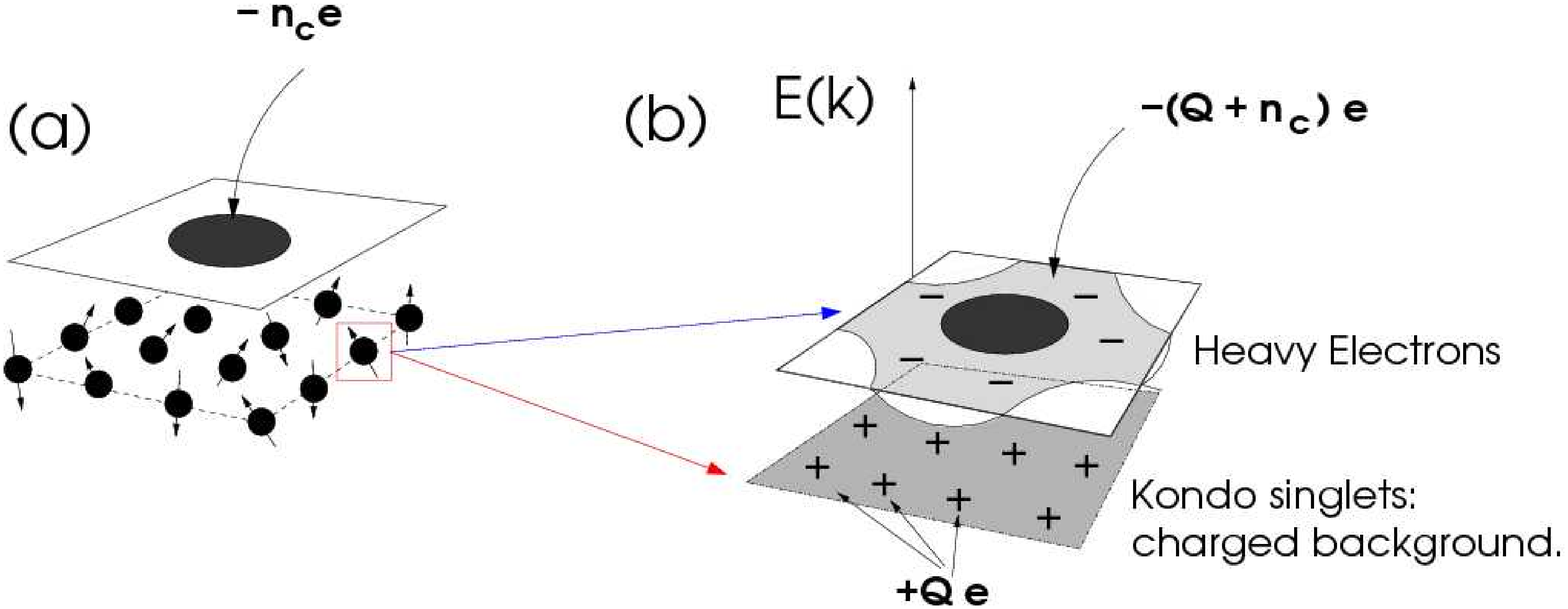}{Schematic diagram from  \citep{indranil05}.(a) High temperature state: small Fermi surface
with a background of spins; (b)Low temperature state where large Fermi
surface develops against a background of positive charge. Each spin
``ionizes'' into $Q$  heavy electrons, leaving behind  a
a Kondo singlet  with charge $+Qe$}{fig21}
\fight=4 truein

To obtain $V$ and $\lambda$ we must compute
the Free energy 
\begin{equation}
\frac{F}{N} = -  T \sum_{\bk ,\pm}\ln \biggl[
1 + e^{-\beta E_{\bk\pm}} \biggr]
+ {\cal N}_{s}\left(\frac{\vert V \vert ^2}{J}- \lambda
q\right).
\end{equation}
At $T=0$, the Free energy converges 
the ground-state energy $E_{0}$, given by 
\begin{equation}
\frac{E_{0}}{N {\cal N}_{s}}= \int_{-\infty }^{0} \rho ^{*} (E)
E + \left(\frac{\vert V \vert ^2}{J}- \lambda
q\right). 
\end{equation}
Using (\ref{dens2} ), the total energy is
\begin{eqnarray}\label{totenerg}
\frac{E_{o}}{N {\cal N}_s}&=& \int_{-D}^{0}d\epsilon\rho EdE
+ \int _{-D}^{0}dE \rho
\vert V \vert ^2\frac{E}{(E-\lambda )^{2}}+ \left(\frac{\vert V \vert ^2
}{J}- \lambda
q\right)\cr
&=& 
\overbrace {
-\frac{D^{2}\rho }{2}}^{E_{c}/ (N{\cal N}_{s})}+ 
\overbrace {
\frac{\Delta }{\pi} \ln \left(\frac{\lambda
e}{T_{K}} \right) - \lambda q}^{E_{K}/ (N{\cal N}_{s})}
\end{eqnarray}
where we have assumed that the upper band is empty and the lower band
is partially filled. $T_{K}=De^{-\frac{1}{J\rho }}$ as before.
The first term in (\ref{totenerg} ) is the conduction electron
contribution to the energy $E_{c}/Nn_{s}$, while the second term is
the lattice ``Kondo'' energy $E_{K}/N_{{\cal N}_{s}}$. 
If now we impose the constraint 
$\frac{\partial
E_{o}}{\partial \lambda }= \langle
n_{f}\rangle -Q=0$ then  $\lambda=\frac{\Delta }{\pi q} 
$
so that the ground-state energy can be written
\begin{equation}\label{}
\frac{
E_{K}}{N {\cal N}_s}= \frac{\Delta }{\pi }\ln \left(
\frac{\Delta e}{\pi q  T_{K}} 
\right).
\end{equation}

This  energy functional has a ``Mexican Hat'' form, with a minimum 
at 
\begin{equation}
\Delta = \frac{\pi q}{e^{2}} T_{K}
\end{equation}
confirming that $\Delta \sim T_{K}$.  
If we now return to the quasiparticle density of states $\rho^{*}$, we
find it has the value
\begin{equation}
\rho ^{*} (0) = \rho + \frac{q}{T_{K}}
\end{equation}
at the Fermi energy so the mass enhancement of the heavy
electrons is then
\begin{equation}
\frac{m^{*}}{m} = 1 + \frac{q}{\rho T_{K}}\sim \frac{q D}{T_{K}}
\end{equation}

\subsubsection{The charge of the f-electron.}

How does the f-electron acquire its charge?  
We have emphasized from the beginning that the 
charge degrees of freedom of the original f-electrons are irrelevant,
indeed, absent from the physics of the Kondo lattice. So how are 
charged f-electrons constructed out of the states of the Kondo
lattice, and how do they end up coupling  to the electromagnetic
field?

The large $N$ 
theory provides an intriguing answer.
The passage from the original Hamiltonian
(\ref{coqblins}) to the mean-field Hamiltonian (\ref{mfham})
is equivalent to the substitution
\begin{equation}\label{factor}
\frac{J}{N} S_{\alpha \beta } (j)
c\dg _{j\beta} c_{j\alpha }
\longrightarrow \bar  Vf\dg_{j\alpha}c_{j\alpha }+
Vc\dg _{j\alpha }f_{j\alpha }.
\end{equation}
In other words, the 
composite combination of spin and
conduction electron are contracted into a single fermi field
\begin{equation}\label{contraction}
\epsfig {file=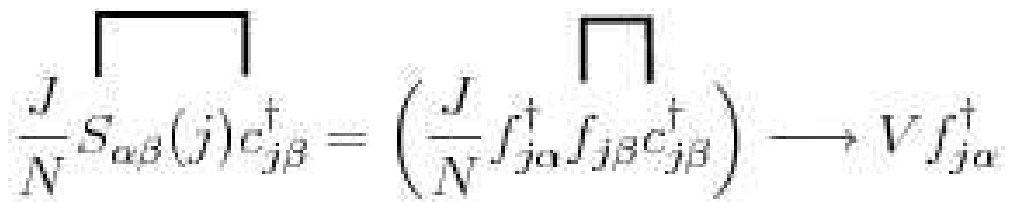,width=7.2cm}.
\end{equation}
%\begin{equation}
%-\frac{J}{N}S_{\alpha \beta} (j)c\dg _{j\beta } =\left( \frac{J}{N}f\dg_{j\alpha }f_{j\beta }c\dg _{j\alpha }
% \right)
%\longrightarrow 
%Vf\dg_{j\alpha }
%\end{equation}
\bxwidth=5.0cm
\upit=-0.35cm 
The amplitude 
%$\frmup{newfigs/contraction2.eps}$ involves
$\frmup{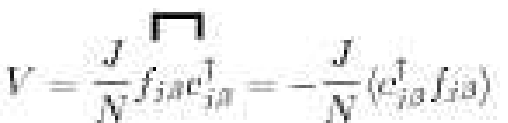}$ involves
electron states  that extend over decades of energy out to the band edges.
%$
%V= \frac{J}{N}c\dg _{j\sigma }f_{j\sigma } = - \frac{J}{N}\langle 
%c\dg _{j\sigma }f_{j\sigma }
%\rangle
%$
In this way the f-electron
emerges as a composite bound-state of a spin and an electron.
More precisely, in the long-time correlation functions,
\begin{equation}\label{}
\langle \bigl[S_{ \gamma\alpha } (i)c_{i\gamma} \bigr ] (t)\ \ 
\bigl [S_{\alpha \beta} (j)c\dg _{j\beta} \bigr ] (t')
\rangle \stackrel{\vert t-t'\vert >> \hbar /T_{K}}
{\relbar\joinrel\relbar\joinrel\relbar\joinrel\relbar\joinrel\relbar\joinrel\relbar\joinrel
\longrightarrow}\frac{N |V^{2}|}{ J^{2}}
\langle f_{i\alpha } (t)f\dg _{j\alpha}(t')\rangle 
\end{equation}
Such ``clustering ''
of composite operators into a single entity  
 is well known 
statistical mechanics
as part of
the operator product expansion \citep{cardy}.
In many body physics, we are used to the clustering 
of fermions pairs  into a
composite boson, as in the BCS model of superconductivity,
\bxwidth=3.9cm
\upit=-0.1cm
%$\frmup{newfigs/contraction3.eps}$.
$\frmup{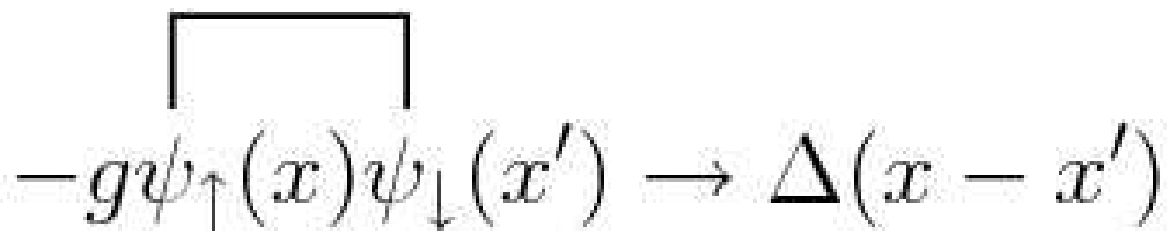}$.
%g\psi_{ \uparrow} (x)\psi_{  \downarrow} (x')\rightarrow \Delta (x-x).\cr
The unfamiliar aspect of the Kondo effect, is the appearance 
of a composite fermion. 

The formation of these composite objects 
profoundly modifies the conductivity and
plasma oscillations of the electron fluid. 
The Read-Newns path integral has two $U (1)$
gauge invariances - an  external electromagnetic gauge invariance associated with the
conservation of charge and an internal gauge invariance associated
with the local constraints. The f-electron couples to the internal gauge 
fields rather than the external electromagnetic fields, so why is it charged?

The answer lies in the broken symmetry associated with
the development of the amplitude $V$.
The phase of $V$ transforms under both internal and external
gauge groups. When $V$ develops an amplitude, its phase does not
actually order, but it does develop a stiffness which is sufficient 
to lock the  internal and external gauge fields 
together so that at low frequencies, they become 
synonymous. 
Written in a schematic long-wavelength form, the gauge-sensitive part of the Kondo lattice 
Lagrangian  is
\begin{eqnarray}\label{l}
{\cal L}&=& \sum_{\sigma}\int d^{D}x\left[ c\dg_{\sigma } (x) (-i \partial_{t}+ e
\Phi (x) + \epsilon_{\bp -e \vec{A}})c_{\sigma } (x) + 
f\dg_{\sigma } (x) (-i \partial_{t}+ \lambda (x)
)f_{\sigma } (x)\right.\cr&+& \left.\biggl (\bar V (x)c\dg_{\sigma } (x)f_{\sigma }
(x)+{\rm  H.c.}\biggr)
\right],
\end{eqnarray}
where $\bp=-i\vec\nabla$.
Suppose $V (x)=\sigma (x)e^{i\phi(x)}$. 
There are two independent gauge transformations that 
that increment the phase $\phi $ of the hybridization. 
In the  external, electromagnetic gauge transformation, the
change in phase is absorbed onto the conduction electron and 
electromagnetic field, so
if $V\rightarrow Ve^{i\alpha }$,
\begin{equation*}
\phi\rightarrow \phi+\alpha,\qquad
c (x)\rightarrow c (x)e^{-i\alpha (x)},\qquad e\Phi(x)\rightarrow
e\Phi{(x)}+ \dot\alpha (x), \qquad e\vec{A}\rightarrow e\vec{A}-\vec {\nabla}\alpha(x).
\end{equation*}
where $(\Phi ,\vec{A})$ denotes the electromagnetic scalar and
vector potential at site $j$ and $\dot \alpha=\partial_{t}\alpha\equiv
-i\partial_\tau\alpha$ denotes the derivative with respect to real
time $t$. By contrast, in the internal gauge transformation, the phase
change of
$V$ is absorbed onto the f-fermion and the internal gauge
field \citep{read},  so if $V\rightarrow Ve^{i\beta }$,
\begin{equation}\label{}
\phi\rightarrow \phi+\beta,\qquad
f (x)\rightarrow f (x)e^{i\beta (x)},\qquad \lambda (x)\rightarrow
\lambda (x)- \dot\beta (x).
\end{equation}
If we expand the mean-field Free energy to quadratic
order in small, slowly varying 
changes in  $\lambda(x)$, then the change in the action is given by 
\[
\delta S = -\frac{\chi_{Q}}{2}\int d^{D}x d\tau \delta \lambda (x)^{2}
\]
where $\chi_{Q}= - \delta^{2}F/\delta \lambda^{2}$ is the f-electron
susceptibility evaluated in the mean-field theory. However, 
$\delta \lambda (x)$ is not gauge invariant, so there must be
additional terms. 
To guarantee gauge invariance under both the internal and external transformation, 
we must replace $\delta \lambda$ by the covariant combination
$\delta \lambda +\dot \phi -e\Phi $. The first two terms are required for
invariance under the internal gauge group, while the last two terms
are required for gauge invariance under the external gauge group.
The expansion of the 
action to quadratic order in the gauge fields
must therefore have the form 
\[
S\sim -\frac{\chi_{Q}}{2}\int d\tau \sum_j (\dot\phi+
\delta \lambda (x)-e\Phi (x))^2,
\]
so the phase $\phi $ acquires a rigidity in time that generates a
``mass'' or energy cost
associated with {\sl difference } of the
external and internal potentials. The minimum energy static configuration
is 
when 
\[
\delta \lambda (\bx)+ \dot\phi(\bx)=e\Phi (\bx),
\]
so when the external potential changes slowly, the internal
potential will track  it.  It is this effect that keeps the Kondo
resonance pinned at the Fermi surface. 
We can always choose the gauge where the phase velocity $\dot\phi$ is
absorbed into the local gauge field $\lambda$. 
Recent work by Coleman, Marston and Schofield \citep{marston} 
has extended this kind of reasoning to the
case where RKKY couplings generate  a dispersion $j_{\bp-{\cal A} }$ for the
spinons, where $\cal A$ is an 
internal vector
potential which suppresses currents of the gauge charge $n_{f}$. 
\fight=3.5 truein
%\fg{newfigs/gaugeflucs.eps}{(a) Spin liquid, or local moment phase, internal field $\cal
\fg{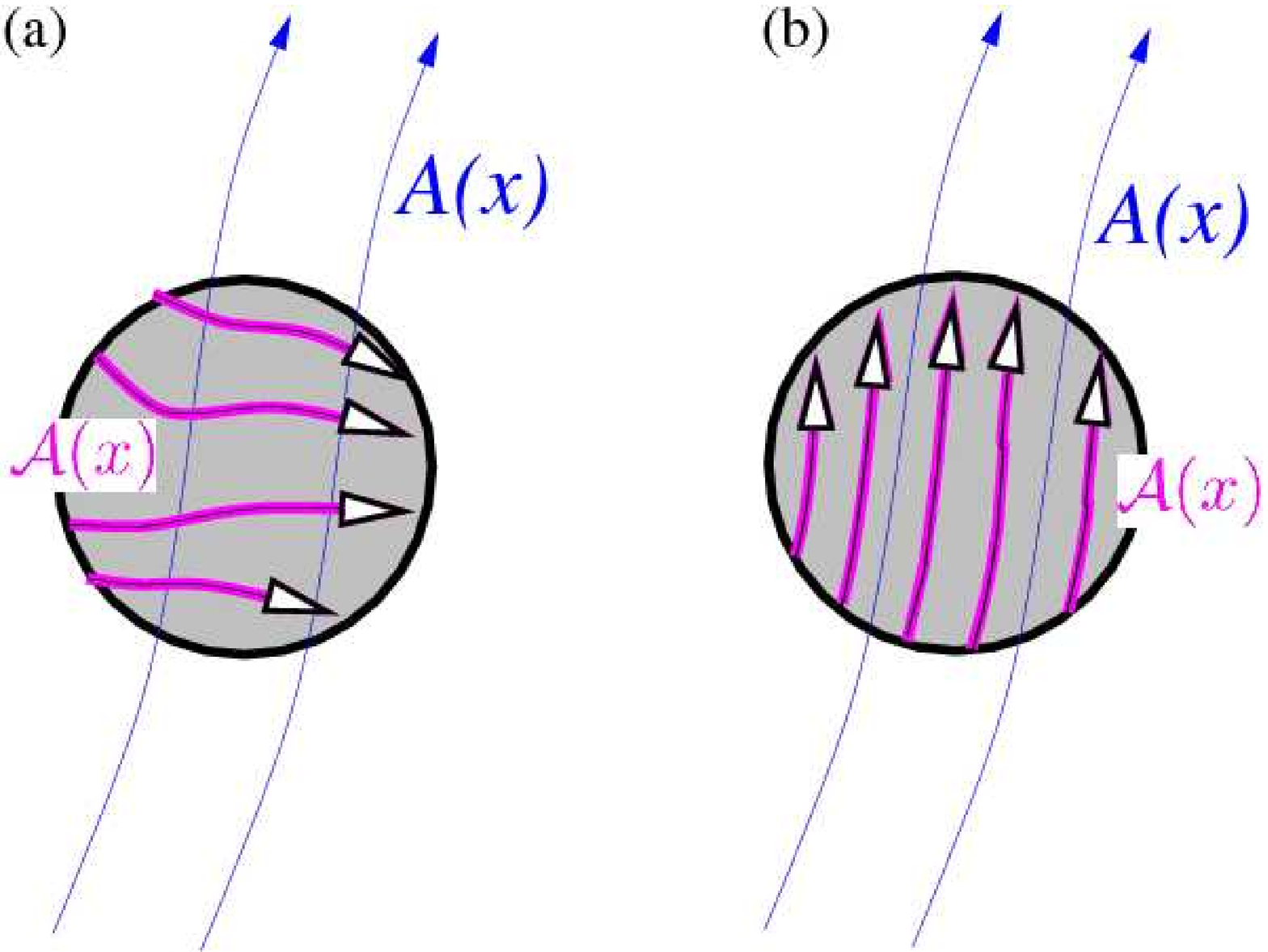}{(a) Spin liquid, or local moment phase, internal field $\cal
\fight=4 truein
A$ decoupled from electromagnetic field (b) Heavy electron phase,
internal gauge field ``locked'' together with electromagnetic
field. Heavy electrons are now charged and difference field $[e\vec{ A}
(x) -{\cal A} (x)]$ is
excluded from the material.   }{fig25}
In this case, the long-wavelength 
action has the form
\[
S = \frac{1}{2}\int d^{3}x d\tau \left[\rho_{s}\left(e\vec{A}+\vec{\nabla
}\phi-\vec{{\cal A}}
 \right)^{2} - \chi_{Q} (e \Phi - \dot\phi-\delta \lambda)^{2}
 \right]
\]
In this general form, heavy electron  physics can be seen to involve a
kind of ``Meissner effect'' that excludes the difference field
$e\vec{A}-\vec{{\cal  A}}$ from
within the metal, locking the internal field to the external electromagnetic
field, so that the f-electrons which couple to it, now become charged
(Fig. \ref{fig25}).

%% Figure showing background density of charge.
%% Plasma oscillations of the heavy electron fluid as a manifestation
%% of f-charge. The f_{1} and f_{2} sum rules.
%%

\subsubsection{Optical Conductivity of the heavy electron fluid.}

One of the interesting interesting consequences of the heavy electron
charge, is a complete renormalization of the electronic plasma
frequency \citep{plasmon}. 
The electronic plasma frequency is related via a f-sum
rule, to the integrated optical conductivity
\begin{equation}
\int_{0}^{\infty } \frac{d\omega }{\pi }\sigma (\omega ) =
f_{1}=\frac{\pi }{2}\left( \frac{n_{c}e^{2}}{m}\right)
\end{equation}
where $n_{e}$ is the density of electrons. \footnote{The f-sum rule is a
statement about the instantaneous, or short-time diamagnetic response of the metal. At
short times $ dj/dt = ( n_{c}e^{2 }/m) E$, so the high frequency
limit of the conductivity is $\sigma (\omega )=
\frac{ne^{2}}{m}\frac{1}{\delta -i\omega }$. But using the Kramer's
Kr{\"o}nig relation 
\[
\sigma (\omega )= \int \frac{dx}{i\pi }\frac{\sigma
(x)}{x-\omega-i\delta }
\]end{equation} 
at large frequencies, 
\[
\omega (\omega )= \frac{1}{\delta -i\omega }\int \frac{dx}{\pi }\sigma
(x)
\]
so that the short-time diamagnetic response implies the f-sum rule.
}
In the absence of local moments, this is the total spectral weight
inside the Drude peak of the optical conductivity. 

When the the heavy electron fluid forms, we need to consider the 
plasma oscillations of the  enlarged Fermi surface.  If the
original conduction sea was less than half filled, then the 
renormalized heavy electron band is more than half filled, forming 
a partially filled hole band. 
 The
density of electrons  in a filled band is $N/a^{D}$, so the effective
density of hole carriers is then
\[
n_{HF}= (N-Q-{\cal N}_{c})/a^{D}= (N-Q)/a^{D}-n_{c}.
\]
The mass of the excitations is also renormalized, $m\rightarrow m^{*}$.
The two effects produce a low frequency 
`quasiparticle'' Drude peak in the conductivity, with a small total weight
\begin{equation}
\int_{0}^{\sim V}{d\omega }\sigma (\omega ) = f_{2}=
\frac{\pi }{2}\frac{n_{HF}e^{2}}{m^{*}}\sim f_{1}\times
\frac{m}{m^{*}}\left(\frac{n_{HF}}{n_{c}} \right) << f_{1}
\end{equation}
Optical conductivity probes the plasma excitations of the electron
fluid at low momenta. The direct gap between the upper and lower bands
of the Kondo lattice are separated by a direct hybridization gap of order
$2V\sim \sqrt{D T_{K}}$.  This scale is substantially larger than the
Kondo temperature, and it defines the 
separation between the thin Drude peak of
the heavy electrons and the high-frequency contribution from the
conduction sea. 

In other words, 
the total spectral weight is divided up into a small ``heavy fermion''
Drude peak, of total weight $f_{2}$, 
where 
\begin{equation}
\sigma (\omega )= \frac{n_{HF}e^{2}}{m^{*}}\frac{1}{(\tau
^{*})^{-1}-i\omega }
\end{equation}
separated off by an energy of order $V\sim \sqrt{T_{K}D}$
from an ``inter-band'' component associated with
excitations between the lower and upper Kondo
bands \citep{millislee,anders}. 
This second term 
carries
the bulk $\sim f_{1}$ of the spectral weight (Fig. \ref{fig26}  ).

\fg{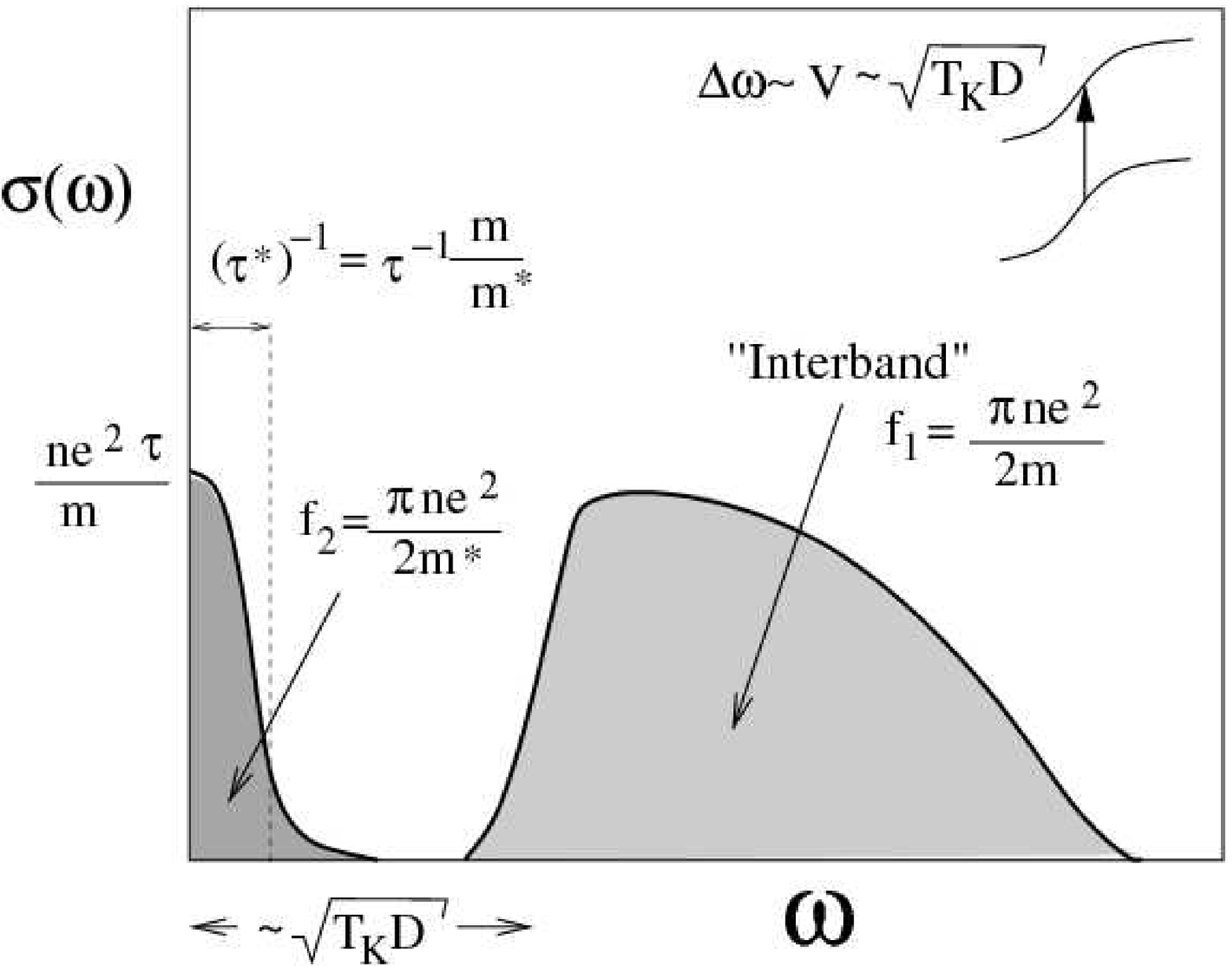}{Separation of the optical sum rule
%\fg{newfigs/fig21.eps}{Separation of the optical sum rule
in a heavy fermion system into a high energy ``inter-band'' component
of weight $f_{2}\sim ne^{2}/m$
and a low energy Drude peak of weight $f_{1}\sim ne^{2}/m^{*}$.
}{fig26}

Simple calculations, based on the Kubo formula confirm
this basic expectation, \citep{millislee,anders} showing that the relationship between the
original relaxation rate of the conduction sea and the heavy electron
relaxation rate $\tau ^{*}$ is 
\begin{equation}\label{}
(\tau ^{*})^{-1} = \frac{m}{m^{*}}(\tau )^{-1} .
\end{equation}
Notice that this means that the residual resistivity
\begin{equation}
\rho _{o}= \frac{m^{*}}{ne^{2}\tau ^{*}}= \frac{m}{ne^{2}\tau }
\end{equation}
is unaffected by the effects of mass renormalization. 
This can be understood by observing that the 
heavy electron Fermi velocity is also renormalized by the effective mass,
$v_{F}^{*}= \frac{m}{m^{*}}$, so that the mean-free path of the
heavy electron quasiparticles is unaffected by the Kondo effect.
\begin{equation}
l^{*}= v_{F}^{*}\tau ^{*}= v_{F}\tau .
\end{equation}

The formation of a narrow Drude peak, and the
presence of a direct hybridization gap, have been seen in optical measurements
on heavy electron systems \citep{zachandzach,gruner88,dordevic}.
One of the interesting features about the hybridization gap of size
$2V$,  is that the mean-field theory predicts that the ratio of the
direct, to the indirect hybridization gap is given by
$\frac{2V}{T_{K}}\sim \frac{1}{\sqrt{\rho T_{K}}}\sim 
\sqrt{\frac{m^{*}}{m_{e}}}
$, 
so that the effective mass of the heavy electrons should scale as
square of the ratio between the hybridization gap and the
characteristic scale $T^{*}$ of the heavy Fermi liquid
\[
\frac{m^{*}}{m_{e}}
\propto 
\left(\frac{2V}{T_{K}}
 \right)^{2}
\]
In practical experiments, $T_{K}$ is replaced by the ``coherence
temperature''  $T^{*}$ where the resistivity reaches a maximum. This
scaling law is broadly followed (see Fig. \ref{fig27}) in measured optical data \citep{dordevic}, and provides 
provides further confirmation of the correctness of the Kondo
lattice picture.
%\fg{newfigs/gap_scale.eps}{
\fg{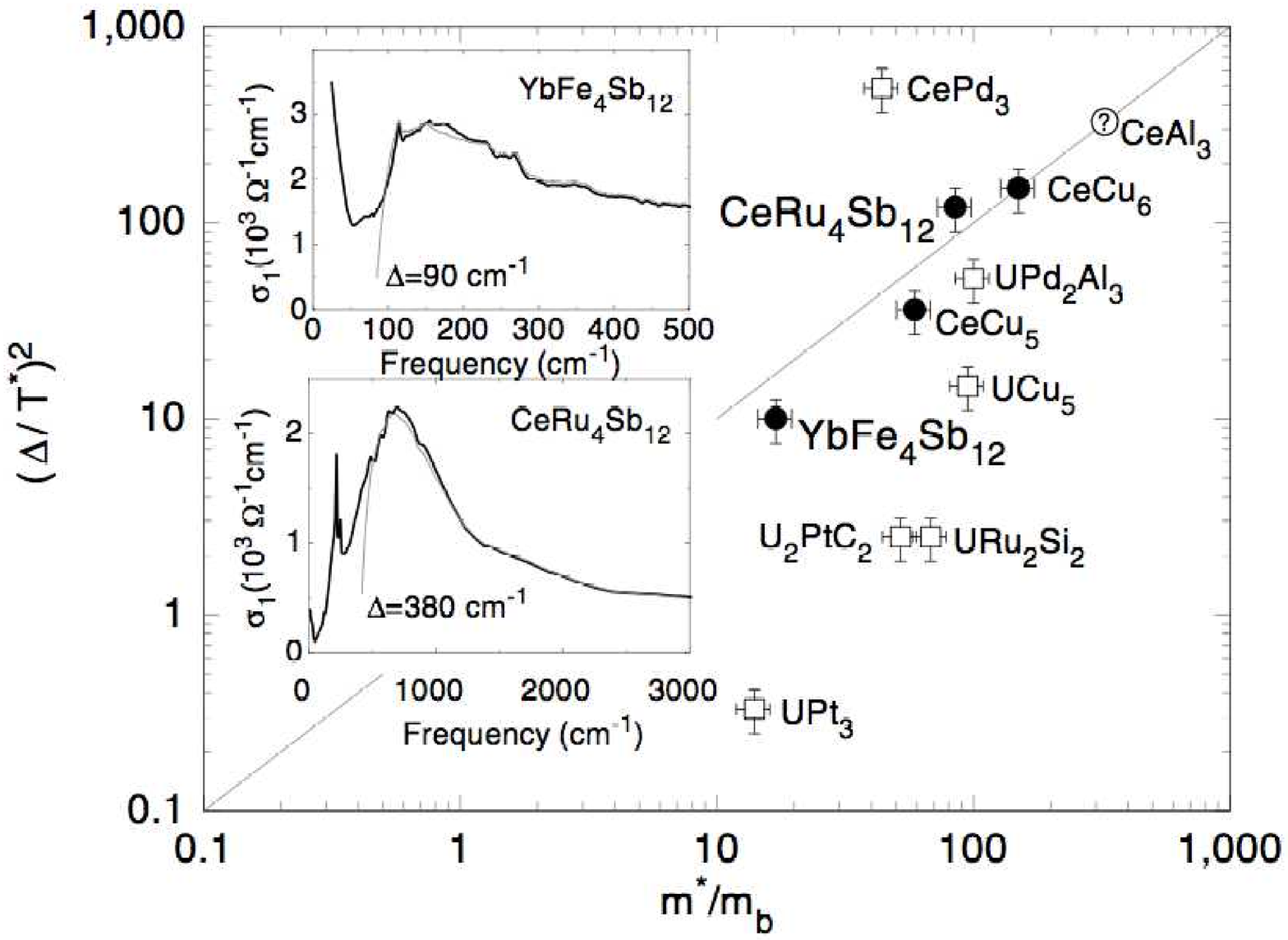}{
Scaling of the effective mass of heavy electrons with the square of
the optical hybridization gap after  \citep{dordevic}
}{fig27}

\subsection{Dynamical Mean Field Theory. }\label{}

The fermionic large $N$ approach to the Kondo lattice provides a invaluable
description of heavy fermion physics, one that can be improved upon
beyond the mean-field level. For example, the fluctuations around the
mean-field theory can be used to compute the interactions, the
dynamical correlation functions and the optical conductivity \citep{coleman_optical,millislee}.
However, the method does face a number of serious outstanding drawbacks:
\begin{itemize}

\item False phase transition. 
In the large $N$ limit, the cross-over between the heavy Fermi liquid
and the local moment physics sharpens into a phase
transition where the $1/N$ expansion becomes singular.  There is no
known way of eliminating this feature in the $1/N$ expansion.

\item Absence of magnetism and superconductivity. 
The large $N$ approach, based on the $SU (N)$
group, can not form a two-particle singlet for $N>2$.  
The $SU (N)$ group is fine for particle physics, 
where baryons are bound-states of $N$ quarks, but for condensed matter
physics, we sacrifice the possibility of forming two-particle or two-spin
singlets, such as Cooper pairs and
spin-singlets.  Antiferromagnetism 
and superconductivity are consequently absent from the mean-field theory.
\end{itemize}

Amongst the  various alternative approaches currently under
consideration, one of particular note is the dynamical mean-field
theory(DMFT). 
The idea of dynamical mean-field theory (DMFT) is to reduce the lattice
problem to the physics of a single magnetic ion  embedded within
a self-consistently determined effective medium \citep{krauth,kotliarrmp2}. 
The effective medium is determined self-consistently
from the self-energies of the electrons that scatter off the single
impurity. In its more advanced form, the single impurity is replaced
by a cluster of magnetic ions. 

Early versions of the dynamical mean-field theory were considered by
Kuromoto \citep{kuromoto}, Grewe and Cox \citep{coxdmft} and others, who used diagrammatic means to extract
the physics of a single impurity. 
The modern conceptual framework for DMFT was  developed by
Metzner and Vollhardt \citep{metzner}, Kotliar and Georges \citep{kotliar}
The basic idea behind DMFT is linked to early work
of Luttinger and Ward \citep{luttingerward,kotliarrmp2}, who found a way of writing the Free energy as a
variational functional of the full electronic 
Green's function 
\begin{equation}\label{}
{\cal G}_{ij}= - \langle T \psi_{i} (\tau )\psi \dg _j(0)\rangle
\end{equation}
Luttinger and Ward showed that the Free energy is a
variational functional of $F[{\cal G}]$ from which Dyson's equation
relating the ${\cal G}$ to the bare Green's function ${\cal G}_{0}$
\[
[{\cal G}_{0}^{-1}
-
{\cal G}^{-1}]_{ij} = \Sigma_{ij}[{\cal G}].
\]
The quantity $\Sigma [{\cal G}]$ is a functional, a machine which takes
the full propagator of the electron and outputs the 
self-energy of the electron.  Formally, this functional is the sum of
the one-particle irreducible Feynman diagrams for the self-energy:
while its output 
depends on the input Greens function, the actual 
the machinery of the functional is determined solely by the
interactions. The only problem is, that we don't know how to calculate
it. 

\fight=4 truein
%\fg{newfigs/dmft.eps}{In the Dynamical Mean Field Theory, the many body
\fg{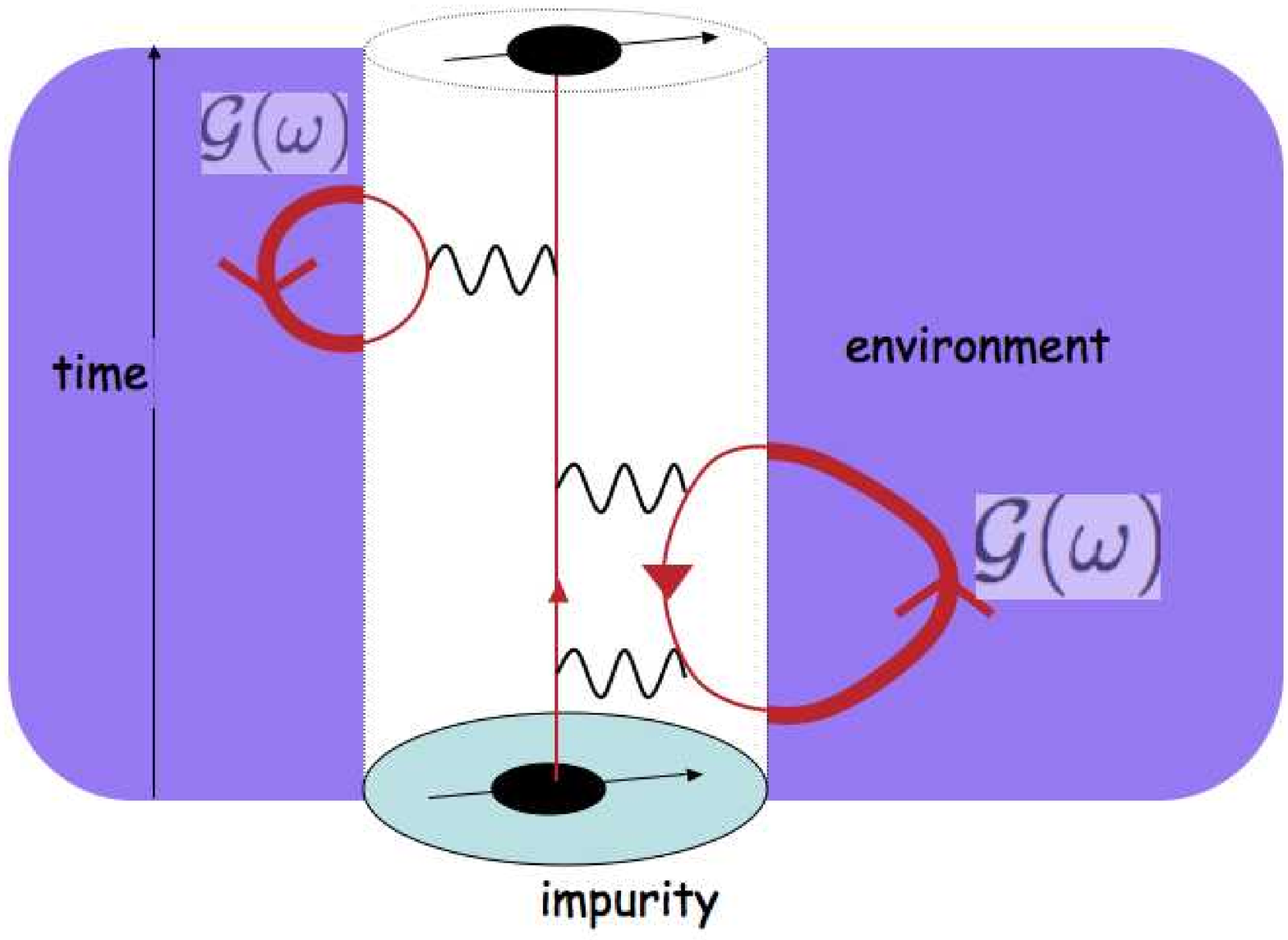}{In the Dynamical Mean Field Theory, the many body
physics of the lattice is approximated by a single impurity in a
self-consistently determined environment. Each time the electron makes
a sortie from the impurity, its propagation through the environment is
described by a self-consistently determined local propagator $\cal G
(\omega)$, represented by the thick red line.}{fig28}
\fight=4 truein

Dynamical mean-field theory solves this problem by 
approximating this functional by that of a single impurity
or a cluster of magnetic impurities ( Fig. \ref{fig28} ).
This is an ideal approximation
for a local Fermi liquid, where the physics is 
highly retarded in time, but local in space.  The local approximation
is also asymptotically exact in the limit of infinite dimensions \citep{metzner}.
If one approximates the input 
Green function to $\Sigma $ by 
its on-site component ${\cal G}_{ij}\approx {\cal G}\delta_{ij}$,
then the functional  becomes the local self-energy functional 
of a single magnetic impurity,
\begin{equation}\label{}
\Sigma_{ij} [{\cal  G}_{ls}] \approx \Sigma_{ij}[{\cal G}\delta_{ls}]\equiv \Sigma_{impurity}[{\cal G}]\delta_{ij}
\end{equation}

Dynamical mean field theory 
extracts the local self-energy by solving an 
Anderson impurity model embedded in an arbitrary electronic
environment. The physics of such a model is described by a path
integral with the action
\[
S= -\int_{0}^{\beta}d\tau d\tau ' f\dg_{\sigma} (\tau ){\cal G}_{0}^{-1}
(\tau -\tau ')f_{\sigma} (\tau ')  + U \int_{0}^{\beta}d\tau n_{\uparrow} (\tau)n_{\downarrow} (\tau )
\]
where $G_{0} (\tau )$ describes  the bare Green's function
of the f-electron, hybridized with its dynamic environment. 
This quantity is self-consistently updated by the 
DMFT. There are by now,
a large number of superb numerical methods to solve an Anderson model for
an arbitrary environment, including the use of exact diagonalization, 
diagrammatic techniques and the use of Wilson's renormalization group \citep{bulla}. 
Each of these methods is able to take an input ``environment'' Green's function
providing as output, the impurity self-energy $\Sigma [{\cal
G}_{0}]=\Sigma (i\omega_n)$. 

Briefly, here's how the DMFT computational cycle works.  One starts with an estimate for the
environment Green's function ${\cal G}_{0}$ and uses this as input to the ``impurity
solver'' to compute the first estimate $\Sigma (i \omega_{n})$
of the local self-energy. The interaction strength is set within the 
impurity solver. 
This local self-energy is used to compute the Green's functions of the
electrons in the environment. 
In an Anderson lattice, the Green's function becomes
\begin{equation}\label{}
G (\bk ,\omega)= 
\left[\omega -
E_{f}-\frac{V^{2}}{\omega-\epsilon_{\bk }}  - \Sigma (\omega)
\right]^{-1}
\end{equation}
where $V$ is the hybridization and $\epsilon_{\bk }$ the dispersion of
the conduction electrons.  It is through this relationship that
the physics of the lattice is fed into the problem. From $G (\bk,\omega)$
the local propagator is computed
\[
{\cal  G} (\omega)= \sum_{\bk}\left[\omega -
E_{f}-\frac{V^{2}}{\omega-\epsilon_{\bk }}  - \Sigma (\omega)
\right]^{-1}
\]
Finally, the new estimate for the bare environment Green's
function ${\cal G}_{0}$, is then obtained by inverting the equation ${\cal G}^{-1}={\cal
G}_{0}^{-1}-\Sigma $, so that 
\begin{equation}\label{}
{\cal G}_{0} (\omega) = \left[G^{-1} (\omega)+ \Sigma (\omega) \right]
\end{equation}
This
quantity is then re-used as the input to an ``impurity solver'' to
compute the next estimate of $\Sigma (\omega)$.  The whole procedure
is then re-iterated to self-consistency.
%\fg{newfigs/cyzcholl.eps}{Resistivity for the Anderson lattice,
\fg{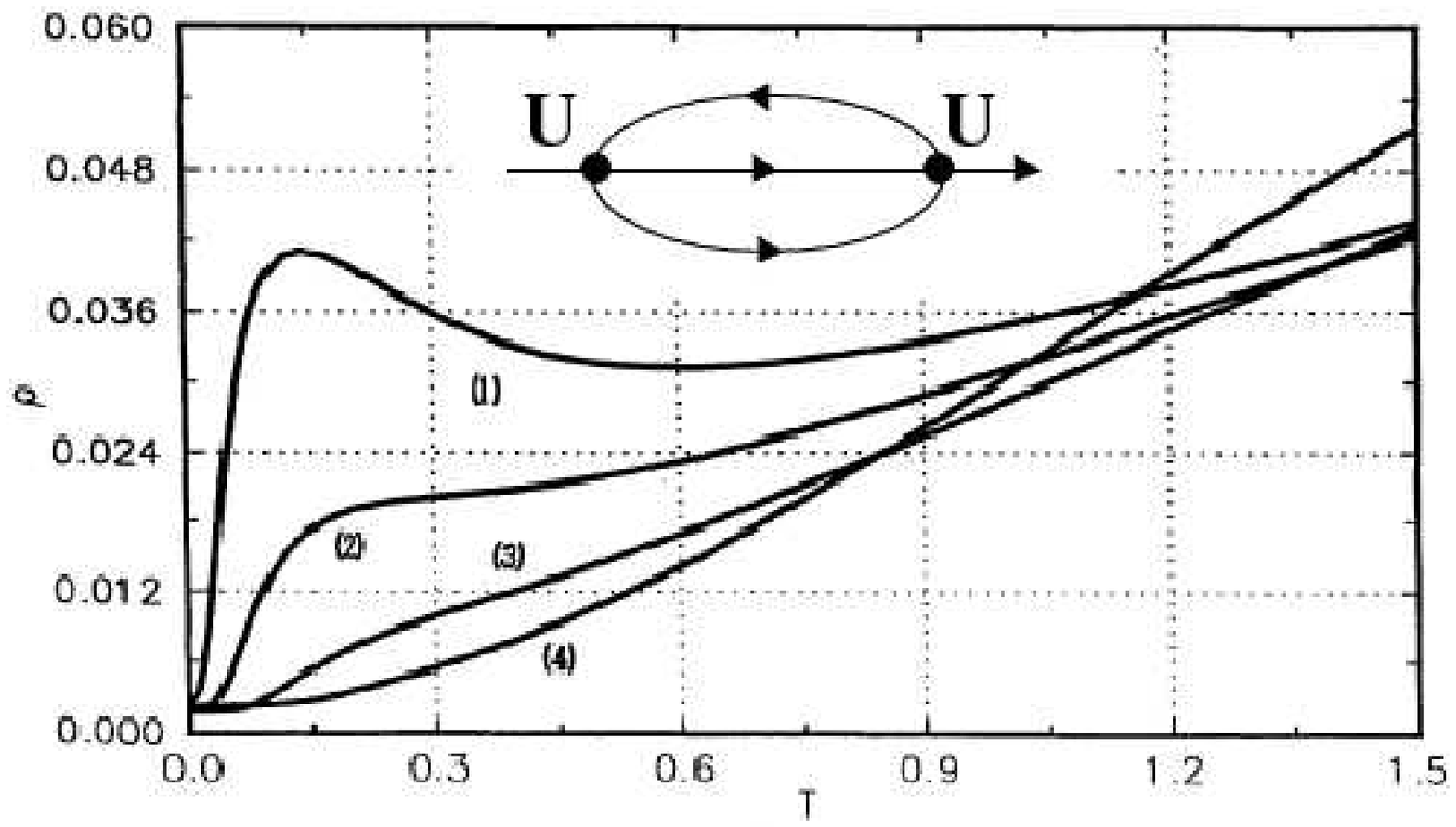}{Resistivity for the Anderson lattice,
calculated using the DMFT, computing the self-energy to order $U^{2}$,
after  \citep{cyzcholl}. (1), (2), (3) and (4) correspond to a
sequence of decreasing electron density corresponding to $n_{TOT}= (
0.8, 0.6, 0.4, 0.2)$ respectively.
}{fig29}
For the Anderson lattice, Cyzcholl \citep{cyzcholl} has shown that remarkably good results are obtained  using
a perturbative expansion for $\Sigma $ to order $
U^{2}$ (Fig. \ref{fig29}). Although this approach is not sufficient to capture the
limiting Kondo behavior, much the qualitative physics of the Kondo
lattice, including the development of coherence at low temperatures is already
captured by this approach. However, to go to the strongly correlated
regime where the ratio of the interaction to the impurity
hybridization width $U/ (\pi \Delta )$ is much larger than unity, one
requires a more sophisticated solver. 

There are many ongoing developments under way using
this powerful new computational tool, including the  incorporation of realistic descriptions
of complex atoms, and the extension to ``cluster DMFT'' involving 
clusters of magnetic moments embedded in a self-consistent
environment. Let me end this brief summary with a list of a few
unsolved issues with this technique
\begin{itemize}

\item There is at present, no way to relate the thermodynamics of the
bulk to the impurity thermodynamics.

\item At present, there is  no natural extension of these methods
to the infinite $U$ Anderson or Kondo models that incorporates the 
Green's functions of the {\sl localized} f-electron degrees of freedom as an integral
part of the dynamical mean-field theory. 

\item The method is largely a numerical black box, making it difficult to
compute microscopic quantities beyond the electron-spectral
functions. At the human level it 
is  difficult for students and researchers
to separate themselves from the ardors of coding the impurity solvers,
and make time to develop new conceptual and qualitative understanding of the physics. 

\end{itemize}

\section{Kondo Insulators}\label{}
\subsection{Renormalized Silicon}\label{}
\fight=3.0 truein
%\fg{newfigs/fesi.eps}{Temperature dependent susceptibility in $FeSi$
\fg{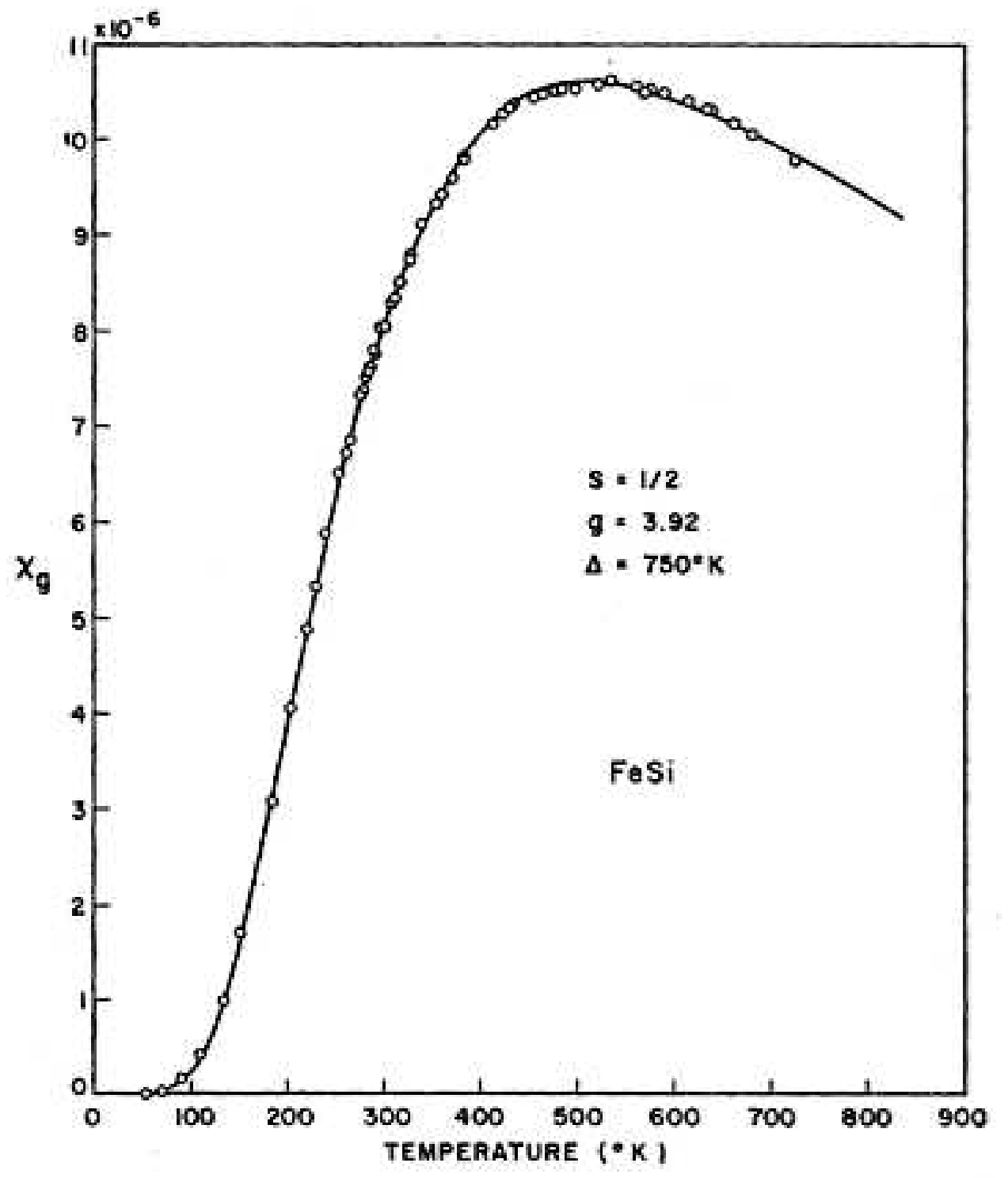}{Temperature dependent susceptibility in $FeSi$
after  \citep{jaccarino}, fitted to the activated Curie form $\chi (T)= ( C/T)e^{-
\Delta / ( k_{B}T)}$, with $C= (g{\mu_{B}})^{2}j (j+1)$, and $g=3.92$, 
$\Delta = 750K$. The Curie tail has been subtracted.
}{fig30}
\fight=4 truein

The ability of a dense lattice of local moments to transform a metal
into an insulator,  a ``Kondo insulator'' is one of the 
remarkable and striking consequences of the dense Kondo effect
 \citep{revfiskaeppli,revki1,revki2}.  
Kondo insulators are heavy electron systems in which the 
the liberation of mobile charge through the Kondo effect 
gives rise to a filled heavy electron band in which the 
chemical potential lies in the middle of the hybridization gap. 
From a quasiparticle perspective, Kondo insulators 
are highly renormalized ``band insulators'' (Fig. \ref{fig31}). The 
d-electron  Kondo insulator  $FeSi$  has been referred to as 
 ``renormalized silicon''.
However, like Mott-Hubbard insulators, the gap in their
spectrum is driven by interaction effects, and they display optical
and magnetic properties that can not be understood with band theory.

There are about a dozen known Kondo insulators, including the
rare earth systems 
$SmB_{6}$ \citep{smb6}, $YB_{12}$ \citep{yb12},
$CeFe_{4}P_{12}$ \citep{cefep}, $Ce_{3}Bi_{4}Pt_{3}$ \citep{ce3bi4pt3}
, $CeNiSn$ \citep{cenisn,cenisn-semimetal,CeNiSn-thermal1} and
$CeRhSb$ \citep{CeRhSb-thermal}, and the d-electron Kondo insulator $FeSi$.
At high temperatures, Kondo insulators
are local moment metals, with classic Curie
susceptibilities, but at low temperatures, as  the Kondo effect
develops coherence, the conductivity and the magnetic
susceptibility drop towards zero. Perfect insulating behavior is
however, rarely observed due to difficulty of eliminating impurity
band formation in ultra-narrow gap systems. 
One of the cleanest examples of Kondo insulating behavior occurs in
the  d-electron system $FeSi$ \citep{jaccarino,fesi}. 
This ``fly-weight'' heavy electron system provides a rather clean realization of the phenomena seen in other
Kondo insulators, with a  spin and
charge gap of about $750K$  \citep{zachandzach}. Unlike its rare-earth
counterparts, the small spin-orbit coupling in this materials eliminates the Van
Vleck contribution to the susceptibility at $T=0$, giving rise 
a susceptibility which almost completely vanishes at low temperatures \cite[]{jaccarino}
(Fig. (\ref{fig30}).

%\fg{newfigs/kondo_ins.eps}{Schematic band picture of Kondo insulator,
\fg{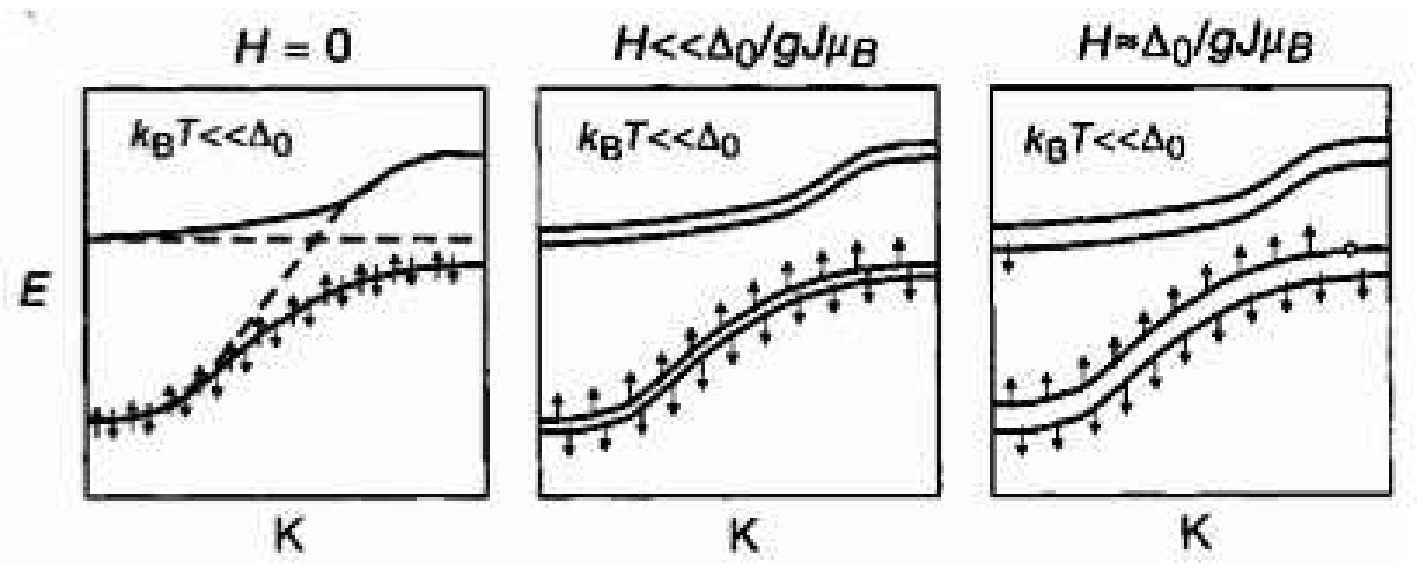}{Schematic band picture of Kondo insulator,
illustrating how a magnetic field drives a metal insulator
transition. Modified from  \citep{revfiskaeppli}.}{fig31}

 Kondo insulators can be understood as ``half-filled'' Kondo lattices
in which each quenched moment liberates a negatively charged heavy
electron, endowing each magnetic ion
an extra unit of positive charge.
There are three good pieces of support for this theoretical picture:
\begin{itemize}
\item 
Each Kondo insulator has its fully itinerant semiconducting analog. For example, 
$CeNiSn$ is iso-structural and iso-electronic
with the semiconductor  $TiNiSi$ containing $Ti^{4+}$ ions, even
though the former contains $Ce^{3+} $ ions  with localized $f$
moments. Similarly, $Ce_{3}Bi_{4}Pt_{3}$, with a tiny gap of order
$10meV$
 is isolectronic with semiconducting
$Th_{3}Sb_{4}Ni_{3}$, with a $70meV$ gap, in which the $5f$ -electrons of the $Th^{4+}$ ion
are entirely delocalized. 

\item Replacing the magnetic site with iso-electronic non-magnetic ions is equivalent
to doping, thus in $Ce_{1-x}La_{x}Bi_{4}Pt_{3}$, each $La^{3+}$ ion,
behaves as an electron donor in a lattice of effective $Ce^{4+}$ ions.
$Ce_{3-x}La_{x}Pt_{4}Bi_{3}$ is in fact, very similar to $CePd_{3}$
which contains a pseudo-gap in its optical conductivity, with a tiny
Drude peak \cite[]{schlesingercepd3}.  

\item The magneto-resistance of Kondo insulators is large and
negative, and the ``insulating gap'' can be closed by the action of
physically accessible fields. Thus in $Ce_{3}Bi_{4}Pt_{3}$ a $30T$ field is sufficient
to close  the indirect hybridization gap.

\end{itemize}
These equivalences support the picture of
the Kondo effect liberating a composite fermion.

Fig. (\ref{fig32} (a) ) shows the sharp rise in the 
resistivity of $Ce_{3}Bi_{4}Pt_{3}$ as the Kondo insulating gap forms.
In Kondo insulators, the complete elimination of carriers at low
temperatures is also manifested in the optical conductivity. 
Fig. (\ref{fig32} (b)) shows the temperature dependence of
the optical conductivity in $Ce_{3}Bi_{4}Pt_{3}$, showing the
emergence of a gap in the low frequency optical response and the loss
of carriers at low energies.

The optical conductivity of the Kondo insulators is of particular
interest.  Like the heavy electron metals, the development of
coherence is marked by the formation of a direct hybridization gap
in the optical conductivity. As this forms, a pseudo-gap develops in
the optical conductivity. In a non-interacting band gap, the lost
f-sum weight inside the pseudo-gap would be deposited above the
gap. 
In heavy fermion metals, a small fraction of this weight is
deposited in the Drude peak - however, most of the weight is sent off
to energies comparable with the band-width of the conduction
band. This is one of the most direct pieces of evidence that the
formation of Kondo singlets involves electron energies that spread out
to the bandwidth. Another fascinating feature of the heavy electron
``pseudo-gap'', is that it forms at a temperature that is significantly
lower than the pseudogap. This is because the pseudogap has a larger
width given by the geometric mean of the coherence temperature and the band-width.
 $2V\sim \sqrt{T_{K }D}$.  The extreme upward transfer of spectral weight in
 Kondo insulators has not yet been duplicated in detailed theoretical
 models.  For example, while 
calculations of the optical conductivity within
 the dynamical mean field theory, do show spectral weight transfer,
it is not yet possible to reduce the indirect band-gap to the point
where  it is radically smaller than the interaction scale $U$ \citep{gabihf}. 
It may be that these discrepancies will  disappear in future
calculations based on the more extreme physics of the Kondo model, but
these
calculations have yet to be carried out. 

%\fg{newfigs/ce343combine.eps}{(a) Temperature dependent resistivity of
\fg{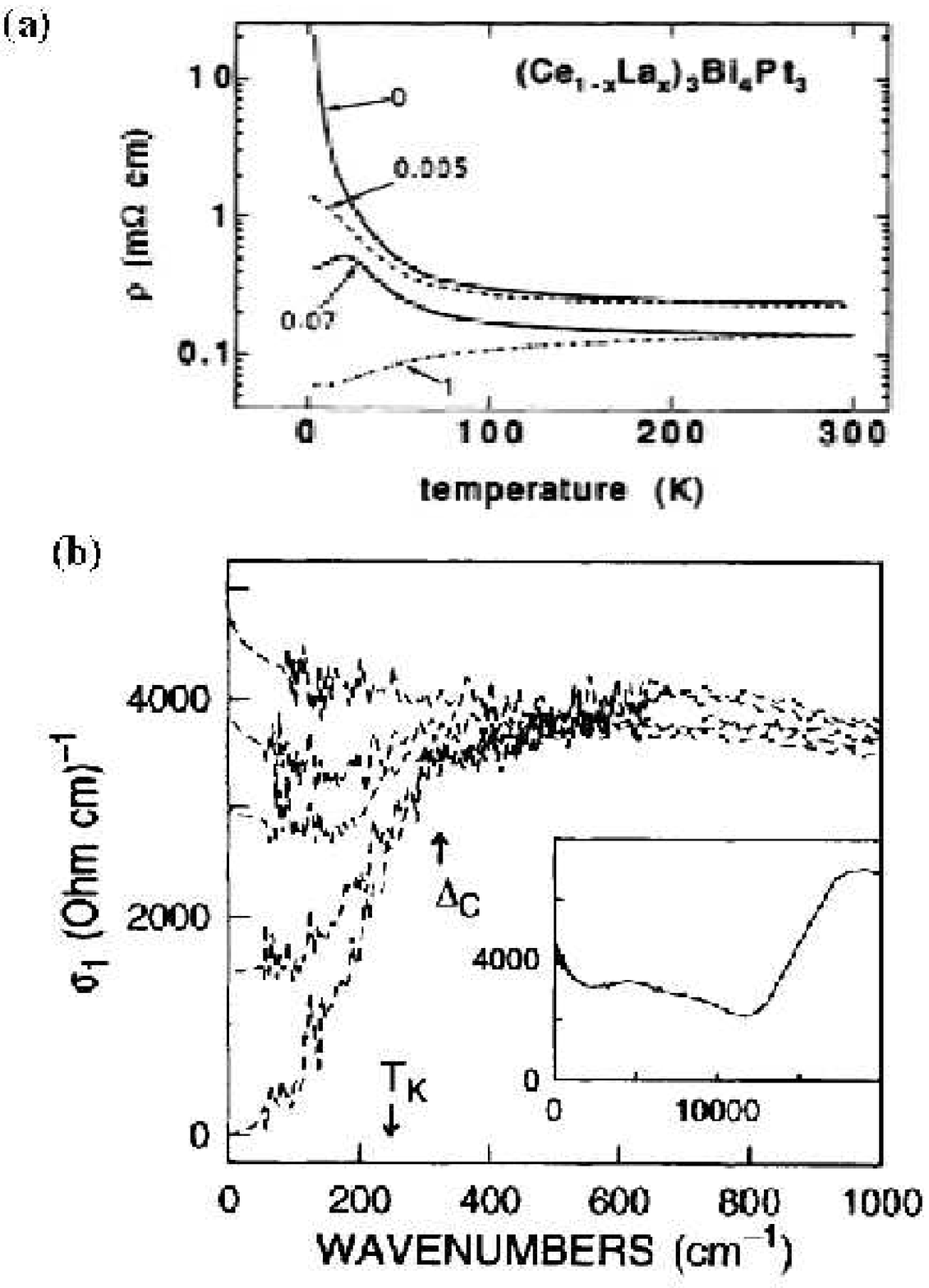}{(a) Temperature dependent resistivity of
$Ce_{3}Pt_{4}Bi_{3}$ showing the sharp rise in resistivity at low
temperatures after \citep{ce3bi4pt3}. Replacement of local moments with spinless $La$ ions
acts like a dopant. (b) Real part of optical rconductivity
$\sigma_{1} (\omega)$ for Kondo insulator $Ce_{3}Pt_{4}Bi_{3}$after
\cite{ce3bi4pt3optical}. The formation of the pseudogap associated with
the direct hybridization gap, leads to the transfer of f-sum spectral
weight to high energies of order the band-width. The pseudogap forms
at temperatures that are much smaller than its width (see text). 
Insert shows $\sigma_{1} (\omega)$ in the
optical range.
}{fig32}

There are however, a number of aspects of Kondo insulators that are
poorly understood from the 
the simple hybridization picture, in particular
\begin{itemize}

\item The apparent disappearance of RKKY magnetic interactions at low
temperatures. 

\item The nodal  character of the hybridization gap that develops in the
narrowest gap Kondo insulators, $CeNiSn$ and $CeRhSb$.

\item The nature of the metal-insulator transition induced by doping. 

\end{itemize}

\subsection{Vanishing of RKKY interactions}\label{}

There are a number of experimental indications that the low-energy magnetic interactions
vanish at low frequencies in a Kondo lattice. The low temperature
product of the susceptibility and temperature
$\chi  T$  reported \citep{revfiskaeppli} to scale with the inverse Hall
constant $1/R_{H}$, representing the exponentially suppressed density
of carriers, so that 
\[
\chi  \sim \frac{1}{R_{H}T}\sim \frac{e^{-\Delta /T}}{T}.
\]
The important point here, is that  the activated part of the
susceptibility has a vanishing  Curie-Weiss temperature. 
A similar conclusion is reached
from inelastic neutron scattering measurements of the magnetic susceptibility
$\chi' (q,\omega)\sim $ in $CeNiSn$ and $FeSi$,  which  appears to lose all of its momentum
dependence at low temperatures and frequencies. 
There is to date, no
theory that can account for these vanishing interactions. 

\subsection{Nodal Kondo Insulators}\label{}
%% Nodal Kondo insulators

The narrowest gap Kondo insulators,
$CeNiSn$ and $CeRhSb$ are effectively 
semi-metals, for although they do display tiny pseudogaps in their
spin and charge spectra, the purest samples of these materials develop
metallic behavior \citep{cenisn-semimetal}. What is particularly peculiar (Fig. \ref{fig33}) about these two
materials, is that the NMR relaxation rate $1/ (T_{1})$shows a $T^{3}$ temperature
dependence from about $10K$ to $1K$, followed by a linear Korringa
behavior at lower temperatures. The usual rule of thumb, is that the
NMR relaxation rate is proportional to a product of the temperature
and the thermal average of the electronic density of states $N^{*} (\omega)$
\begin{eqnarray}\label{l}
\frac{1}{T_{1}}\sim T \overline{{N (\omega)}^{2}}\sim T [N (\omega \sim T)]^{2}
\end{eqnarray}
where $\overline{N (\omega)^{2}}= \int d\epsilon \left(- \frac{\partial f
(\omega)}{\partial \omega} \right) N (\omega)^{2}$ is the thermally
smeared average of the squared density of states. 
This suggests that the electronic density of states in these materials
has a ``$V$'' shaped form, with a finite value at $\omega = 0$.
Ikeda and Miyake \citep{ikeda} have proposed that the Kondo insulating state in these materials
develops in a crystal field state with an axially symmetric
hybridization vanishing along a single crystal axis.
In such a picture, the 
finite density of states does not derive from a Fermi surface, but 
from the angular average of the 
coherence peaks in the density of states.
The odd
thing about this proposal, is that $CeNiSn$ and $CeRhSb$ are
monoclinic structures, and the low-lying Kramers doublet of the f-state
can be any combination of the $\vert \pm \frac{1}{2}\rangle $, $\vert
\pm \frac{3}{2}\rangle $ or $\vert \pm \frac{5}{2}\rangle $ states:
\[
\vert \pm = b_1\vert \pm 1/2\rangle + b_2 \vert \pm 5/2\rangle  
+b_3 \vert \mp 3/2\rangle  
\]
where where $\hat b= (b_1,b_2,b_3)$ could in principle, point anywhere on  
the unit sphere, depending on details of the monoclinic crystal field.  
The Ikeda Miyake model  
corresponds to three symmetry-related 
points in 
the space of crystal field ground states,  
\bea 
\hat b= \left\{ \begin{array}{lr} 
(\mp \frac{\sqrt{2}}{4},-\frac{\sqrt{5}}{4},\frac{3}{4}) \cr 
(0,0,1) 
\end{array} 
\right. 
\eea 
where a node develops 
along the $x$, $y$ or $z$ axis respectively.  But the nodal crystal field
states are isolated ``points'' amidst a continuum of fully gapped
crystal field states.   Equally strangely, 
neutron  scattering results show  no  
crystal field satellites in the dynamical spin susceptibility  
of CeNiSn, suggesting that 
the crystal electric fields are quenched \citep{Alekseev}, 
and that the nodal physics 
is  a many body effect \citep{Prokof'ev,juana}. One idea, 
is that Hund's interactions provide the driving force for this
selection mechanism. 
Fulde and Zwicknagl  \citep{zwicknagl}have suggested that Hund's coupling's select the
orbitals in multi f-electron heavy electron metals such as
$UPt_{3}$. Moreno and Coleman \citep{juana} propose a similar idea, in which 
virtual valence fluctuations into the $f^{2}$ state generate a
many-body, or Weiss effective field that couples to the orbital
degrees of freedom, producing an effective 
{\sl crystal} field 
which adjusts itself in order to minimize the
kinetic energy of the f-electrons. This hypothesis is consistent with the
observation that the Ikeda Miyake  state
corresponds to the Kondo insulating state
with the lowest kinetic energy, providing
a rational for the selection of the nodal configurations. Moreno and
Coleman also found another nodal state with a more marked  $V$-shaped
density of states that may fit the observed properties more precisely.
This state is also a local minimum of the electron Kinetic energy.
These ideas are still in their infancy, and more work needs to be done
to examine the controversial idea of a Weiss crystal field, both in
the insulators and in the metals.

\fight=\textwidth
%\fg{newfigs/nodal.eps}{(a) NMR relaxation rate $1/T_{1}$ in $CeRhSb$ and
\fg{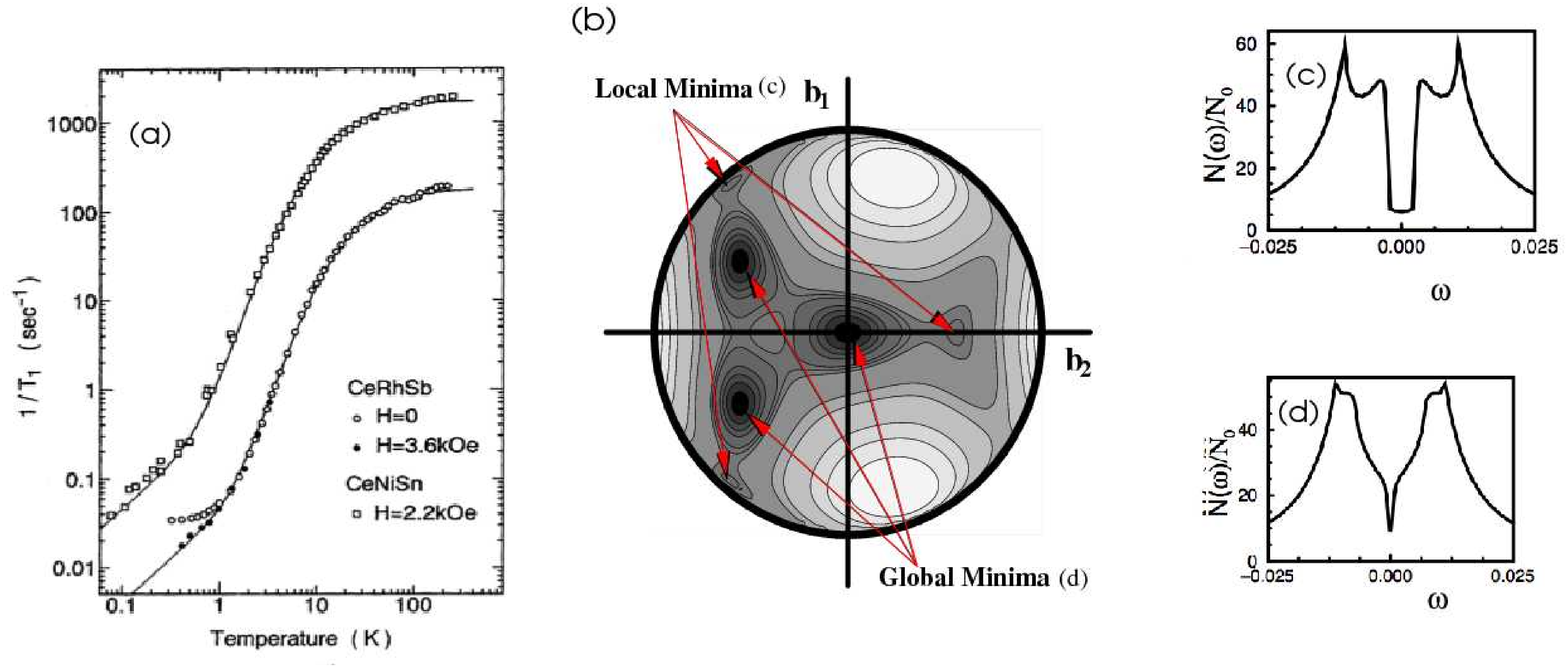}{(a) NMR relaxation rate $1/T_{1}$ in $CeRhSb$ and
$CeNiSn$, showing a $T^{3}$ relaxation rate sandwiched between a low,
and a high temperature $T$-linear Korringa relaxation rate, suggesting
a $V$- shaped density of states after \citep{1Kondo-NMR}
(b)
Contour plot of the ground-state energy in mean-field theory for the 
narrow gap Kondo insulators after  \cite[]{juana}, 
as a function of the two first components of the unit vector
$\hat b$ (the third one is taken as positive).
% as $b_3=\sqrt{1-b_1^2-b_2^2}$).
The darkest regions correspond to lowest values of the free-energy.
Arrows point to the three global and three local minima that
correspond to nodal Kondo insulators.(c) Density of states of Ikeda
Miyake \citep{ikeda} state that appears as the global minimum of the Kinetic energy
(d)
Density of states of the MC state \citep{juana} that appears as a local
minimum of the Kinetic energy, with more pronounced ``V'' shaped
density of states. }{fig33}
 
%% Kondo insulators - spin and charge gaps. Two mysteries: Vanishing of
%% interactions. The mystery of the point nodes. 

%% Heavy electron superconductors - phenomenology - lines. Line
%% structure - heat current, ultrasound.  Microscopics - spin fluctuations;
%% RVB type model - spin liquid plus Kondo effect. The peculiar case
%% of the 115 systems.

%% Quantum criticality - general points, 1/T is a finite size
%% Singularity in the phase diagram
%% Damped phi^{4} theory Millis Hertz - only works for a few cases.
%% Hard quantum critical points - scale collapses; omega/T scaling rho
%% ~T
%% New ideas - local quantum criticality.
%% Deconfinement
%% SP (N)
%%
%% Conclusions: early phase is now over - quantum criticality; heavy
%% electron electrodynamics; magnetism, pairing.  Major challenge to
%% the current generation of condensed matter physicists. Possible
%% cosmological connections, if there is, as it seems a truly new
%% universality class of phase transition awaiting discovery. 
%%

%
%
%
%
%
%

\section{Heavy Fermion Superconductivity}\label{}

\subsection{A quick tour}\label{}

Since the discovery \citep{steglich} of superconductivity in
$CeCu_{2}Si_{2}$, the list of known
heavy fermion superconductors has grown to include more a
dozen \citep{heavyscrev} materials with a great diversity of
properties \citep{revhfsc1,revhfsc2}.
In each of these materials, the jump in the specific
heat capacity at the superconducting phase transition 
is comparable  with the normal
state specific heat
$$
{\rm( C_v^s-C_v^n) \over C_{V}}\sim 1-2,
\eqno(7)$$
and the integrated entropy beneath the $C_v/T$ curve of the superconductor
matches well with the corresponding area for the normal phase obtained
when superconductivity is suppressed by disorder or fields
$$
\int_0^{T_c} dT  {(C_v^s-C_v^n) \over T} = 0.
\eqno(8)$$
Since the normal state entropy is derived from the f-moments, it
follows that these same degrees of freedom are involved in the
development of the superconducting state.
With the exception of a few anomalous  cases, ($UBe_{13}$,
$PuCoGa_{5}$ and $CeCoIn_{5}$)
heavy fermion superconductivity develops out of the 
coherent, paramagnetic heavy Fermi liquid, so heavy 
fermion superconductivity can be said to involve the pairing of heavy
f-electrons.  

Independent confirmation of the ``heavy'' nature of the pairing
electrons comes from observed size of the 
London penetration depth $\lambda_L$ and superconducting coherence length $\xi$ in
these systems, both of which reflect the enhanced effective mass, The
large mass renormalization enhances the penetration depth, whilst
severely
contracting the coherence length, making these extreme type-II
superconductors. 
The London penetration depth of heavy fermion superconductors agree well with
the value expected on the assumption that only spectral weight
in the quasiparticle Drude peak condenses to form a superconductor
by
$$
{1 \over \mu_o \lambda_L^2}  = {ne^2 \over m^* }
= \int_{\om \in { \rm D. P }}{d \om \over \pi} \si(\om) << {ne^2 \over m}
\eqno(9)$$
London penetration depths in these compounds are a factor of
$20-30$ times {\sl longer} \citep{broholm} than in  superconductors, 
reflecting the large enhancement in effective mass.  By contrast, the coherence lengths
$\xi \sim  v_{F}/\Delta \sim h k_{F}/ ( m^{*} \Delta )$ are severely contracted in a
heavy fermion superconductor.  The orbitally limited 
upper critical field is determined by the condition that an area  $\xi^{2}$ contains a flux
quantum $\xi^{2}B_{c}\sim \frac{h}{2e}$. In $UBe_{13}$, a
superconductor with  $0.9K$ transition temperature, the upper critical
field is about $11$ Tesla, a value about 20 times larger than a
conventional superconductor of the same transition temperature.

Table B. shows a selected list of heavy
fermion superconductors. 
``Canonical'' heavy fermion superconductors, such as
$CeCu_{2}Si_{2}$ and $UPt_{3}$, develop superconductivity 
out of a  paramagnetic Landau Fermi liquid.
``Pre-ordered'' superconductors, such as
$UPt_2Al_3$ \citep{steglich2,steglich3}, 
$CePt_{3}Si$ and $URu_{2}Si_{2}$, develop 
another kind of order  before going 
superconducting at a lower temperature. In the case of
$URu_{2}Si_{2}$, the nature of the upper ordering transition is 
still unidentified. but in the  other examples,
the upper transition involves the development of antiferromagnetism.
``Quantum critical'' superconductors, including $CeIn_{3}$ 
and $CeCu_{2}( Si_{1-x}Ge_{x})_{2}$ develop superconductivity when 
pressure tuned close to a quantum critical point. $CeIn_{3}$ develops
superconductivity at the pressure-tuned antiferromagnetic quantum
critical point at $2.5GPa$ ($25kbar$).  $CeCu_{2} (Si,Ge)_{2}$
has two islands, one associated with antiferromagnetism at low
pressure, and a second at still higher pressure, thought to be
associated with critical valence fluctuations \citep{cecu2sige}.

\centerline{\bf Table. B. Selected Heavy Fermion Superconductors.}
\vskip 0.1 truein

\begin{center}
\begin{tabular}{|l||c|c|c|c|c|c|}
\hline
\multirow{2}{15mm}{\bf Type} & Material & $T_{c}$ & Knight Shift& Remarks & \renp{2}{18mm}{Gap Symmetry}& Ref.\\
& & & (Singlet)& & &\\
\hline
\hline\multirow{4}{25mm}{\bf ``Canonical''}
& \renp {2}{30mm}{$CeCu_{2}Si_{2}$} 
& \renp
{2}{20mm}{0.7} & \renp{2}{20mm}{Singlet} & 
\renp{2}{23mm}{First HFSC} 
& 
\renp{2}{20mm}{Line nodes} & \renp{2}{14mm}{[1]}
\\
&&&&&&\\
 \cline{2-7}
& \renp {2}{30mm}{$UPt_{3}$} 
& \renp
{2}{20mm}{$\ 0.48 K$} &\renp{2}{20mm}{?} 
&  \renp{2}{38mm}{Double Transition to\\
T-violating state}&\renp{2}{20mm}{Line and point nodes} & \renp{2}{14mm}{[2]}\\
&&&&&&\\
\hline
\hline
\multirow{6}{25mm}{\bf Pre-ordered }
&\renp {2}{30mm}{$UPd_{2}Al_{3}$} 
& \renp
{2}{20mm}{$\ 2K$} & \renp{2}{20mm}{Singlet} & 
\renp{2}{20mm}{N\' eel AFM $T_{N}=14K$} 
& 
\renp{2}{20mm}{Line nodes \\
 $\Delta\sim \cos 2\chi $} & \renp{2}{14mm}{[3]}
\\
&&&&&&\\
\cline{2-7}
&\renp {2}{30mm}{$URu_{2}Si_{2}$}& \renp
{2}{20mm}{$\ 1.3K$} & \renp{2}{20mm}{Singlet} & 
\renp{2}{30mm}{Hidden order at $T_{0}=17.5K$} 
& 
\renp{2}{20mm}{Line nodes}
& \renp{2}{14mm}{[4]}
\\
&&&&&&\\
\cline{2-7}
&\renp {2}{30mm}{$CePt_{3}Si$} 
& \renp
{2}{20mm}{$\ 0.8K$} & \renp{2}{20mm}{Singlet and Triplet} & 
\renp{2}{38mm}{Parity-violating xtal.\\
$T_{N}=3.7K$} 
& 
\renp{2}{20mm}{Line nodes} 
& \renp{2}{14mm}{[5]}
\\
&&&&&&\\
\hline
\hline
\multirow{4}{25mm}{\bf Quantum Critical}
&\renp {2}{30mm}{$CeIn_{3}$} 
& \renp
{2}{20mm}{$\ 0.2K $ (2.5GPa)} & \renp{2}{20mm}{Singlet} & 
\renp{2}{38mm}{First quantum critical HFSC} 
& 
\renp{2}{20mm}{Line nodes} 
& \renp{2}{14mm}{[6]}
\\
&&&&&&\\
\cline{2-7}
&\renp {2}{32mm}{$CeCu_{2} (Si_{1-x}Ge_{x})_{2}$}& \renp
{2}{24mm}{\small $\ 0.4K $ (P=0)\\
$0.95K$(5.4GPa)} & \renp{2}{20mm}{Singlet} & 
\renp{2}{38mm}{\small Two islands of SC as function of pressure} 
& 
\renp{2}{20mm}{Line nodes.} 
& \renp{2}{14mm}{[7]}
\\
&&&&&&\\
\cline{2-7}
\hline
\hline\multirow{2}{25mm}{\bf Quadrupolar}
&\renp {2}{30mm}{$PrOs_{4}Sb_{12}$}& \renp
{2}{20mm}{$\ 1.85K$} & \renp{2}{20mm}{Singlet} & 
\renp{2}{38mm}{Quadrupolar fluctuations} 
& 
\renp{2}{20mm}{Point nodes} 
& \renp{2}{14mm}{[8]}
\\
&&&&&&\\
\hline
\hline
\multirow{6}{25mm}{\bf ``Strange'' 
}
&\renp {2}{30mm}{$CeCoIn_{5}$} 
& \renp
{2}{20mm}{$\ 2.3K$} & \renp{2}{20mm}{Singlet} & 
\renp{2}{20mm}{Quasi-2D \\
$\rho_{n}\sim T$ } 
& 
\renp{2}{20mm}{Line nodes $d_{x^{2}-y^{2}}$} 
& \renp{2}{14mm}{[9]}
\\
&&&&&&\\
\cline{2-7}
&\renp {2}{30mm}{$UBe_{13}$}& \renp
{2}{20mm}{$\ 0.86K$} & \renp{2}{20mm}{?} & 
\renp{2}{38mm}{\small Incoherent metal at $T_{c}$.} 
& 
\renp{2}{20mm}{Line nodes} 
& \renp{2}{14mm}{[10]}
\\
&&&&&&\\
\cline{2-7}
&\renp {2}{30mm}{$PuCoGa_{5}$}& \renp
{2}{20mm}{$18.5K$} & \renp{2}{20mm}{Singlet} & 
\renp{2}{38mm}{\small Direct transition Curie metal
$\rightarrow $
 HFSC  } 
& 
\renp{2}{20mm}{
Line nodes} 
& \renp{2}{14mm}{[11]}
\\
&&&&&&\\
\hline
\end{tabular}
\vskip 0.1in
\end{center}
{\sl 
\small [1] \citep{steglich},
[2] \citep{upt3},
[3] \citep{steglich3,upd2al3pairing,upd2al3nqr},
[4] \citep{uru2si2,nexus},
[5] \citep{cept3sn},
[6] \citep{mathur},
[7] \citep{cecu2sige},
[8] \citep{pros4sb4},
[9] \citep{sarrao},
[10] \citep{ott},
[11] \citep{sarrao2}.
}

``Strange'' 
superconductors, which 
include $UBe_{13}$, the 115 material $CeCoIn_{5}$ and $PuCoGa_{5}$, condense
into the superconducting state out of an incoherent or strange metallic state.
$UBe_{13}$ has a resistance of order $150\mu\Omega cm$ at its
transition temperature. 
$CeCoIn_{5}$ bears superficial resemblance to a high temperature superconductor, with 
a linear temperature  resistance in its normal state, while 
its cousin, $PuCoGa_{5}$ transitions directly from a
Curie paramagnet of unquenched f-spins into an anisotropically paired,
singlet superconductor.
These particular materials 
severely challenge our theoretical understanding, for 
the heavy electron quasiparticles appear to form
as part of the condensation process, and we are forced to address
how the f-spin degrees of freedom incorporate into
the superconducting parameter.

%\vfill \eject 

\subsection{Phenomenology}\label{}

The main body of theoretical work on heavy electron systems is driven
by experiment, and focuses directly on the phenomenology of the
superconducting state.
For these purposes, it is generally sufficient to 
assume a Fermi
liquid of pre-formed mobile heavy electrons, an electronic analog of superfluid
Helium-3, in which the quasiparticles interact through a
phenomenological BCS model. For most purposes, Landau Ginzburg theory
is sufficient. I regret that in this short review, I do tno have time to properly
represent and discuss the great wealth of fascinating phenomenology -
the wealth of multiple phases, and the detailed models that have been
developed to describe them. I refer the interested reader to reviews
on this subject. \citep{revhfsc1}.

On theoretical grounds, the strong Coulomb interactions of the f-electrons that lead to 
moment formation in heavy fermion
compounds are expected to heavily suppress
the formation of conventional  s-wave pairs
in these systems.  
A large body of evidence
favors the idea  that the gap function and the anomalous Green
function between paired heavy electrons 
$F_{\alpha \beta} ( x)= \la c\dg_{\alpha}(x) c\dg_{\beta}(0)\ra$
is spatially anisotropic, forming either p-wave triplet, or d-wave
singlet pairs.

In BCS theory, the superconducting quasiparticle
excitations are determined by a one-particle Hamiltonian of the form
\[
H = \sum_{\bk,\sigma}\epsilon_{\bk}f\dg_{\bk\alpha}f_{\bk\alpha}
+ \sum_{\bk }[
f\dg _{\bk\alpha}\Delta_{\alpha \beta
}(\bk)f\dg_{-\bk\beta}
+
f_{-\bk\beta}\bar \Delta_{\beta \alpha}(\bk)f_{\bk\alpha}]
\]
where 
\[
\Delta_{\alpha \beta} (\bk )= \left\{
\begin{array}{cl}
\Delta (\bk )(i \sigma_{2})_{\alpha \beta }
&\qquad \hbox{(singlet),}\cr
\vec{d}(\bk )\cdot (i \sigma_{2}\vec{\si})_{\alpha \beta }
&\qquad \hbox{(triplet).}
\end{array}
 \right.  
\]
For singlet pairing, $\Delta (\bk )$ is an even parity function
of $\bk $ 
while for triplet pairing, $\vec{d} (\bk )$ is a an odd-parity function
of $\bk $ with three components. 

The excitation spectrum of an anisotropically paired singlet superconductor is given by
\[
E_{\bk } = \sqrt{\epsilon_{\bk }^{2}+ \vert \Delta_{\bk}\vert^{2}},
\]
This expression can also be
used for a triplet superconductor that does not break time-reversal
symmetry, by making  the replacement $\vert \Delta_{\bk }\vert^{2}\equiv 
\vec{d}\dg_{\bk}\vec{d}_{\bk }$,

Heavy electron superconductors are  anisotropic
superconductors, in which the gap function  vanishes at points,
or more typically,
along lines on the Fermi surface. 
Unlike s-wave superconductors, magnetic and
non-magnetic impurities are equally effective at pair-breaking
and suppressing $T_{c}$ in these materials. 
A node in the gap is the result of sign changes
in the underlying gap function. 
If the gap function vanishes
along surfaces in momentum space,   and the intersection of these
surfaces with the Fermi surface produces ``line nodes''  of gapless
quasiparticle excitations.  As an example, consider $UPt_{3}$, where,
according to one set of models \citep{Upt3theory,Upt3theory2,Upt3theory3,Upt3theory4,Upt3theory5}, pairing involves a
complex d-wave gap 
\[
\Delta_{\bk }\propto k_{z} (\hat k_{x}\pm i k_{y}), \qquad \vert
\Delta_{\bk }\vert^{2}\propto k_{z}^{2} (k_{x}^{2}+k_{y}^{2}).
\]
Here $\Delta_{\bk }$ vanishes along the basal plane $k_{z}=0$, producing a line of nodes
around the equator of the Fermi surface, and along the z-axis,
producing a point node at the poles of the Fermi surface. 

One of the defining properties of line nodes on the Fermi surface
is a quasiparticle 
density  of states that vanishes linearly with energy
\[
N^{*} (E)= 2 \sum_{\bk}\delta (E-E_{\bk })
\propto E
\]
The quasiparticles surrounding the 
line node have a ``relativistic'' energy spectrum. 
In a plane perpendicular to the node, 
$E_{\bk }\sim \sqrt{(v_{F}k_{1
})^{2}+ (\alpha k_{2})^{2}}$ where $\alpha = d\Delta /dk_{2}$ is the
gradient of the gap function and $k_{1} $ and $k_{2}$ the momentum
measured in the plane perpendicular to the line node. 
For a two dimensional relativistic
particle with dispersion $E=ck$, the density of states is given by
$N (E)= \frac{|E|}{4\pi c^{2}}$. For the anisotropic case, we
need to replace $c$ by the geometric mean of $v_{F}$ and $\alpha $, so 
$c^{2}\rightarrow v_{F}\alpha $. This result must then be 
doubled to take account of the spin degeneracy and 
averaged over each line node:
\[
N (E) = 2 \sum_{\hbox{nodes}}\int \frac{dk_{\parallel }}{2
\pi}\frac{|E|}{4 \pi v_{F}\alpha } = |E| \times \sum_{\hbox{nodes}}
\left(\int
\frac{dk_{\parallel }}{4 \pi^{2}v_{F }\alpha } \right)
\]
In the presence of pair-breaking impurities, and in a vortex state the 
quasiparticle nodes are smeared, adding a small constant component to
the density of states at low energies.

This linear density of states 
is manifested in a variety of 
power laws in the temperature dependence of experimental properties, most notably
\begin{itemize}

\item Quadratic temperature dependence of specific heat $C_{V}\propto
T^{2}$, since
the specific heat coefficient is proportional to the thermal average
of the density of states
\[
\frac{C_{V}}{T}\propto  \overbrace {\ \ 
\overline{N (E)}\ \ }^{\propto T}\sim T
\]
where $\overline{N (E)}$ denotes the thermal average of $N (E)$.

\item A ubiquitous $T^{3}$ temperature dependence in the nuclear
magnetic and quadrapole relaxation (NMR/NQR) 
relaxation rates $1/T_{1}$. 
The nuclear relaxation rate is proportional 
to the thermal average
of the squared density of states, so for a superconductor with line
nodes, 
\[
\frac{1}{T_{1}} \propto T \overbrace {\ \ \overline{ N (E)^{2}}\ \ }^{\propto T^{2}} \sim T^{3}
\]
Fig. \ref{fig34} shows the $T^{3}$ NMR relaxation rate of the Aluminum
nucleus in $UPd_{2}Al_{3}$.

\end{itemize}
\fight=0.5\textwidth
%
%\fg{newfigs/upd2al3nqr.eps}{Temperature dependence of the $^{27}Al$ NQR relaxation
\fg{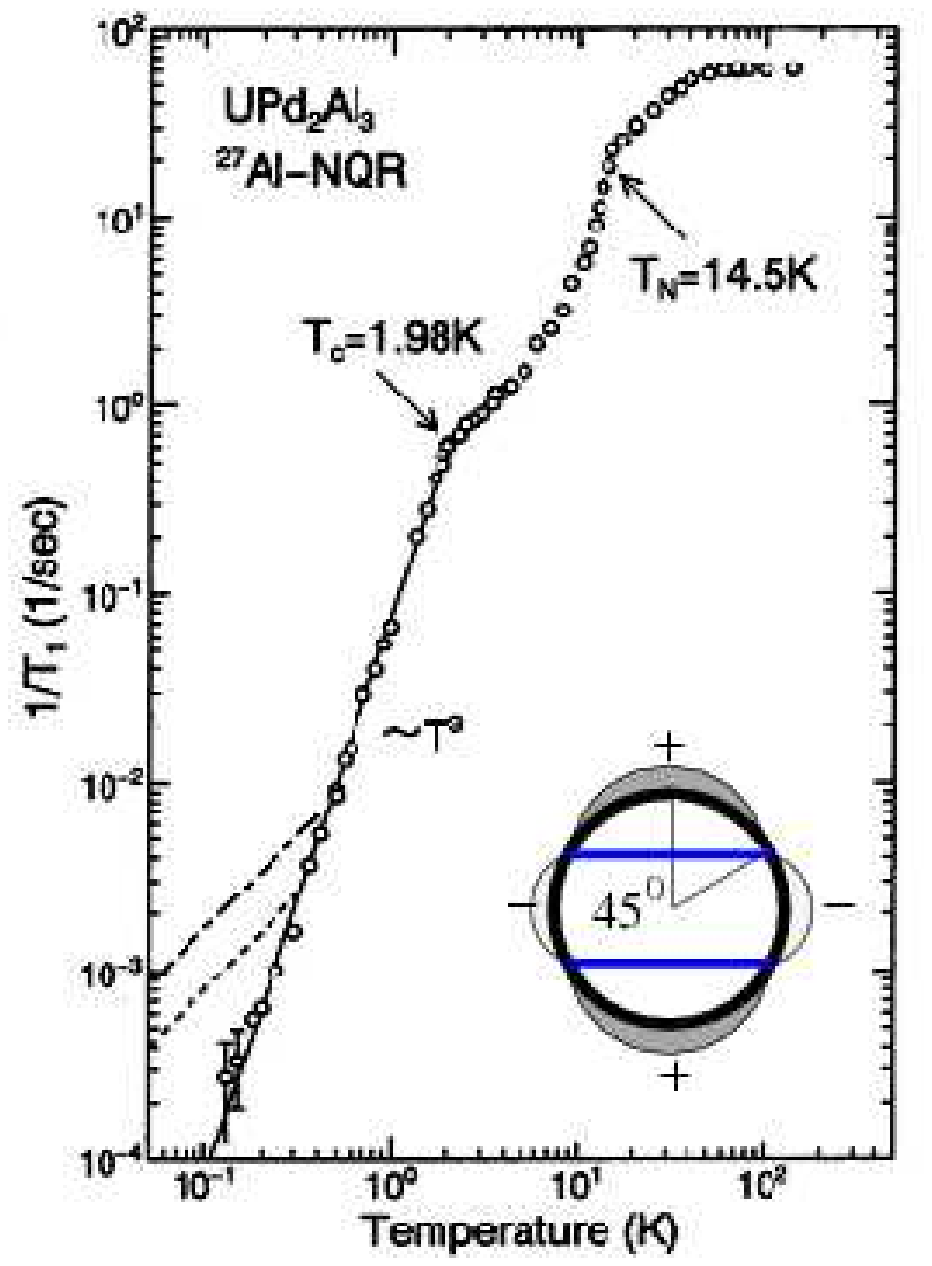}{Temperature dependence of the $^{27}Al$ NQR relaxation
rate $1/T_{1}$ for $UPd_{2}Al_{3}$ after  \citep{upd2al3nqr} showing
$T^{3}$ dependence associated with lines of nodes. Inset showing nodal
structure $\Delta \propto \cos (2\theta )$ 
proposed from analysis of anisotropy of thermal conductivity in  \citep{maki}.}{fig34}

Although power-laws can distinguish line and point nodes, they
do not provide any detailed information about the triplet or singlet 
character of the order parameter, or the location of the nodes.
The observation of upper critical fields that are ``Pauli limited''
(set by the spin coupling, rather
than the diamagnetism), and the observation of a Knight shift
in most heavy fermion superconductors, indicates that they 
are anisotropically singlet paired. Three notable
exceptions to this rule are $UPt_{3}$, $UBe_{13}$ and $UNi_{2}Al_{3}$,
which do not display either a Knight shift or a Pauli-limited upper
critical field, are the best candidates for
odd-parity triplet pairing. In the special case of $CePt_{3}Sn$, the
crystal structure lacks a center of symmetry and the resulting parity
violation is must give a mixture of triplet and singlet pairs. 

Until comparatively recently, very little was known about the
positions of the line nodes in heavy electron  superconductors. 
In one exception, experiments carried out almost twenty years ago 
on  $UPt_3$ observed marked anisotropies in the ultrasound attenuation length
and the penetration 
depth,  \citep{usound,broholm} that appear to support
a line of nodes in the basal plane.
The ultrasonic attenuation $\alpha_s(T)/ \alpha_n$
in single crystals of $UPt_3$ 
has a $T$ linear dependence when the polarization lies in the
basal plane of the gap nodes but a $T^3$ dependence when the polarization is
along the c-axis.

An interesting advance in the experimental analysis of nodal gap 
structure has recently occurred, thanks to new insights
into the behavior of the nodal excitation spectrum in the flux phase
of heavy fermion superconductors. 
In the nineties, Volovik \citep{volovic} observed that 
the energy of heavy electron quasiparticles in a flux lattice is ``Doppler shifted''
by the superflow around the vortices, giving rise to 
a finite density of quasiparticle states around the gap nodes.
The Doppler shift in the
quasiparticle energy resulting from superflow 
is given by 
\[
E_{\bk }\rightarrow E_{\bk }+ \vec{p}\cdot \vec{v}_{s} = E_{{\bk }} + \vec{v}_{F}\cdot \frac{\hbar }{2}\vec{\nabla}\phi
\]
where $\vec{v}_{s}$ is the superfluid velocity and $\phi$ the
superfluid phase.  This has the effect of shifting quasiparticle
states by an energy of order $\Delta E\sim \hbar \frac{v_{F}}{2R}
$,
where $R$ is the average distance between vortices in the flux
lattice. Writing $\pi H R^{2}\sim \Phi_{0}$, and $\pi H_{c2}\xi^{2}\sim
\Phi_{0}$
where $\Phi_{0}= \frac{h}{2e}$ is the flux quantum, $H_{c2}$ is the upper critical
field and $\xi$ is the coherence length. it follows that
$\frac{1}{R}\sim \frac{1}{\xi}
\sqrt{\frac{H}{H_{c2}}}$. Putting $\xi\sim v_{F}/\Delta $, where
$\Delta $ is the typical size of the gap,
the typical shift in the energy of nodal quasiparticles
is of order
$E_{H}\sim \Delta \sqrt{\frac{H}{H_{c2}}}$.
Now since the density of states is of order $N (E)= \frac{\vert E\vert}{\Delta }
N (0)$, 
where $N (0)$ is the density of states in the normal phase. it follows
that 
the smearing of the nodal quasiparticle energies  will
produce a density of states of order 
\[
N^{*} (H)\sim N(0)\sqrt{\frac{H}{H_{c2}}}
\]
This effect, the ``Volovik effect'', produces a linear component to
the specific heat $C_{V}/T \propto \sqrt{\frac{H}{H_{c2}}}
$. This enhancement of the density of states 
is largest when the group velocity $\vec{V}_{F}$ at the node is 
perpendicular to the applied field $\vec{H}$, and when the field is
parallel to $\vec{v}_{F}$ at a particular node, the node is unaffected
by the vortex lattice (Fig. \ref{fig35}). 
 This gives  rise to
an angular dependence in the specific heat coefficient and thermodynamics
that can be used to measure the gap anisotropy.
In practice, the
situation is complicated at higher fields where the Andreev
scattering of quasiparticles by vortices becomes 
important.  The case of $CeCoIn_{5}$ is of particular current interest.
Analyses of the field-anisotropy of the thermal conductivity in this material
was interpreted early on in terms of 
a gap structure with $d_{x^{2}-y^{2}}$, while the anisotropy
in the specific heat appears to suggest a $d_{xy}$
symmetry.  Recent theoretical work by Vorontsev and Vekhter \citep{vorontsov}
 suggests
that the discrepancy between the two interpretations can be resolved
by taking into account the effects of the vortex quasiparticle
scattering  that were ignored in the specific heat interpretation.
They predict that at lower fields, where
vortex scattering effects are weaker, the sign of the anisotropic term
in the specific heat will reverse, accounting for the discrepancy

\fight=0.6\textwidth
%\fg{newfigs/vekhter.eps}{Schematic showing how the nodal quasiparticle
\fg{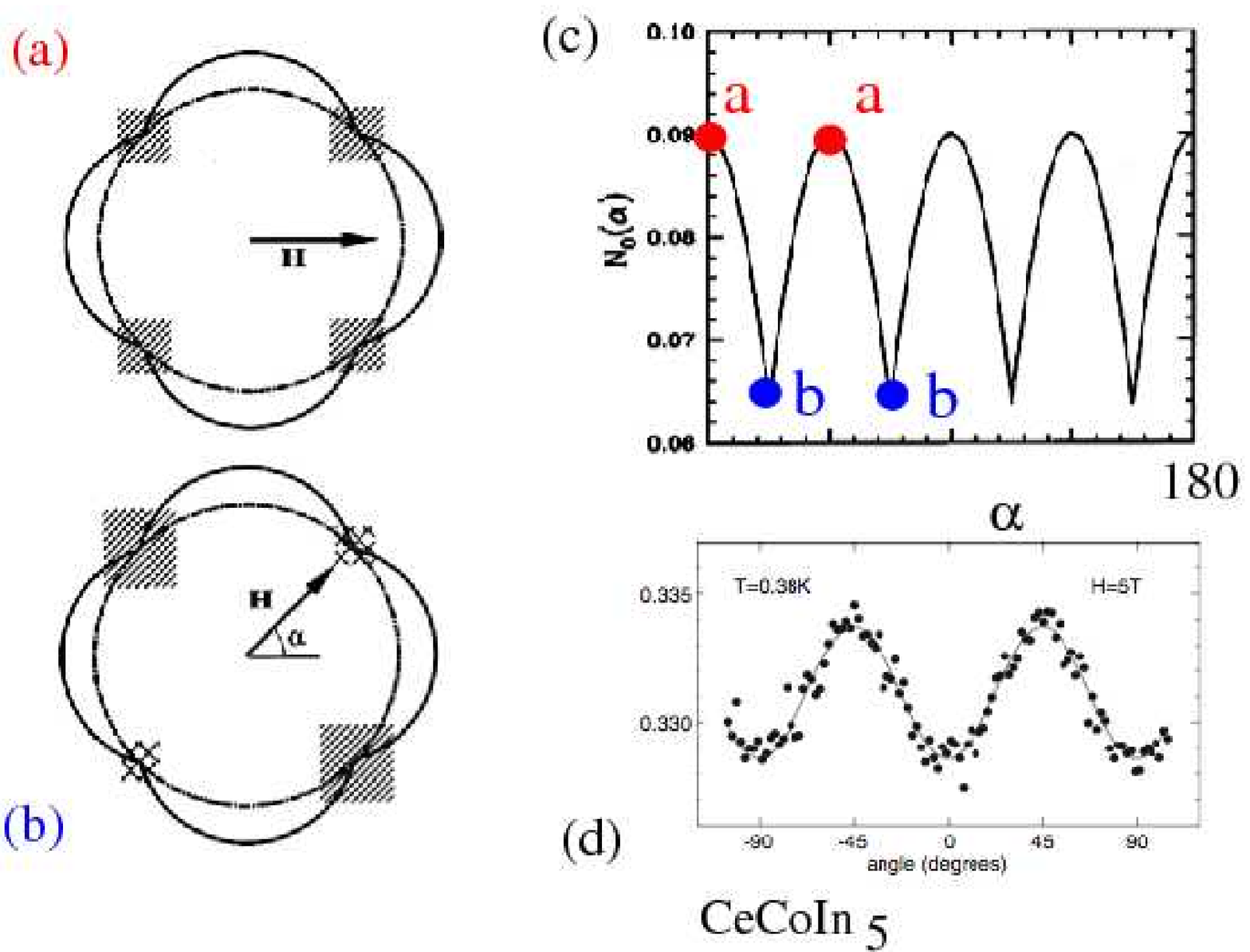}{Schematic showing how the nodal quasiparticle
density of states depends on field orientation after  \citep{vekhter}.
(a) Four nodes are activated when the field points towards an
antinode, creating  a maximum
in density of states. (b) Two nodes activated when the field points
towards a node,
creating  a minimum in the density of states. (c) Theoretical dependence of
density of states on angle, after  \citep{vekhter}. (d) Measured angular
dependence of $C_{v}/T$ after  \citep{aoki} is $45^{0}$ out of phase
with prediction. This discrepancy is believed to be due to vortex
scattering, and is expected to vanish at lower fields. }{fig35}

It is clear that, despite the teething problems in the interpretation
of field-anisotropies in transport and thermodynamics, 
this is an important emerging tool for the
analysis of gap anisotropy,  and to date, it has been used to give
tentative assignments to the gap anisotropy of 
$UPd_{2}Al_{3}$, $CeCo{In}_{5}$ and $PrOs_{4}Sb_{12}$. 

\subsection{Microscopic models}\label{}

\subsubsection{Antiferromagnetic fluctuations as a pairing force.}\label{}

The classic theoretical models for heavy fermion superconductivity
treat the  heavy electron
fluids as a Fermi liquid  with antiferromagnetic
interactions amongst their quasiparticles \cite[]{bealmonod,scalapino,
monthoux}.
$UPt_{3}$ provided the  experimental inspiration for early theories of
heavy fermion superconductivity, for its
superconducting state
forms from within a well-developed Fermi liquid. Neutron scattering
on this material shows signs of antiferromagnetic spin
fluctuations \citep{aeppliupt3}, making it natural to presuppose that these might be the
driving force for heavy electron pairing. 

Since the early seventies, theoretical models had predicted that strong ferromagnetic
spin fluctuations, often called ``paramagnons'', could induce p-wave pairing,
and it this mechanism was widely held to be the driving force for
pairing in superfluid $He-3$.  
An early proposal
that antiferromagnetic interactions could provide the driving force
for  anisotropic singlet
pairing was made by Hirsch \citep{hirsch}. Shortly thereafter, three seminal papers,  by
B\' eal-Monod, Bourbonnais and Emery \citep{bealmonod} (BBE),  Scalapino, Loh and
Hirsch  \citep{scalapino} (SLH) and by Miyake,
Schmitt-Rink and Varma \citep{miyake} (MSV), solidified this idea 
with a concrete demonstration that antiferromagnetic interactions 
drive an attractive BCS interaction in the d-wave pairing channel.
It is a fascinating thought that at the same time that
this set of authors were forging the foundations of our current
thoughts on the link between antiferromagnetism and d-wave superconductivity,
Bednorz and Mueller were in the process of discovering high
temperature superconductivity. 

The BBE and SLH papers develop a paramagnon theory 
for d-wave pairing
in a Hubbard model with a contact interaction $I$, having in mind a
system, which in the modern context, would be said to be close to an
antiferromagnetic quantum critical point. 
The MSV paper
starts with a model with a pre-existing antiferromagnetic interaction,
which in the modern context would be associated with the ``t-J''
model. It is this approach that I sketch here.  The MSV model is written
\begin{equation}\label{msvmod}
H = \sum \epsilon_{\bk }a\dg_{\bk \sigma}a_{\bk \sigma }+ H_{int}
\end{equation}
where 
\begin{equation}\label{msvint}
H_{int} = 
\frac{1}{2}\sum_{\bk,\bk'} \sum_{\bf q} J (\bk - \bk  ')
\vec{\sigma }_{\alpha \beta }
\cdot
\vec{\sigma }_{\gamma\delta} \times 
\left( a\dg_{\bk  + \bq  /2 \alpha }
a\dg_{-\bk  + \bq  /2 \gamma}
\right)
\left(\phantom{a\dg }\hskip -0.1in
a_{-\bk'  + \bq  /2 \delta}
a_{\bk'  + \bq  /2 \beta } \right)
\end{equation}
describes the antiferromagnetic interactions. 
There are a number of interesting points to be made here:
\begin{itemize}

\item The authors have in mind a strong coupled model, such as the
Hubbard model at large $U$ where the interaction can not be simply
derived from paramagnon theory. 
In a weak-coupled Hubbard model a contact interaction $I$ and 
bare susceptibility $\chi_{0} (q)$,
the induced magnetic interaction can be calculated in an RPA approximation \citep{miyake} as
\[
J (q)= -\frac{I}{2[1 + I \chi_{0} (q)]}.
\]
MSV make the point that 
the detailed mechanism that links the low-energy antiferromagnetic
interactions to the microscopic interactions is poorly described by a
weak-coupling theory, and quite likely to involve other processes, such as
the RKKY interaction, and the Kondo effect
that lie outside this treatment. 

\item Unlike phonons, magnetic interactions in heavy fermion systems
can not generally be regarded
as retarded interactions, for they 
extend up to an energy scale
$\omega_{0}$  that is comparable
with the heavy electron band-width $T^{*}$.
In a classic BCS treatment, the electron energy 
are restricted to lie within a Debye energy of the Fermi energy. 
But here, $\omega_{0}\sim T^{*}$, so  all momenta are
involved in magnetic interactions, 
and the interaction can transformed to 
real  space as 
\begin{equation}\label{tjmsv}
H = \sum \epsilon_{\bk }a\dg_{\bk \sigma}a_{\bk \sigma }+ 
\frac{1}{2}\sum_{i,j} J (\bR_{i}-\bR_{j})
\vec{\sigma }_{i }
\cdot
\vec{\sigma }_{j}
\end{equation}
where $J (\bR )= \sum_{\bq} e^{i\bq\cdot\bR}J (\bq )$ is the Fourier transform
of the interaction and $\vec{\sigma}_{i}= a\dg_{i\alpha }\vec{\sigma
}_{\alpha \beta }a_{i\beta }$ is the spin density at site $i$.
Written in real space, the MSV model is seen to be an early  predecessor of the $t-J$
model used extensively in the context of high temperature superconductivity. 

\end{itemize}
To see that antiferromagnetic interactions favor d-wave pairing, 
one can use the, let us decouple the interaction in real space in
terms of triplet and singlet pairs. Inserting 
the identity
\footnote{
To prove this identity, first note that any two dimensional
matrix, $M$ can be expanded as
$M=m_{0}\sigma_{2}+\vec{m}\cdot\sigma_{2}\vec{\sigma}$,
($b= (1,3)$) where 
$m_{0}=\frac{1}{2}{\rm Tr}[M \sigma_{2}]$ and 
$\vec{m}=\frac{1}{2}{\rm Tr}[M \vec{\sigma }\sigma_{2}
]$, so that in index notation
\[
M_{\alpha \gamma}= \frac{1}{2} 
{\rm Tr}[M \sigma_{2}](\sigma_{2})_{\alpha \gamma}
+
\frac{1}{2} {\rm Tr}[M \vec{\sigma}\sigma_{2}]\cdot (\sigma_{2}\vec{\sigma})_{\alpha \gamma}
\]
Now if we apply this relationship to the $\alpha \gamma$ components 
of $\vec{\sigma }_{\alpha \beta }\cdot \vec{\sigma }_{\gamma \delta}
$, we obtain
\[
\vec{\sigma }_{\alpha \beta }\cdot \vec{\sigma }_{\gamma \delta}
=
\frac{1}{2}
\left(\vec{\sigma}^{T}\sigma_{2}\vec{\sigma } \right)_{\delta \beta }(\sigma_{2})_{\alpha \gamma}
+ 
\frac{1}{2}\sum_{b=1,3}
\left(\vec{\sigma}^{T}\sigma_{2}\sigma_{b}\ \vec{\sigma }
\right)_{\delta \beta }(\sigma_{2}\sigma_{b})_{\alpha \gamma}
\]
If we now use the
relation $\vec{\sigma}^{T}\sigma_{2}=-\sigma_{2}\vec{\sigma} $, together
with $\vec{\sigma }\cdot\vec{\sigma}=3$
and $\vec{\sigma }\sigma_{b}\vec{\sigma }=-\sigma_{b}$, we obtain
\[
\vec{\sigma }_{\alpha \beta }\cdot \vec{\sigma }_{\gamma \delta}=
-\frac{3}{2}(\sigma_{2})_{\alpha \gamma}
(\sigma_{2})_{\delta \beta }
+
\frac{1}{2}(\vec{\sigma }\sigma_{2})_{\alpha \gamma}
\cdot\left(\sigma_{2} \vec{\sigma }\right)_{\delta \beta }
\]
}
\begin{equation}\label{spinid}
\vec{\sigma }_{\alpha \beta }
\cdot
\vec{\sigma }_{\gamma\delta  } = -\frac{3}{2} (\sigma_{2})_{\alpha
\gamma} (\sigma_{2})_{\beta \delta }+
\frac{1}{2} (\vec{\sigma }\sigma_{2})_{\alpha
\gamma}\cdot(\sigma_{2}\vec{\sigma })_{\beta \delta }.
\end{equation}
into 
(\ref{tjmsv} ) gives 
\begin{equation}\label{}
H_{int}
=-\frac{1}{4}\sum_{i,j} J_{ij}
\left[3\Psi\dg _{ij }
\Psi_{ij}-
\vec{\Psi}\dg _{ij }\cdot
\vec{\Psi}_{ij}
 \right]
\end{equation}
where
\begin{eqnarray}\label{l}
{\Psi}\dg _{ij}
&=& 
\left(a\dg _{i\alpha } 
(-i\sigma)_{\alpha \gamma }\ 
a\dg _{j \gamma} 
 \right)\cr
\vec{\Psi}\dg _{ij }
&=& \left(a\dg _{i \alpha }  
 (-i\vec{\sigma }\sigma_{2}
)_{\alpha \gamma  }\ 
a\dg _{j\gamma}
\right)
\end{eqnarray}
create singlet and triplet pairs with electrons located at sites $i$
and $j$ respectively.
In real-space it is thus quite clear that an
antiferromagnetic interaction $J_{ij}>0$ induces attraction in the
singlet channel, and repulsion in the triplet channel.
Returning to momentum space, substitution of (\ref{spinid} ) into (\ref{msvint}) gives
\begin{equation}\label{}
H_{int}
=-\sum_{\bk,\bk'} J (\bk -\bk ')
\left[3\Psi\dg _{\bk ,\ \bq }
\Psi_{\bk' ,\ \bq }-
\vec{\Psi}\dg _{\bk ,\ \bq }\cdot
\vec{\Psi}_{\bk' ,\ \bq }
 \right]
\end{equation}
where 
${\Psi}\dg _{\bk ,\bq }
= 
\frac{1}{2}\left(a\dg _{\bk +\bq /2\  \alpha } 
(-i\sigma_{2})_{\alpha \gamma }\ 
a\dg _{-\bk - \bq /2\  \gamma} 
 \right)
$
and 
$\vec{\Psi}\dg _{\bk ,\bq }
= \frac{1}{2}
\left(a\dg _{\bk +\bq /2 \ \alpha } 
(-i\vec{\sigma }\sigma_{2})_{\alpha \gamma }\ 
a\dg _{-\bk - \bq /2 \ \gamma} 
 \right)
$
create singlet and triplet pairs at momentum $\bq$ respectively.  
Pair condensation is described by the zero momentum component of this
interaction, which gives 
\begin{equation}\label{}
H_{int}
=\sum_{\bk,\bk'} 
\left[V^{(s)}_{\bk ,\bk '}
\Psi\dg _{\bk }
\Psi_{\bk'}+V^{(t)}_{\bk ,\bk '}
\vec{\Psi}\dg _{\bk }\cdot
\vec{\Psi}_{\bk' }
 \right]
\end{equation}where 
${\Psi}\dg _{\bk }
= 
\frac{1}{2}\left(a\dg _{\bk \alpha } 
(-i\sigma_{2})_{\alpha \beta }\ 
a\dg _{-\bk  \beta} 
 \right)
$
and 
$\vec{\Psi}\dg _{\bk ,\bq }
= 
\frac{1}{2}\left(a\dg _{\bk \alpha } 
(-i\vec{\sigma }\sigma_{2})_{\alpha \beta }\ 
a\dg _{-\bk  \beta} 
 \right)
$
create Cooper pairs  and 
\begin{eqnarray}\label{l}
V^{s)}_{\bk ,\bk '}&=& - 3 [
J (\bk -\bk ')+J (\bk +\bk ')]/2\cr
V^{(t)}_{\bk ,\bk '}&=& \ \ [
J (\bk -\bk ')-J (\bk +\bk ')]/2
\end{eqnarray}
are the BCS pairing potentials in the singlet, and triplet channel
respectively. (Notice how the even/odd parity symmetry of the triplet
pairs pulls out the corresponding symmetrization of $J (\bk -\bk ')$)

For a given choice of $J (\bq)$,  it now becomes possible to decouple
the interaction in singlet and triplet channels. For example, on a
cubic lattice of side length,  if the magnetic interaction has the form
\[
J (\bq ) = 2J (\cos(q_xa)+\cos(q_ya) + \cos(q_za))
\]
which generates soft antiferromagnetic fluctuations at  the staggered
$\bQ $ vector 
 $\bQ= (\pi/a,\pi/a,\pi/a)$, 
then the pairing interaction can
be decoupled into singlet and triplet components, 
\begin{eqnarray}\label{l}
V^{s}_{\bk ,\bk '}&=& - \frac{3J}{2} \bigl[
s (\bk) 
s(\bk ')+
d_{x^{2}-y^{2}}(\bk )
d_{{x^{2}-y^{2}}} (\bk ')+
d_{2z^{2}-r^{2}} (\bk )
d_{2z^{2}-r^{2}} (\bk' )\bigr]
\cr
V^{t}_{\bk,\bk'}&=&\frac{J}{2} \sum_{i=x,y,z}p_{i} (\bk)p_{i} (\bk')
\end{eqnarray}
where
\begin{equation}\label{}
\begin{array}{ll}
s (\bk) 
=\sqrt{\frac{2}{3}} (\cos (k_{x}a)+\cos (k_{y}a)+\cos
(k_{z}a))& (\hbox{extended s-wave})\cr
\left.
\begin{array}{l}
d_{x^{2}-y^{2}}(\bk )
=(\cos(k_xa) - \cos(k_ya), \cr
d_{2z^{2}-r^{2}} (\bk )=\frac{1}{\sqrt{3}}
(\cos(k_xa) +\cos(k_ya)- 2\cos(k_za))
\end{array}
 \right\}
&(\hbox{d-wave})
\end{array}
\end{equation}
are the gap functions for singlet pairing and 
\begin{equation}\label{}
p_{i} (\bq)
= \sqrt{2}\sin (q_{i}a), \qquad (i=x,y,z), \qquad \qquad (\hbox{p-wave})
\end{equation}
describe three triplet gap functions. 
For $J>0$, this particular BCS  model will then give rise to extended
s and d-wave superconductivity with approximately the same transition
temperatures, given by the gap equation
\[
\sum_{\bk}\tanh \left(\frac{ \epsilon_{{\bk }}}{2T_{c}} \right)\frac{1}{\epsilon_{\bk }}\left\{\begin{array}{c}
s (\bk)^{2}\cr
d_{{x^{2}-y^{2}}} (\bk )^{2}
\end{array} \right\}= \frac{2}{3J}
\]

\subsubsection{Towards a unified theory of HFSC}\label{}

Although the spin fluctuation approach described provides a good starting point for
the phenomenology of heavy fermion superconductivity, 
it leaves open a wide range of questions that suggest this problem is
only partially solved:
\begin{itemize}

\item How can we reconcile
heavy fermion superconductivity 
with the local moment origins of the heavy electron quasiparticles? 

\item How can the 
the incompressibility of the heavy electron fluid be incorporated 
into the theory?
In particular, extended s-wave solutions are expected to produce a
large singlet f-pairing amplitude, giving rise to a large Coulomb
energy. Interactions are
expected  to significantly depress, if not totally eliminate such extended s-wave solutions. 

\item Is there a controlled limit where a model of heavy electron
superconductivity can be solved?

\item What about the ``strange'' heavy fermion superconductors $UBe_{13}$,
$CeCoIn_{5}$ and $PuCoGa_{5}$, where $T_{c}$ is comparable with the Kondo
temperature?  In this case, the superconducting order parameter must involve the
f-spin as a kind of ``composite'' order parameter. What is the nature
of this order parameter, and what physics drives $T_{c}$ so high that
the Fermi liquid forms at much the same time as the superconductivity
develops? 

\end{itemize}

One idea  that may help to understand the heavy fermion pairing mechanism 
is Anderson's resonating valence bond (RVB)
theory \citep{science} 
of high temperature superconductivity.
Anderson proposed \citep{science,baskaran,kotliarrvb} that the parent state of 
the high temperature superconductors is a two-dimensional spin liquid of
resonating valence bonds between spins,  which becomes
superconducting upon doping with holes.  
In the early nineties, Coleman and Andrei \citep{colemanandrei} adapted
this theory to a Kondo lattice. 
Although an RVB spin-liquid 
is unstable with respect to antiferromagnetic order in three dimensions, 
in situations close to a magnetic instability, 
where the energy of the antiferromagnetic
state is comparable with the Kondo temperature,  $E_{AFM}\sim T_{K}$, 
conduction electrons will partially spin-compensate the spin liquid, 
stabilizing it
against magnetism (Fig. \ref{fig36} (a)).  
In the Kondo-stabilized spin liquid,  the Kondo effect induces some
resonating valence bonds in the f-spin liquid to escape into the
conduction fluid where they pair charged electrons to form
a heavy electron superconductor.  
\fight=\textwidth
%\fg{newfigs/andreicoleman.eps}{Kondo stabilized spin liquid, diagram from
\fg{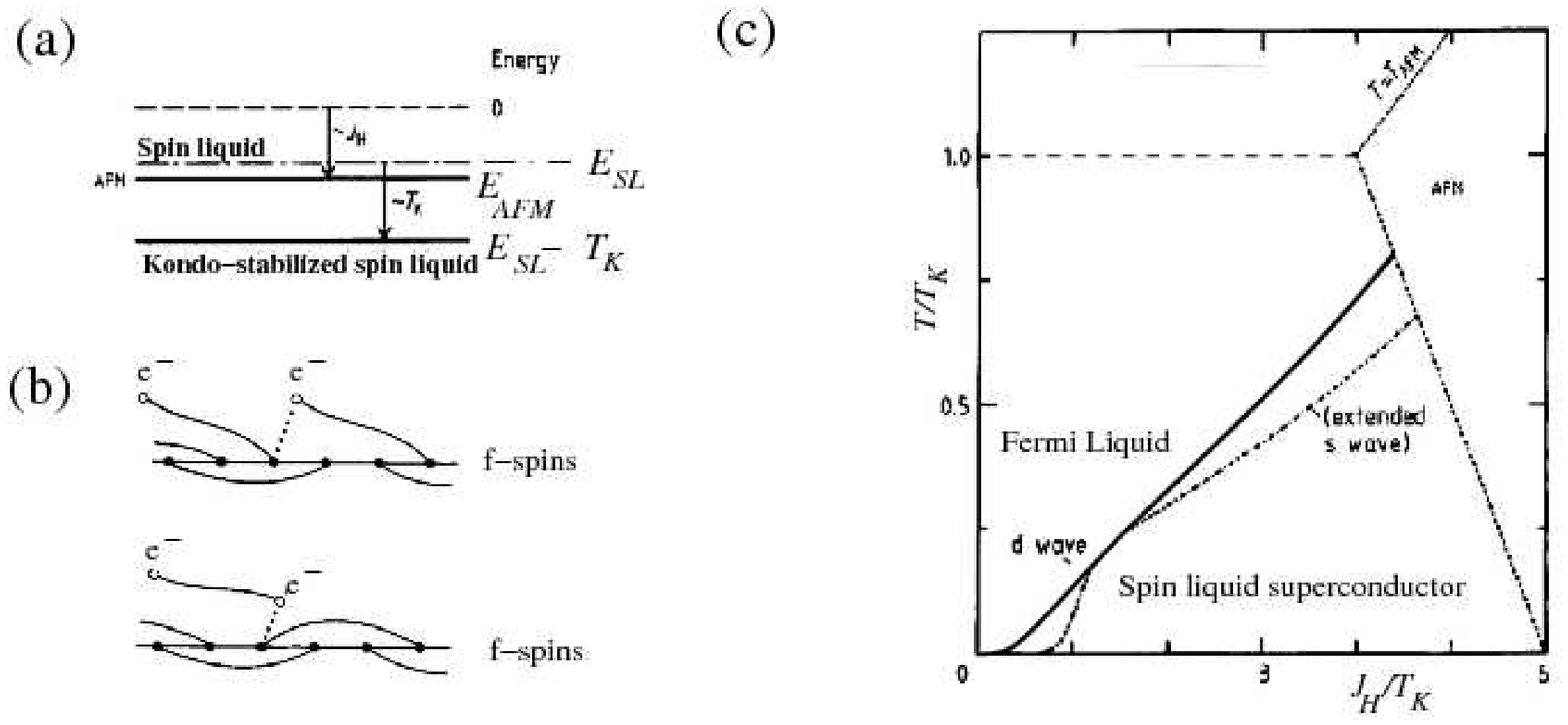}{Kondo stabilized spin liquid, diagram from
 \citep{colemanandrei}(a)Spin liquid stabilized by Kondo effect, (b)
Kondo effect causes singlet bonds to form between spin liquid and
conduction sea. Escape of these bonds into the conduction sea induces
superconductivity. (c) Phase diagram computed using $SU (2)$ mean-field theory
of Kondo Heisenberg model. }{fig36}

A key observation of 
RVB theory is that when charge fluctuations are removed to form a  spin
fluid, there is no distinction between particle and hole \citep{affleck}.  The
mathematical consequence of this, is that the 
the spin-$1/2$ operator
\[
\vec{S}_f = f\dg _{i \alpha} \left(\frac{{\vec \sigma } 
}{2}
\right)_{\alpha\beta}
f\dg_{\alpha}, \qquad n_{f}=1
\]
is not only invariant under a change of phase $f_{\sigma }\rightarrow
e^{i\phi }f_{\sigma }$, but it also possesses a continuous
particle hole symmetry 
\begin{equation}\label{}
f\dg_{\sigma}\rarrow
\cos \theta f\dg_{\s} +{\rm sgn} \si sin\theta f_{-\s}.
\end{equation}
These two symmetries combine to create a {\sl local $SU (2)$ gauge
symmetry}.  
One of the implications is that the
constraint $n_{f}=1$ associated with the spin operator, is
actually part of a triplet of Gutzwiller constraints
\begin{equation}\label{}
f\dg_{i\up}f_{i\up}-f_{i\dw}f\dg_{i \dw}=0 , \quad
f\dg_{i \up}f\dg_{i \dw} = 0 , \quad f_{i \dw}f_{i\up}=0.
\end{equation}
If we introduce the Nambu spinors
\begin{equation}\label{}
\tilde{f}_{i}\equiv 
\begin{pmatrix} f_{i \uparrow}\cr f\dg_{i \downarrow }\end{pmatrix}
,\qquad 
\tilde{f}\dg_i=(f\dg_{i \up}, f_{i\dw}).
\end{equation}
then this means that all three components of the ``isospin'' of the
f-electrons vanish, 
\begin{equation}\label{l}
\tilde{f}\dg_i \vec \tau \tilde{f}_i= (f\dg_{i \up},
f_{i\dw})\left[\begin{pmatrix}0 & 1 \cr 1 & 0\end{pmatrix}
,\begin{pmatrix}0 & -i \cr i & 0\end{pmatrix}
,\begin{pmatrix}1 & 0 \cr 0 & -1\end{pmatrix}
 \right]\begin{pmatrix} f_{i \uparrow}\cr f\dg_{i \downarrow }\end{pmatrix}=0,
\end{equation}
where $\vec{\tau} $ is a triplet of
Pauli spin operators that act on the f-
Nambu spinors.
In other words, in the incompressible
f-fluid, there can be no
$s-wave$ singlet pairing. 

This symmetry is preserved in spin-1/2 Kondo models.  When applied to the
Heisenberg Kondo model 
\begin{equation}\label{}
H=\sum_{\vk \sigma} \epsilon_{\vk}c\dg_{\vk \sigma} c_{\vk \sigma}
+ J_H\sum_{(i,j)} \Sfi \cdot \Sfj + J_K \sum_j
c\dg_{j \s  }{\vec \sigma } _{\s \s'}c_{j\s'   }
\cdot
\Sfj 
\end{equation}
where $\Sfi= f\dg _{i \alpha} \left(
\frac{{\vec \sigma }}{2}
\right) _{\alpha \beta} f_{i\beta}$ represents an
f-spin at site i, it leads to an SU (2) gauge theory for the Kondo
lattice with Hamiltonian
\begin{eqnarray}\label{l}
H &= \sum_{\vec k} \epsilon _{\vk}
\tilde{c}\dg_{\vec k}
\tau_3 \tilde{c}_{\vec k}+\sum \vec{\lambda}_{j}\tilde{f}_{j}\dg \vec{\tau }\tilde{f}_{j}+
\sum_{(i,j)}[\tilde{f}\dg_i U_{ij}\tilde{f}_j + H.c] +  { 1\over J_H }
\Tr [U\dg_{ij}U_{ij}] \cr
&+ \sum_i [ \tilde{f}\dg_i V_i \tilde{c}_i + H.c] +  { 1\over J_K}
\Tr [ V \dg_i V_i] . 
\end{eqnarray}
where $\lambda_{j}$ is the Lagrange multiplier that imposes the
Gutzwiller constraint $\vec{\tau }=0$ at each site,
$\tilde{c}_{k}=\left({c_{k \uparrow}\atop c\dg _{-k\downarrow }}
\right)$ 
and 
$\tilde{c}_{j}=\left({c_{j \uparrow}\atop c\dg _{j\downarrow }}
\right)$ are Nambu conduction electron spinors in the momentum, and
position basis, respectively, while
\[
U_{ij}= \mat {h_{ij}& \Delta_{ij}\cr \bar \Delta_{ij}& - \bar h_{ij}},
\qquad V_{i}= \mat{V_i& \bar \alpha_{i}\cr \alpha_{i}&- \bar V_{i}}
\]
are matrix order parameters  associated with the Heisenberg and Kondo
decoupling, respectively.  This model has the local gauge invariance
$\tilde{f}_{j}\rightarrow g_{j}\tilde{f}_j$, $V_{j}\rightarrow g_{j}V_{j}$
$U_{ij}\rightarrow g_{i}U_{ij}g\dg_{j}$ where $g_{j}$ is an $SU (2)$ matrix.
In this kind of model, one can ``gauge fix'' the model so that the
Kondo effect occurs in the particle-hole channel ($\alpha_{i}=0$). When one does so
however,  the spin-liquid instability takes place preferentially 
in an anisotropically paired Cooper channel. Moreover, the constraint on the f-electrons not
only suppresses singlet s-wave pairing, it also suppresses extended
s-wave pairing (Fig. \ref{fig36} ). 

One of the initial difficulties with both the RVB and the 
Kondo stabilized-spin liquid approaches is that in its original formulation, it could not be
integrated into a large   $N$ approach. Recent  work 
indicates  that both the fermionic RVB and the Kondo
stabilized spin liquid picture can be formulated as a controlled $SU
(2) $ gauge theory by carrying out a large $N$ expansion using 
the group $SP (N)$ \citep{sp2n}, originally
introduced by Read and Sachdev for problems in frustrated magnetism,  in place of the group $SU (N)$.
The local particle-hole symmetry associated with the spin operators in
$SU (2)$ is intimately related to the 
symplectic property of Pauli spin operators
\begin{equation}\label{}
\sigma_{2}\vec{\sigma }^{T}\sigma_{2}= - \vec \sigma 
\end{equation}
where $\vec{\sigma}^{T}$ is the transpose of the spin
operator. This relation, which represents the sign reversal of spin operators
under time-reversal, is only satisfied by a subset of the $SU (N)$ spins for
$N>2$. This subset defines the generators of the symplectic 
subgroup of $SU (N)$, called $SP (N)$. 
\fight=0.5\textwidth
%\fg{newfigs/pucoga5.eps}{Temperature dependence of the magnetic
\fg{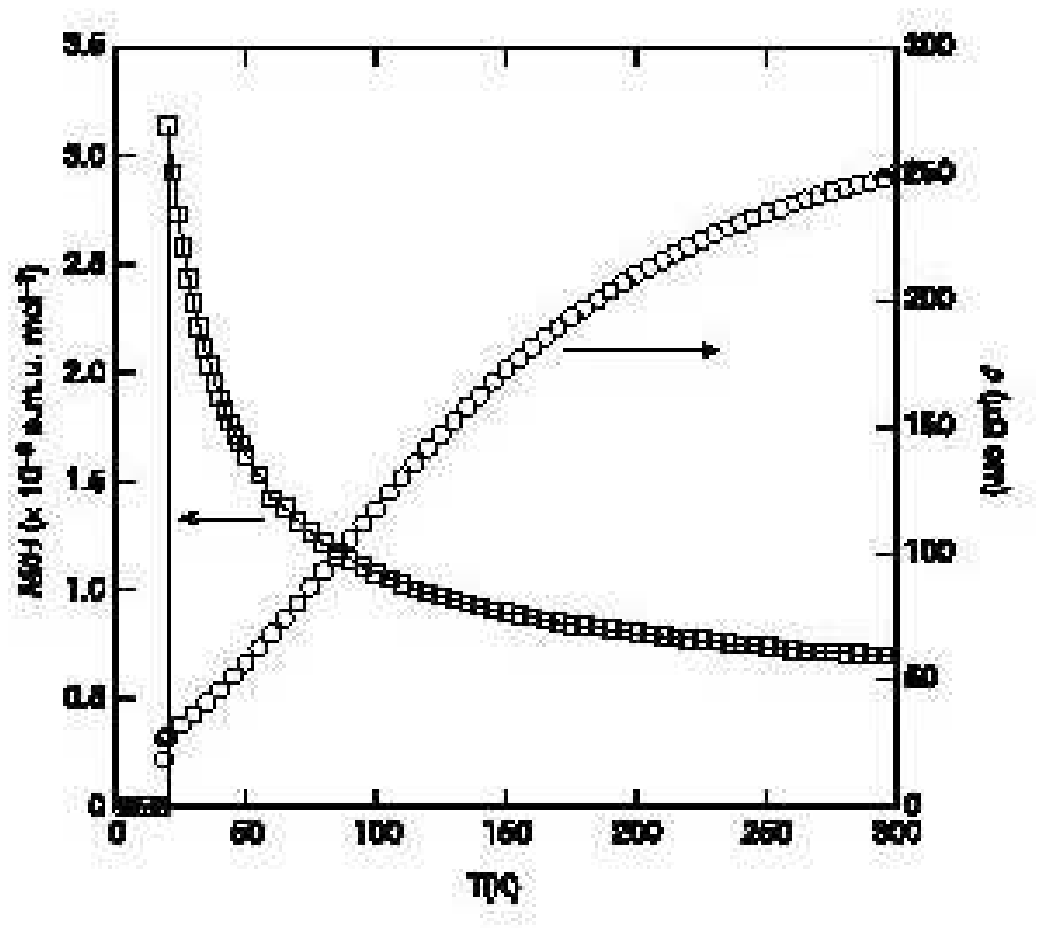}{Temperature dependence of the magnetic
susceptibility of $PuCoGa_{5}$ after  \citep{sarrao2}. The susceptibility
shows a direct transition from Curie Weiss paramagnet into heavy
fermion superconductor, without any intermediate spin quenching.
}{fig37}

Concluding this section, I want to
briefly mention the challenge posed by the highest $T_{c}$
superconductor, $PuCoGa_{5}$ \citep{sarrao2,curro}. This material, discovered some four years
ago at Los Alamos, undergoes a direct transition from a Curie
paramagnet into a heavy electron superconductor at around
$T_{c}=19K$ (Fig.\ref{fig37}).
The Curie paramagnetism is also seen in the Knight-shift, which scales
with the bulk susceptibility \citep{curro}. 
The remarkable feature of this
material, is that the specific heat anomaly has the large  size
($110mJ/mol/K^{2}$ \citep{sarrao2})
characteristic of heavy fermion superconductivity, yet there there are
no signs of saturation in the susceptibility as a precursor to
superconductivity, suggesting that the heavy quasiparticles do not
develop from local moments until the transition. This aspect of the
physics can not be explained by spin-fluctuation theory\citep{bang},
and suggests
that the Kondo effect takes place simultaneously with the pairing mechanism.
One interesting possibility here, is that the development of coherence between
the Kondo effect in two different channels created by the different
symmetries of the valence fluctuations into the $f^{6}$ and $f^{4}$ states
might be the driver of
this intriguing new superconductor \citep{coxhitc,keeetal}. 

%% Odd frequency pairing, composite pairing.

%% RVB + HFSC

%\section{Non-Fermi liquids}\label{}

%
%
%
%
%
%

\section{Quantum Criticality}\label{}

\subsection{Singularity in the Phase diagram}\label{}

Many heavy
electron systems can be tuned, with pressure, chemical doping or
applied magnetic field, to a point where their antiferromagnetic
ordering temperature is driven continuously to zero to produce a
``quantum critical point'' \citep{revqcp1,revqcp1addendum,revqcp2,revvarma2001,revqcp3,revqcp4}. The remarkable transformation
in metallic properties, often referred to as ``non-Fermi liquid
behavior'',  that is induced over a wide range of
temperatures above the quantum critical point (QCP), 
together with the marked tendency 
to develop superconductivity in the vicinity of such ``quantum
critical points'' has given rise to a resurgence of interest in
heavy fermion materials.  

The experimental realization of quantum criticality returns us
to central questions left unanswered since the first
discovery of heavy fermion compounds. In particular:
\begin{itemize}

\item What is the fate of the Landau quasiparticle
when interactions become so large
that the ground state is no longer adiabatically connected
to a non-interacting system? 

\item What is the mechanism by which the antiferromagnet transforms
into the heavy electron state? Is there a break-down of the Kondo effect,
revealing local moments at the quantum phase transition, or is the transition
better regarded as a spin density wave transition?

\end{itemize}
Fig. \ref{fig38}. illustrates quantum criticality in $YbRh_{2}Si_{2}$\citep{custers}, a
material with a $90mK$ magnetic transition that can be tuned
continuously to zero by  a modest magnetic field.  In wedge-shaped
regions either side of the
transition, the resistivity displays the  $T^{2}$ dependence
$\rho (T)=\rho_{0} + AT^{2}$ (colorcoded blue) that is the hallmark of Fermi
liquid behavior. Yet in a tornado shaped region that stretches far above the
QCP to about $20K$,  the resistivity follows a {\sl linear}
dependence} over more than three decades.  The
QCP thus represents a kind of ``singularity'' in the
material phase diagram.

\fight=0.8\textwidth
%\fg{newfigs/qc0.eps}{(a) Color coded plot of the logarithmic derivative
\fg{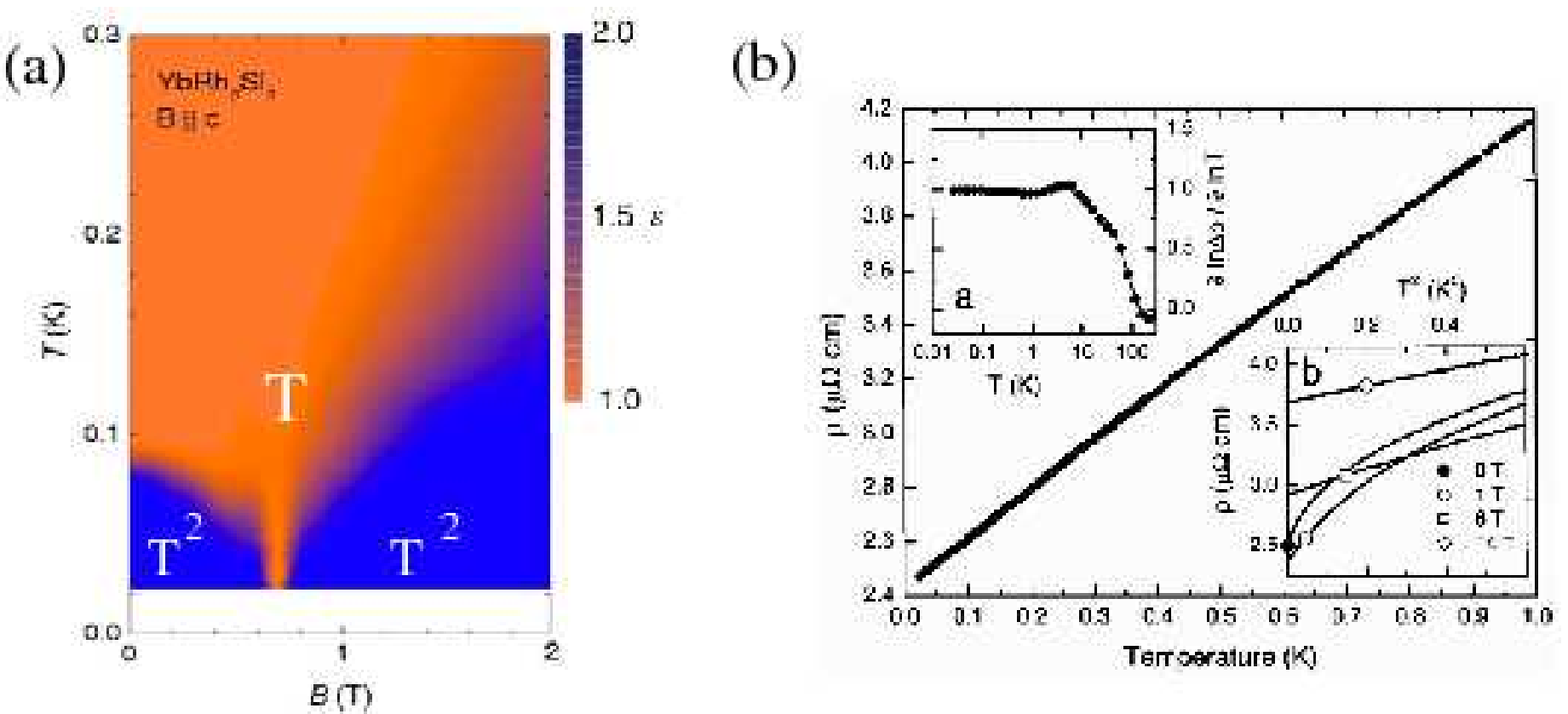}{(a) Color coded plot of the logarithmic derivative
of resistivity $d\ln \rho /d\ln  T$ after  \citep{custers}(b) Resistivity of $YbRh_{2}Si_{2}$
in zero magnetic field, after  \citep{trovarelli}. Inset shows
logarithmic derivative of resistivity}{fig38}

Experimentally, quantum critical heavy electron materials 
fall between two extreme limits that I shall call ``hard'' and
``soft'' quantum criticality. 
``Soft'' quantum critical systems 
are moderately well described in terms 
quasiparticles interacting with the soft quantum spin fluctuations
created by a spin-density wave instability.
Theory predicts \citep{moriya} that in a three dimensional metal, 
the quantum spin density wave fluctuations give rise
to a weak   $\sqrt{T}$ 
singularity in the low temperature behavior of the specific heat  coefficient
\[ 
\frac{C_{V}}{T}=\gamma_0-\gamma_{1}\sqrt{T}
\]
Examples of such behavior include
$CeNi_{2}Ge_{2}$ \citep{grosche,ceni2ge2} chemically doped $Ce_{2-x}La_{x}Ru_{2}Si_{2}$ and
``A''-type antiferromagnetic phases of  $CeCu_{2}Si_{2}$ at a
pressure-tuned QCP. 

At the other extreme, in ``hard'' quantum critical heavy materials, 
many aspects of the physics
appear consistent with a break-down of the Kondo effect
associated with a re-localization 
of the f-electrons into ordered, 
ordered local moments beyond the QCP.   
Some of the most
heavily studied examples of this behavior occur in 
the chemically tuned
QCP in  $CeCu_{6-x}Au_{x}$ \citep{hilbert,hilbert2,schroeder,schroeder2}.
and
$YbRh_{2}Si_{2-x}Ge_{x}$ \citep{custers,custers2}
and the field-tuned 
tuned QCP s of  $YbRh_{2}Si_{2}$ \citep{trovarelli} and $YbAgGe$\citep{budko,niklowitz,budko2,fak}
Hallmarks of hard quantum criticality 
include a logarithimically diverging specific heat coefficient at the
QCP, 
\begin{eqnarray}\label{l}
\frac{C_{v}}{T}&\sim & 
\frac{1}{T_{0}}\ln
\left(\frac{T_{0}}{T} \right)
\end{eqnarray}
and a quasi-linear resistivity
\begin{equation}\label{}
\rho (T)\sim T^{1+\eta }
\end{equation}
where $\eta $ is in the range $0-0.2$.  The most impressive results to
date have been observed at field-tuned quantum critical points 
in $YbRh_{2}Si_{2}$ and $CeCoIn_{5}$, where linear
resistivity has been seen to extend 
over more than two decades of temperature at the field-tuned
quantum critical point \citep{steglich,paglione,paglione2,capan}.
Over the range where 
linear , where the ratio between
the change in the size of the
resistivity $\Delta \rho $ to the zero temperature 
(impurity driven) resistivity $\rho_{0}$
\[
\Delta \rho/\rho_{0} >> 1.
\]
$CeCoIn_{5}$ is particularly interesting, for in this case, this
resistance ratio exceeds $10^{2}$ for current flow along the c-axis \citep{paglionethesis}.
This observation excludes any explanation of
that attributes the unusual resistivity to an interplay between
spin-fluctuation scattering and impurity scattering
 \citep{roschy}. 
Mysteriously, $CeCoIn_{5}$ also exhibits a $T^{3/2}$ resistivity for 
resistivity for current flow in the basal plane below about $2K$ \citep{paglionethesis}.
Nakasuji, Fisk and Pines\cite{nakasuji} have proposed that this kind of behavior
may derive from a {\sl two fluid} character to the underlying
conduction fluid. 

In quantum critical
$YbRh_{2}Si_{2-x}Ge_{x}$, the specific heat coefficient
develops a $1/T^{1/3}$ divergence at the very lowest temperature.
In the approach to a quantum critical point, Fermi liquid behavior 
is confined to an ever-narrowing range of temperature. Moreover, both the linear
coefficient of the specific heat and the 
the quadratic coefficient $A$ of  the resistivity  appear to diverge
 \citep{devisser,trovarelli}.
Taken together, these results suggests that the Fermi temperature renormalizes
to zero and the quasiparticle effective masses diverge 
\begin{equation}\label{}
T_{F}^{*}\rarrow 0, \qquad \frac{m^{*}}{m}\rarrow \infty 
\end{equation}
at the QCP of these three dimensional materials.
A central property of the Landau
quasiparticle, is the existence of a finite overlap ``$Z$'', or ``wavefunction
renormalization''  between a single quasiparticle state, denoted by
$\vert \hbox{qp}^{-}\ra$ and a bare electron state 
denoted by $\vert e^{-
}\ra = c\dg _{{\bf k }\sigma}\vert 0\ra$, 
\begin{equation}\label{}
Z= \vert \la e^{- }\vert \hbox{qp}^{-}\ra \vert ^{2} \sim \frac{m}{m^{*}}
\end{equation}
If the quasiparticle mass diverges,  the overlap
between the quasiparticle and the electron state from which it is
derived is driven to zero, signaling a complete break-down in the
quasiparticle concept at a ``hard'' quantum critical point \citep{revvarma2001}.

\vspace{0.1 truein}
\newcommand{\entre}[7]{
 &\renp {2}{27mm}{#1} 
& \renp{2}{22mm}{#2} 
&\renp{2}{18mm}{\small #3} 
&\renp{2}{25mm}{\small #4} 
&\renp{2}{15mm}{#5}
&\renp{2}{25mm}{#6} 
&\renp{2}{15mm}{#7 } 
\\
&&&&&&&\\}
\begin{center}
\centerline{\bf Table. C. Selected Heavy Fermion compounds with quantum critical points.}
\vskip 0.1 truein
\begin{tabular}{|r||c|c|c|c|c|c|c|}
\hline
&Compound & $x_c$/$H_c$ & $\frac{C_v}{T}$ & $\rho\sim T^a $ &
$\Gamma(T)=\frac{\alpha}{C_P}$& Other& Ref.\\
\hline
\hline
\multirow{10}{15mm}{\\
$\ $
 \\
$\ $
\\
\bf ``hard''
 \\
$\ $
\\
$\ $
\\
$\ $
 \\
$\ $
\\
$\ $
 \\
$\ $
\\
{\bf\  ``soft''}  \\
$\ $
}
\entre{$CeCu_{6-x}Au_x $}{$x_c=0.1$}{ 
$\frac{1}{T_0} \ln \left(
\frac{T_o}{T}\right)
$}{$ T + c $ }{-}{$\chi''_{{\bf Q_{0} }}
 (\omega,T)=\frac{1}{T^{0.7}}F\left[\frac{\omega}{T} \right]$}{[1]}
\cline{2-8}
\entre{$YbRh_2Si_{2}$}{$B_{c\parallel}=0.66T$}{}{ $ T$ }{-}{Jump in Hall constant.}{[2]}
\cline{2-8}
\entre{$YbRh_2Si_{2-x}Ge_{x}$}{$x_{c}=0.1$}{
$\frac{1}{T^{1/3}}\leftrightarrow \frac{1}{T_0} \ln \left(
\frac{T_o}{T}\right)$}{
 $ T$}{$T^{-0.7}$}{}{[3]}
\cline{2-8}
\entre{$YbAgGe$}{$B_{c\parallel}=9T$ $B_{c\perp}=5T$}{
$ \frac{1}{T_0} \ln \left(
\frac{T_o}{T}\right)$}{
 $ T$}{-}{NFL over range of fields}{[4]}
\cline{2-8}
\entre{$CeCoIn_5$ }{ $ B_c=5T$  }{ $\frac{1}{T_0}
\ln\left(\frac{T_0}{T}\right)$ }{  $T$/ $T^{1.5}$ }{ -
}{$\rho_{c}\propto T $,\ $\rho_{ab}\propto T^{1.5}$}{[5]}
\cline{2-8}
\entre{$CeNi_2Ge_2$ }{ $P_c =0$  }{$\gamma_0- \gamma_1\sqrt{T}$}{ $T^{1.2-1.5}$ }{ $T^{-1}$ }{}{[6]}
\hline
\hline
\end{tabular}
\end{center}
\vspace{0.1 truein}
{\sl \small
[1]  \citep{hilbert,hilbert2,schroeder,schroeder2},
[2]  \citep{trovarelli,silke},
[3]  \citep{custers,custers2},
[4]  \citep{budko,niklowitz,budko2,fak}
[5]  \citep{paglione,paglione2,capan,paglionethesis},
[6]  \citep{grosche,ceni2ge2}.
}

Table C. shows a tabulation of selected quantum critical materials. 
One  interesting variable that exhibits singular behavior at
both hard and soft quantum critical points, is the 
Gr\" uneisen
 ratio. This quantity, defined
as the ratio
\[
\Gamma = \frac{\alpha }{C} = - \frac{1}{V}
\left. 
\frac{\partial \ln
T}{\partial P}\right|_{S} \propto \frac{1}{T^{\epsilon}}
\]
of the thermal
expansion coefficient $\alpha = \frac{1}{V}\frac{dV}{dT}$ to the
specific heat $C$, diverges at a QCP.  The Gr\" uneisen
ratio is a sensitive measure of the rapid acquisition of entropy on
warming away from QCP. Theory predicts that 
$\epsilon=1$ at a 3D SDW QCP
, as seen in $CeNi_{2}Ge_{2}$. In the ``hard''
quantum critical material $YbRh_{2}Si_{2-x}Ge_{x}$, $\epsilon=0.7$
indicates a serious departure from a 3D spin density wave instability \citep{ceni2ge2}.

\subsection{Quantum, versus classical criticaltiy. }\label{}

Figure
\ref{fig39}. illustrates some key distinctions
between classical and quantum criticality \citep{subirchapter}. 
Passage through a classical 
second-order phase transition is achieved by tuning the temperature. Near the transition, 
the imminent arrival of order 
is signaled by the growth of droplets  of 
nascent order whose typical size 
$\xi$ diverges at the critical point.
Inside  each droplet, fluctuations of the order parameter 
exhibit a universal powerlaw dependence on distance
\begin{equation}\label{}
\langle \psi (x)\psi (0) \rangle \sim 
\frac{1}{x^{d-2+\eta}},\qquad  (x<< \xi).
\end{equation}
Critical matter ``forgets'' about its microscopic
origins, for its  thermodynamics, scaling laws and
correlation exponents associated with critical matter are so robust and universal that they recur
in such diverse contexts as the Curie point of iron or the critical
point of water. 
At a conventional critical point order parameter fluctuations
are ``classical'', for 
the  characteristic energy of the 
critical modes $\hbar \omega (q_{0})$, evaluated at a wavevector $q_{0}\sim \xi^{-1}$, 
innevitably drops below the thermal energy $\hbar \omega (q_{0})<< k_{B}T_{c}$ as $\xi\rightarrow \infty $.

In the seventies various authors, notably  Young \citep{young} and 
 Hertz \citep{hertz} recognized that  when if the transition
 temperature of a continuous phase transition can be depressed to
 zero, the critical modes become quantum mechanical in nature.
The partition function for a quantum phase transition  is described 
by a Feynman integral over order parameter configurations {$\{\psi (x,\tau
)  \}$}
in both space {\bf and} imaginary time  
 \citep{subirchapter,hertz}
\begin{eqnarray}\label{l}
Z_{quantum}= \sum_{\hbox{\small space-time configs}}\hskip -0.3in e^{-S[\psi ]},
\end{eqnarray}
where the action 
\begin{equation}\label{}
S[\psi ] = \int_{0}^{\frac{\hbar}{k_{B}T}}d\tau  \int _{-\infty}^{\infty}d^d{x}L[\psi (x,\tau )],
\end{equation}
contains an integral of the Lagrangian $L$
over an infinite range in space, but a {\sl finite } time interval
\[
l_{\tau }\equiv \frac{\hbar }{k_{B}T}.
\]
Near a QCP, bubbles of quantum critical matter
form within a metal, with  finite size $\xi_{x}$ and duration $\xi_{\tau }$. 
(Fig. \ref{fig39}).  These two quantities diverge as the quantum
critical point is approached, but 
the rates of divergence
are related by a dynamical critical exponent \citep{hertz}, 
\[
\xi_{\tau }\sim (\xi_{x})^{z}
\]
One of the consequences of this scaling behavior, is that time counts
as $z$ spatial dimensions, $[\tau ]=[L^z]$ in general. 

\fight=0.8\textwidth
\fg{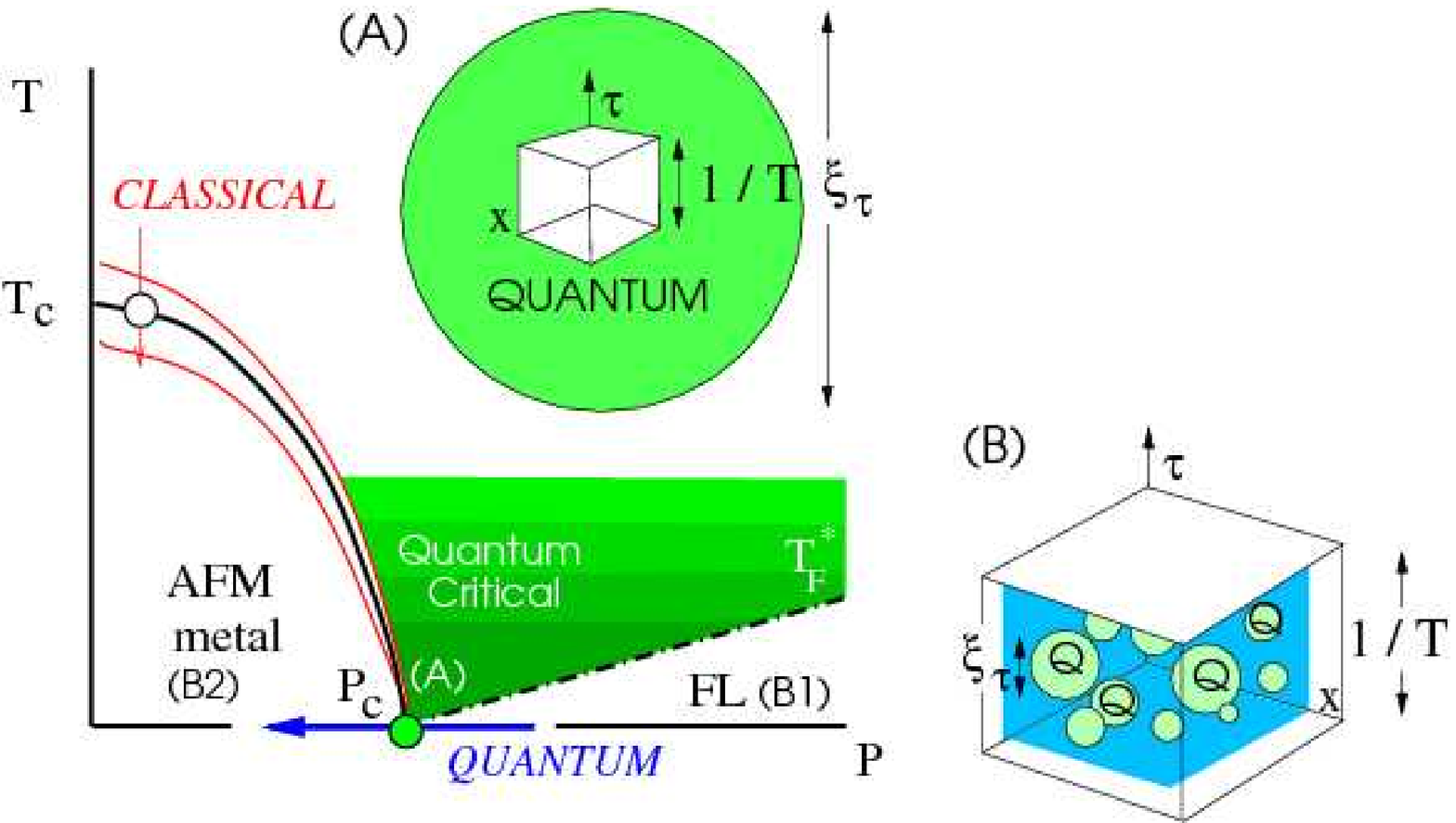}{
%\fg{newfigs/qcz.eps}{
Contrasting classical and quantum criticality in heavy electron
systems. At a QCP an external  parameter $P$, such as pressure or magnetic
field replaces temperature as the ``tuning parameter''. 
Temperature assumes the new role of a finite size
cutoff $l_{\tau }\propto 1/T$
on the temporal extent of quantum fluctuations.
(A) Quantum critical regime, where $l_{\tau }< \xi_{tau}$ probes the
interior of the quantum critical matter
(B) Fermi liquid regime, where $l_{\tau }>\xi_{\tau }$, where like
soda, bubbles of quantum critical matter fleetingly form within a
Fermi liquid that is paramagnetic (B1), or antiferromagnetically ordered (B2). }{fig39}

At a classical critical point, temperature is a tuning parameter that
takes one through the transition. 
The role of temperature is fundamentally different at a quantum
critical point: for it sets the scale
$l_{\tau}\sim 1/T$ in the time direction, introducing a {\sl finite
size correction} to the quantum critical point.
When the temperature is raised, $l_{\tau }$ reduces and the quantum
fluctuations are probed  on shorter and shorter time-scales.
There are then two regimes to the phase diagram, 
\begin{equation}\label{l}
(A)\qquad  \hbox{Quantum critical: }\qquad l_{\tau } << \xi_{\tau }   \qquad \qquad 
\end{equation}
where the physics probes the ``interior'' of the quantum 
critical bubbles, and 
\begin{equation}\label{l}
(B)\qquad \hbox{Fermi liquid/AFM } \qquad l_{\tau } >> \xi_{\tau }   \qquad \qquad 
\end{equation}
where the physics probes the quantum fluid ``outside'' the quantum
critical bubbles. 
The quantum fluid that forms in this region is a sort 
of ``quantum soda'', containing
short-lived bubbles of quantum critical matter surrounded by a
paramagnetic (B1)   or antiferromagnetically ordered (B2) Landau Fermi liquid.
Unlike  a classical phase transition, in which the critical
fluctuations are confined to a narrow region either side of the
transition, in a quantum critical region (A), fluctuations persist
up to temperatures where $l_{\tau }$ becomes comparable the 
with the microscopic short time cut-off
in the problem \citep{kopp} (which for heavy electron systems is most
likely, the single-ion Kondo temperature $l_{\tau }\sim \hbar /T_{K}$). 

\subsection{Signs of a new universality.}\label{}

\fight=\textwidth
%\fg{newfigs/dhva.eps}{(a) Hall crossover line for sudden evolution of
\fg{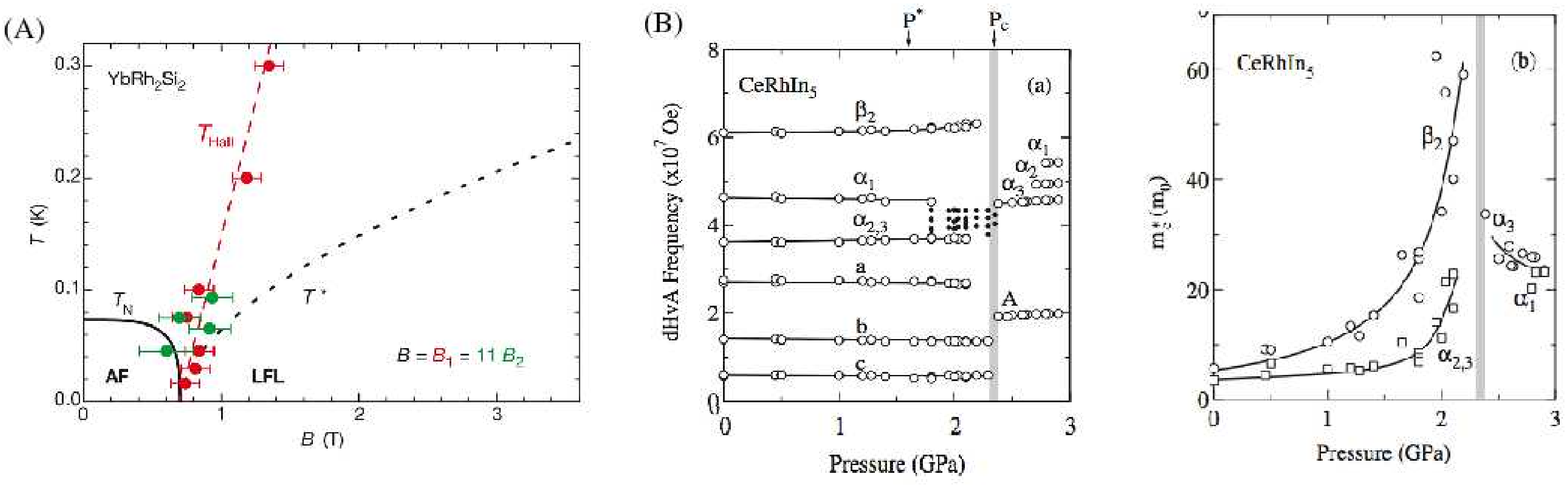}{(a) Hall crossover line for sudden evolution of
Hall constant in $YbRh_{2}Si_{2}$ after
 \citep{silke}. (b) Sudden change in dHvA frequencies and divergence of
quasiparticle effective masses at pressure tuned, finite
field QCP in $CeRhIn_{5}$ after  \citep{cerhindhva}.}{fig40}

The discovery of quantum criticality in heavy electron systems
raises an alluring possibility of {\bf quantum critical matter}, a
universal state of matter that like its classical counterpart, forgets
its microscopic, chemical and electronic  origins.  There are three
pieces of evidence that are particularly fascinating in this respect:
\begin{enumerate}

\item Scale invariance, as characterized by 
$E/T$ scaling in the quantum-critical inelastic spin fluctuations observed in 
$CeCu_{1-x}Au_{x}$ \citep{schroeder,schroeder2}. ($x=x_{c}=0.016$),
\[
\chi''_{\bQ_{0}} (E,T) = \frac{1}{T^{a}} F (E/T)
\]
where $a\approx 0.75$ and $F[x]\propto (1-ix)^{-a}$. Similar behavior
has also been seen in powder samples of $U Cu_{5-x} Pd_x$ \citep{aronson}.

\item A jump in the Hall constant of
$YbRh_{2}Si_{2}$ when field tuned through its quantum critical point \citep{silke}.
(see Fig. \ref{fig40} (a)).

\item  A sudden  change in the area of the extremal Fermi surface
orbits observed by de Haas van Alphen at a pressure tuned QCP in $CeRhin_{5}$ \citep{cerhindhva}.
(see Fig. \ref{fig40} (b)).

\end{enumerate}
Features (2) and (3) suggest
that the Fermi surface 
jumps from a ``small'' to ``large'' Fermi surface as 
the magnetic order is lost, as if 
the phase
shift associated with the Kondo effect collapses to zero at the
critical point, as if the f-component of the electron fluid
Mott-localizes at the transition. 
To reconcile a sudden change in the Fermi
surface with a second-order phase transition, 
we are actually forced to infer that 
the quasiparticle weights vanish at the QCP. 

These features are quite incompatible with a spin density wave QCP. In
a spin density wave scenario, 
the Fermi surface and Hall constant are expected to evolve continuously through a
QCP. Moreover, in an SDW description,
the dynamical critical exponent is $z=2$ so time counts as $z=2$
dimensions in the scaling theory, and the effective
dimensionality $D_{eff}= d+2>4$ lies above the upper critical dimension,
where mean-field theory is applicable and scale invariant behavior
is no longer expected. 

These observations have ignited a ferment of theoretical
interest in the nature of heavy fermion criticality. 
We conclude with a brief discussion  of some 
of the competing ideas currently under consideration.

\subsubsection{Local quantum criticality }\label{}

One of the intriguing observations \citep{schroeder} in $CeCu_{6-x}Au_{x}$, is that 
the uniform magnetic susceptibility, 
$\chi^{-1}\sim T^{a}+ C $, $a=0.75$  displays the same  powerlaw dependence on
temperature observed in the inelastic neutron scattering at the
critical wavevector $\bQ_{0}$.
A more detailed set of measurements by Schroeder et al. \citep{schroeder2}  revealed that the scale-invariant component
of the 
dynamical spin susceptibility appears to be momentum
independent, 
\begin{equation}\label{lab1}
\chi^{-1} ({\bf q},E) =  T^{a}[\Phi (E/T)]+\chi _{0}^{-1} ({\bf q}).
\end{equation}
This behavior suggests that the critical behavior associated with the heavy fermion QCP
contains some kind of {\sl local} critical excitation \citep{schroeder,sces98}. 

One possibility, is that this  local
critical excitation is the spin itself, so that 
 \citep{sces98,sachdevje,sengupta}
\begin{equation}\label{}
\langle S (\tau )S (\tau ')\rangle =\frac{1}{(\tau -\tau ')^{2-\epsilon
}},
\end{equation}
is a power-law, but 
where $\epsilon \ne 0$ signals non- Fermi liquid behavior.
This  is the basis 
of the 
```local quantum criticality'' theory developed by 
Si, Ingersent, Rabello and Smith \citep{smith,qmsi,qmsi2}. 
This theory requires that 
the local spin susceptibility $\chi_{loc}=
\sum_{\bq}\chi(\bq,\omega)_{\omega=0}$ diverges at a heavy fermion
QCP. Using an extension of the methods of dynamical mean-field
theory \citep{krauth,kotliarrmp2}, Si et al. find that it is possible 
to account for the local scaling form of the dynamical susceptibility, obtaining
exponents that
are consistent with the observed properties of $CeCu_{6-x}Au_{x}$ \citep{grempel}.

However, there are some significant difficulties with this theory.
First, as a local theory, the quantum
critical fixed point of this model is 
expected to possess a finite zero-point entropy per spin, 
a feature that is to date, inconsistent with thermodynamic measurements \citep{custers}.
Second, the requirement of a divergence in the local spin
susceptibility imposes the requirement that the surrounding spin fluid
behaves as layers of decoupled two dimensional spin fluids.
By expanding $\chi_{0}^{-1} (\bq )$ (\ref{lab1})
about the critical wavevector $\bQ $, one finds that the singular
temperature dependence in  the local 
susceptibility is given by 
\begin{equation}\label{}
\chi _{loc} (T)\sim \int d^{d }q \frac{1}{({\bf q}-{\bf  Q})^{2} +
T ^{\alpha
}}\sim T^{(d-2)\alpha /2}, 
\end{equation}
requiring that $d\le 2$.

My sense, is that the validity of the original scaling by Schroeder et
al still stands and that these difficulties stem from a  mis-identification
of the critical local modes driving the scaling seen by neutrons.
One possibility, for example, is that the right soft variables are not spin per-se,
but the fluctuations of the phase shift associated with the Kondo effect.
This might open up the way to an alternative formulation of local criticality.

\subsubsection{Quasiparticle fractionalization and deconfined criticality.}\label{}

One of the competing sets of ideas under consideration at present, is
the idea that in the process of localizing into an ordered magnetic
moment, the composite heavy electron breaks
up into constituent spin and charge components. Loosely speaking,
\begin{equation}\label{}
e^{-}_{\sigma }\rightleftharpoons s_{\sigma } + h^{-}
\end{equation}
where $s_{\sigma }$ represents a neutral spin-1/2 excitation, or ``spinon''.
This has led to  proposals \citep{revqcp2,vojta,pepin} 
that gapless spinons develop at the quantum critical
point. This  idea is faced with a
conundrum, for even if free neutral spin-1/2 excitations can exist
at the QCP, they must surely  be confined as one tunes away from this
point, back into the Fermi liquid. According to the model of
``deconfined criticality'' proposed 
by Senthil,Vishwanath, Balent, Sachdev and
Fisher \citep{deconfinedcrita}, 
the spinon confinement  scale $\xi_{2}$ introduces
a second diverging length scale to
the phase transition, 
where $\xi_{2}$ diverges more
rapidly to infinity than $\xi_{1}$. 
One possible realization
of this proposal is the quantum melting
of two dimensional $S=1/2 $ Heisenberg antiferromagnet,
where the smaller  correlation length $\xi_1$ is associated with the
transition from antiferromagnet to spin liquid, and the second
correlation length $\xi_{2}$ is associated with the confinement of spinons to
form a valence bond solid (Fig. \ref{fig41}).
\fight=0.6\textwidth
%\fg{newfigs/senthil.eps}{``Deconfined
\fg{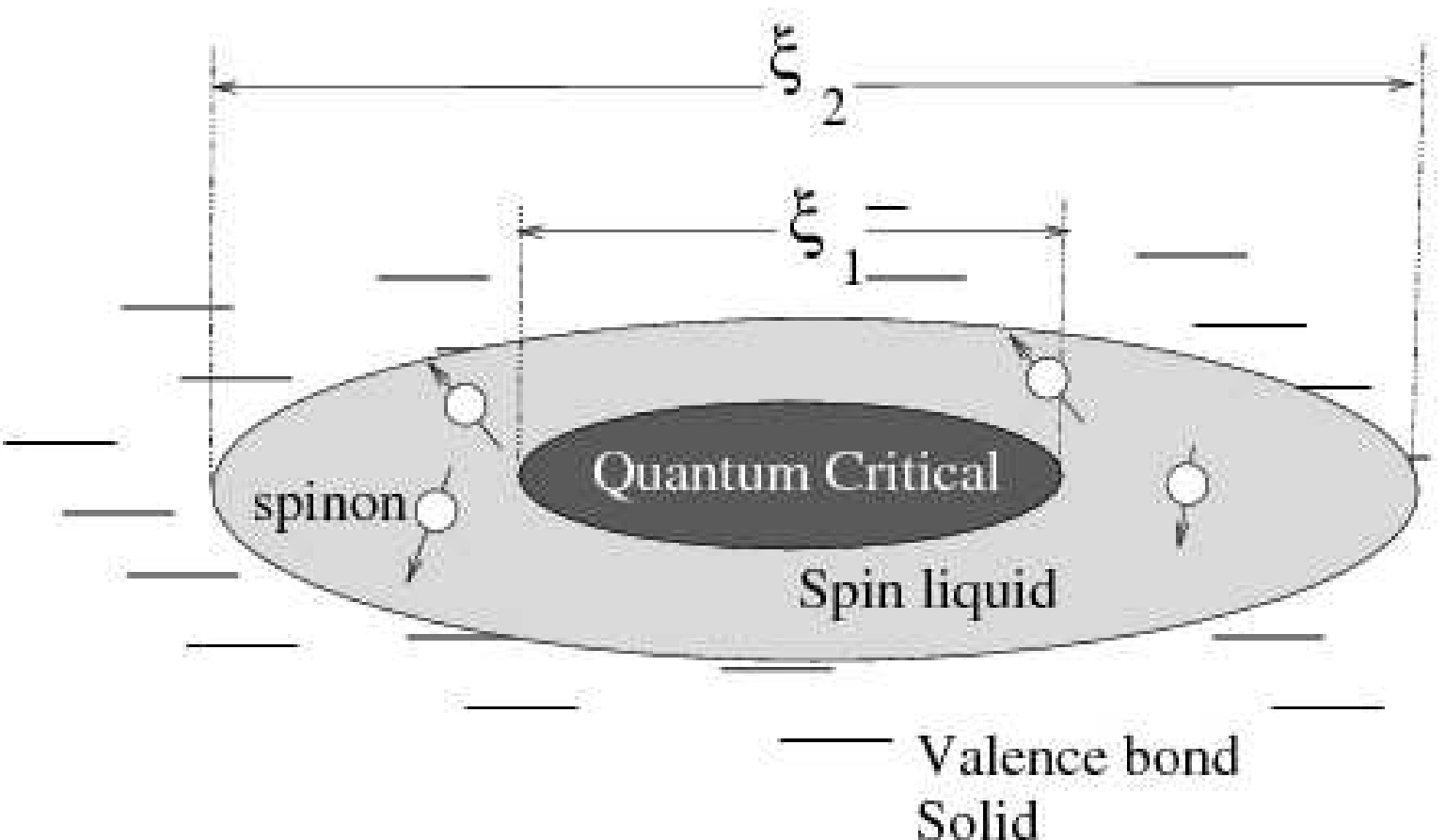}{``Deconfined
criticality'' \citep{deconfinedcrita}.
The quantum critical droplet is defined by two divergent length scales - $\xi_{1}$ governing the spin
correlation length, $\xi_{2}$ on which the spinons confine, in the
case of the Heisenberg model, to form a valence bond solid. 
}{fig41}

It is not yet clear how this scenario will play out for heavy
electron systems. Senthil, Sachdev and Vojta \citep{deconfinedcritb}
have proposed that in a heavy electron system, the intermediate spin
liquid state may involve a Fermi surface of neutral (fermionic) spinons co-existing with a
small Fermi surface of conduction electrons which they call an
FL$^{*}$ state.  In this scenario, the quantum critical point involves
an instability of the heavy electron fluid to the FL$^{*}$ state,
which is subsequently unstable to antiferromagnetism. Recent work
suggests that Hall constant can indeed jump at such a
transition \citep{marston}.

\subsubsection{Schwinger Bosons}\label{}

A final approach to quantum criticality 
currently under development, attempts to forge a kind of ``spherical
model'' for the antiferromagnetic quantum critical point 
through the use of a large $N$ expansion in which the
spin is described by Schwinger bosons, rather than
fermions \citep{AA88,PG97}, 
\[
S_{ab}= b\dg _ab_b-\delta_{ab}\frac{n_{b}}{N}
\]
where the spin $S$ of the moment is determined by the constraint 
$n_{b}=2S$ on the total number of bosons per site.
Schwinger bosons are well-suited to describe low dimensional magnetism \citep{AA88}.  However,
unlike fermions, only one boson can enter a Kondo singlet. To obtain
an energy that grows with $N$, Parcollet and
Georges proposed a new class of large $N$ expansion based
around the multi-channel Kondo model with $K$ channels \citep{PG97}, 
where $k=K/N$ is kept fixed. The Kondo interaction takes the form
\begin{equation}\label{}
H_{int } = \frac{J_{K}}{N}\sum_{\nu=1,K, \ \alpha, \ \beta }
S_{\alpha \beta }c\dg _{\nu \beta\mu}c_{\nu\alpha}
\end{equation}
where the channel index $\nu$ runs from one to $K$.  When written in
terms of Schwinger bosons, this interaction can be factorized in terms
of a charged, but spinless exchange fermion $\chi_{\nu}$ (``holon''), as follows
\begin{equation}\label{l}
H_{int}\rightarrow \sum_{\nu\al}
\frac{1}{\sqrt{N}}\left[( c\dg _{\nu\alpha }b_{\alpha} )
\chi\dg _{\nu}
+{\rm  H.c.} 
\right] + \sum_{\nu}\frac{\chi\dg _{\nu}\chi_{\nu}}{J_{K}}.
\end{equation}
Parcollet and
Georges originally used this method to study the over-screened Kondo
model \citep{PG97}, where $K>2S$. 

Recently, it has proved possible to find the Fermi liquid
large $N$ solutions to the fully screened  Kondo impurity model, where the
number of channels is commensurate with the number of
bosons ($K=2S$) \citep{rech,lebanon2}.  One of the intriguing features of these
solutions, is the presence of a gap for spinon excitations, roughly
comparable with the Kondo temperature.
Once antiferromagnetic interactions are introduced,
the spinons pair-condense,
forming a state with a
{\sl large} Fermi surface, but one that co-exists with gapped spinon (and holon)
excitations \citep{indranil05}.   

\fight=0.9\textwidth
\fg{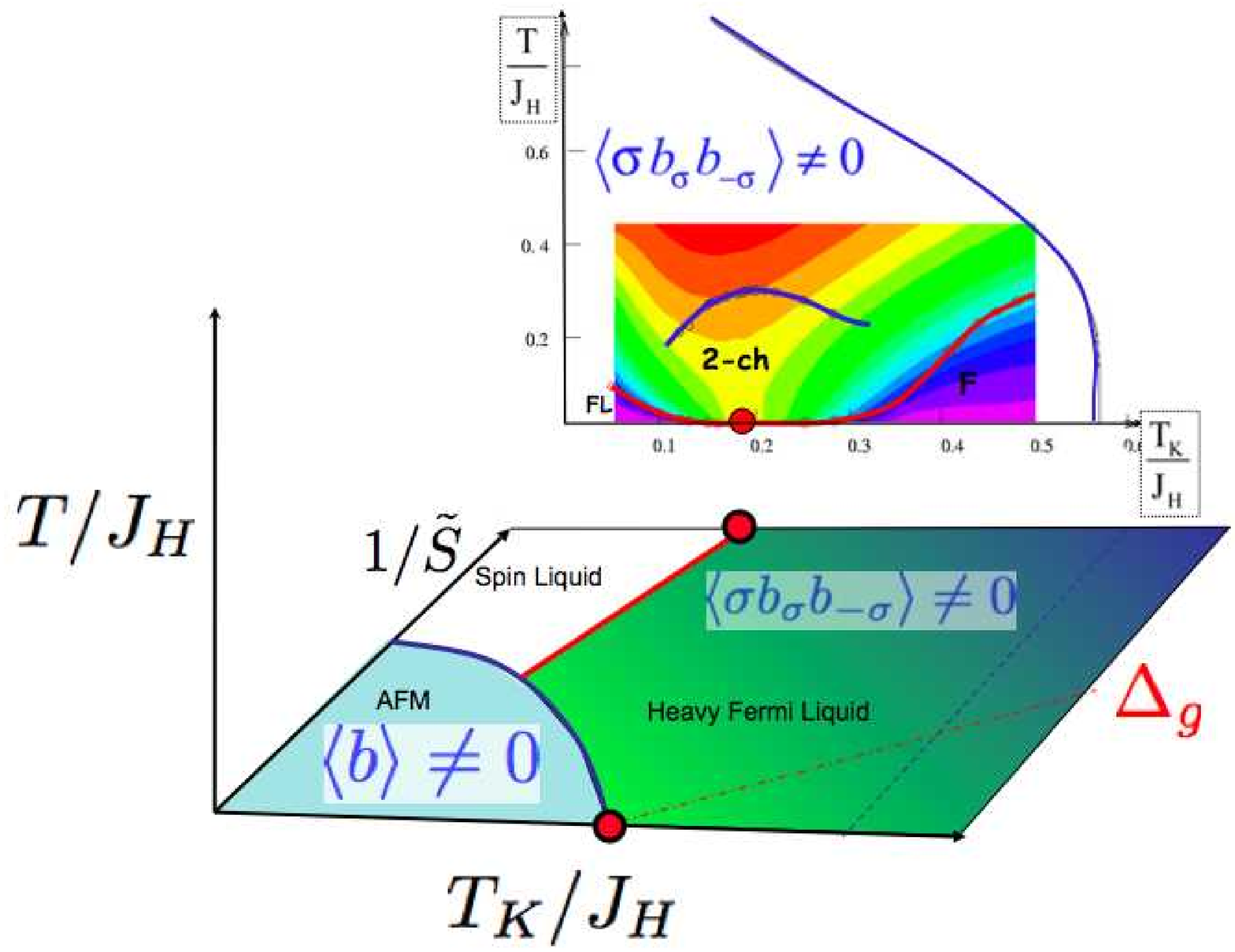}{Proposed phase diagram for the large $N$ limit
%\fg{newfigs/bigdream.eps}{Proposed phase diagram for the large $N$ limit
of the two impurity and Kondo lattice models.  Background- the two
impurity model, showing contours of constant entropy as a function of
temperature and the ratio of the Kondo temperature to Heisenberg
coupling constant, after  \citep{rech}. Foreground, proposed phase
diagram of the fully screened, multi-channel Kondo lattice, where $\tilde{S}$  is the spin of the
impurity. At small $\tilde{S}$, there is a phase transition between a
spin liquid and heavy electron phase.  At large $\tilde{S}$, a phase
transition between the antiferromagnet and heavy electron phase. If
this phase transition is continuous in the large $N$ limit, then both the
spinon and holon gap are likely to close at the quantum critical
point \citep{lebanon}. }{fig42}
The gauge symmetry associated with
these particles guarantees that if the gap for the  spinon goes to
zero continuously, then the gap for the holon must also go to zero. This
raises the possibility that gapless charge degrees of
freedom may develop at the very same time as magnetism (Fig. \ref{fig42} ). 
In the two impurity model, Rech et al have recently shown that the
large $N$ solution  contains a ``Jones Varma'' quantum critical point
where a static valence bond forms between the Kondo impurities. At
this point, the holon and spinon excitations become gapless. Based on this
result, Lebanon and Coleman \citep{lebanon} have recently proposed that the holon spectrum
may become gapless at the magnetic quantum critical point (Fig. \ref{fig42}.) in three
dimensions.

\section{Conclusions and Open Questions}\label{}

I shall end this chapter with a brief list of open questions in the
theory of heavy fermions. 

\begin{enumerate}

\item To what extent does the mass enhancement in heavy electron
materials owe its size to the vicinity to a nearby quantum phase
transitions? 

\item What is the microscopic origin of heavy fermion
superconductivity and the in extreme cases $UBe_{13}$ and
$PuCoGa_{5}$ how does the pairing relate to both spin quenching and
the Kondo effect? 

\item What is the origin of the linear resistivity and the logarithmic
divergence of the specific heat at a ``hard'' heavy electron quantum critical
point?  

\item What happens to magnetic interactions in a Kondo insulator, 
and why do they appear to vanish? 

\item In what new ways can the physics of heavy electron systems be
interfaced with the tremendous current developments in mesoscopics?  
The
Kondo effect is by now a well-established feature of Coulomb blockaded
quantum dots \citep{glazman}, but there may be many other ways in which
we can learn about local moment physics from mesoscopic experiments.
Is is possible, for example, to observe voltage driven quantum phase transitions in
a mesoscopic heavy electron wire?  This is an area ripe with
potential. 

\end{enumerate}
It should be evident that I believe there is tremendous  prospect for 
concrete progress on many of
these issues in the near future. I hope that in some ways, I have whet
your appetite enough to encourage you too try your hand at their 
at their future solution.

{\bf Acknowledgments:} This research was supported by the
National Science Foundation grant  DMR-0312495. I'd like to thank
E. Lebanon and T. Senthil for discussions related to this work. 
I'd also like to thank the Aspen Center for Physics, where part of the work
for this chapter was carried out. 

\bibliography{coleman_wiley}
\bibliographystyle{apsrmp}
%\begin{thebibliography}

%\end{thebibliography}
\end{document}